\documentclass{aa}

\usepackage{graphicx}
\usepackage{txfonts}
\usepackage[T1]{fontenc} 
\usepackage{siunitx}
\usepackage{amsmath}
\usepackage{amssymb}
\usepackage{longtable}  
\usepackage{tabularx} 
\usepackage[para]{footmisc}
\usepackage{lscape}
\usepackage[export]{adjustbox}
\usepackage{subcaption}
\usepackage{perpage}
\usepackage{alphalph}
\usepackage{hyperref}
\usepackage[capitalise]{cleveref} 
\usepackage{float}
\usepackage{xcolor}
\usepackage{caption}
\usepackage{threeparttablex}

\newcommand{\kms}{km.s$^{\rm -1}$}

\newcommand{\Mjup}{M$_{\rm Jup}$}

\newcommand{\vsini}{$v\sin{i}$}
\newcommand{\Msun}{M$_{\sun}$}

\newcommand{\Mstar}{M$_{\rm \star}$}
\newcommand{\bv}{$B-V$}
\newcommand{\msini}{$m_{\rm p}\sin{i}$}
\newcommand{\rhk}{log$R'_{\rm HK}$}
\newcommand{\safir}{S{\small AFIR}}
\newcommand{\sophie}{S{\small OPHIE}}
\newcommand{\harps}{H{\small ARPS}}
\newcommand{\hipp}{H{\small IPPARCOS}}

\newcommand{\sphere}{S{\small PHERE}}

\DeclareSIUnit\year{yr}
\DeclareSIUnit\arcsecond{as}
\DeclareSIUnit\MJ{M_{Jup}}
\DeclareSIUnit\au{au}
\DeclareSIUnit\jdb{d}
\DeclareSIUnit\day{days}
\DeclareSIUnit\hour{h}
\DeclareSIUnit\night{nights}
\DeclareSIUnit\arcsec{arcsec}
\DeclareSIUnit\pc{pc}
\DeclareSIUnit\msun{M_{\odot}}
\DeclareSIUnit\rearth{R_{\oplus}}
\makeatletter
\newcommand\footnoteref[1]{\protected@xdef\@thefnmark{\ref{#1}}\@footnotemark}
\makeatother

\begin{document} 

   \title{HARPS radial velocity search for planets in the Scorpius-Centaurus association}

   \author{A. Grandjean
          \inst{1}
          \and
          A.-M. Lagrange\inst{2,1,3}
        \and
        N. Meunier\inst{1}
                \and
        G. Chauvin\inst{1}
                \and
        S. Borgniet\inst{2}
                \and
        S. Desidera \inst{4}
                \and
        F. Galland \inst{1}
                \and
        F. Kiefer \inst{2}
                \and
        S. Messina \inst{5}
                \and 
        Iglesias, D. \inst{6}
                \and
        B. Nicholson\inst{7,8}
                \and
        B. Pantoja \inst{9}
        \and
        P. Rubini\inst{10}
                \and
        E.  Sedaghati \inst{11}
                \and
        M. Sterzik \inst{12}
        \and
        N. Zicher  \inst{7}
          }

   \institute{
Univ. Grenoble Alpes, CNRS, IPAG, 38000 Grenoble, France
 \\
\email{Antoine.Grandjean1@univ-grenoble-alpes.fr}
\and 
LESIA, Observatoire de Paris, Université PSL, CNRS, Sorbonne Université, Université de Paris, 5 place Jules Janssen, 92195 Meudon, France
\and
IMCCE – Observatoire de Paris, 77 Avenue Denfert-Rochereau, 75014 Paris, France
\and
INAF-Osservatorio Astronomico di Padova, Vicolo dell’Osservatorio 5, Padova, Italy, 35122-I
\and
INAF–Osservatorio Astrofisico di Catania, via Santa Sofia, 78 Catania, Italy 
\and
School of Physics and Astronomy, Sir William Henry Bragg Building, University of Leeds, Leeds LS2 9JT, UK
\and
Sub-department of Astrophysics, Department of Physics, University of Oxford, Denys Wilkinson Building, Keble Road, Oxford,OX1 3RH, UK
\and
University of Southern Queensland, Centre for Astrophysics, WestStreet, Toowoomba, QLD 4350 Australia
\and
Departamento de Astronomía, Universidad de Chile, Camino al Observatorio, Cerro Calán, Santiago, Chile
\and
Pixyl, 5 Avenue du Grand Sablon, 38700 La Tronche, France
\and
European Southern Observatory, Alonso de Córdova 3107, Santiago, Chile
\and
European Southern Observatory, Karl-Schwarzschild-Str 1, D-85748 Garching, Germany
             }


   \date{Received 1 May 2021 / Accepted 26 August 2022}

 
  \abstract
   {
   The Scorpius-Centaurus (Sco-Cen) young and nearby massive star-forming region is particularly well suited for extrasolar planet searches with both direct imaging and radial velocity (RV) techniques. The RV search, however,  is challenging, as the stars are faster rotators on average than their older stellar counterparts of similar spectral types. Moreover, the RV time series show strong signatures of stellar variability (spots and faculae) and/or stellar pulsations. 
 }
   {Our aim is to search for giant planets (GPs) and brown dwarfs at short orbital distances around star members of the Sco-Cen association. We also aim at using these data together with others available on young stars to estimate the GP occurrence rate for young stars for periods of up to $\SI{1000}{\day}$. }
   {We used the High Accuracy Radial velocity Planet Searcher (\harps) spectrograph on the $\SI{3.6}{\meter}$ telescope at the La Silla Observatory to monitor $88 \ A-F$ Sco-Cen stars. 
 To improve our statistics and analysis, we combined this survey with two previous surveys that focused on young nearby stars (YNS) to compute companion occurrence rates from a sample of $176$ young $A-M$ stars.}
   {We report the discovery of a massive hot-Jupiter candidate around HD 145467, together with the discovery of one probable short-period ($P<\SI{10}{\day}$) brown dwarf around HD 149790.
In addition, we confirm the binary nature  of eight single-line binaries: HD 108857, HD 108904, HD 111102, HD 114319, HD 121176, HD 126488, HD 126838, and HD 133574.

From our sample, we obtain a GP ($m_c\in[1;13]\si{\MJ}$) occurrence rate of  $0.7_{-0.2}^{+1.6} \ \%$ for periods between $1$ and  $\SI{1000}{\day}$ and a brown dwarf ($m_c\in[13;80]\si{\MJ}$) occurrence rate  of $0.6_{-0.2}^{+1.4} \ \%$,  in the same period range. In addition, we report a possible lack of close ($P\in[1;1000]\ \si{\day}$) GPs around young F-K stars compared to their older counterparts, with a confidence level of $95\%$.
 }
   {}


   \maketitle
%

\section{Introduction} 

The discovery of the $5000$\footnote{\url{exoplanet.eu}} exoplanets and brown dwarf (BD) companions known to date has shown the importance of early stages of planet formation and evolution. This is the case, for instance, for hot Jupiters (HJs),  which did not form in situ, and moved in a second step toward the star. While models predict that giant planets (GPs)  form beyond a few $\si{\au}$ \citep{Pollack}, disk-planet interactions \citep{Migration-disk} or gravitational interactions with a third body may cause significant planet migration \citep{High_excentricity_migration} and may thus explain the HJs that are observed around solar- to late-type main-sequence (MS) stars. Radial velocity (RV) studies of young planets in particular can help constrain the timescales associated with planet migration.

Young stars have been poorly studied with RV techniques, however, because activity or pulsations usually induce high  RV variations (jitters), with  amplitudes up to a few $\si{\kilo\meter\per\second}$ \citep{Lagrange,Grandjean_HARPS}, which is much higher than the planet-induced signals. Activity has led to several false planet detections in the past \citep{Huelamo,Figueira,Soto}. Moreover, young stars are generally faster rotators than their older counterparts \citep{Stauffer_2016, Rebull_2016, Gallet_2015}, which leads to broader stellar absorptions, and hence, lower precision on the measured RV.

Previous RV surveys carried out on young stars ($<\SI{300}{\mega\year}$) showed no evidence for young HJs \citep{Esposito,Paulson,Grandjean_HARPS,Grandjean_SOPHIE}. Detections of young HJs were reported using spectropolarimetric techniques, but the detections are  still debated \citep{Carleo,Donati_2020,Damasso}. On the other hand, young HJs have been found by transit photometry \citep{Cameron,Tanimoto,Deleuil,Mann,Alsubai,David,Rizzuto}, and some of them were confirmed with RV methods \citep{Deleuil,Alsubai}.
Finally, no BDs with periods shorter than $\SI{10}{\day}$ were found with RV around young stars, although one was discovered via transit \citep{James}. Therefore, the occurrence rates of young HJs and young short-period BDs still need to be determined with accuracy.

We carried out three RV surveys on young stars from A to M types with the final aim of coupling RV data with direct-imaging  data. This will permit  exploring the stellar environments from a fraction of au to hundreds of au and then to estimate GP and BD occurrence rates in this semimajor axis (sma) range. The first, southern, survey was performed with the \harps\footnote{High Accuracy Radial velocity Planet Searcher} \ spectrograph on young nearby stars (YNS); its results are presented in \cite{Grandjean_HARPS}. The second, northern, survey was performed with the \sophie \   spectrograph and is reported in  \cite{Grandjean_SOPHIE}. The latter paper also presents a  statistical analysis that combines the results of the two surveys. Our third survey, also performed with the  \harps \ instrument, focuses\ on star members of the Scorpius-Centaurus (Sco-Cen) association. Focusing on a given association ensures a homogeneous sample in age, and allows us to study specific stages in the formation of planetary systems. At an age of $3-\SI{18}{\mega\year}$ \citep{Pecaut_16}, Sco-Cen is well suited for studying the early stages of planet formation. 
It is also of pivotal importance because : 
Several planets were imaged in the association: HD106906 b \citep{Bailey}, HD95086 b \citep{Rameau_95086}, UScoCTIO 108 b \citep{Bejar}, GSC06214-00210 b \citep{Ireland}, 1RXS J160929.1-210524 b \citep{Lafreniere_2008}, HD 116434 b \citep{Chauvin}, PDS 70 b and c \citep{Keppler_PDS70,Haffert}, TYC 8998-760-1 b and c \citep{Bohn_20a, Bohn_20b}, and TYC 8984-2245-1 \citep{bohn21}, as well as numerous BDs \citep{Hinkley};
 Moreover, numerous stars of this association present infrared (IR) excesses. In a growing number of  cases, disks that caused these excesses have been resolved;  they exhibit structures (gaps, rings, and spirals) that might be indicative of the presence of as yet unseen planets  \citep{Pecaut_16,Bonnefoy_17,Matthews_17,Keppler_PDS70,Garufi,Bohn_19};
 In addition, its relative proximity (Upper Scorpius: $\SI{145}{\pc}$; Upper Centaurus-Lupus: $\SI{140}{\pc}$; Lower Centaurus-Crux: $\SI{140}{\pc}$;  \cite{Zeeuw_scocen})  permits searching for GPs in direct imaging down to  $15-\SI{20}{\au}$  with current instrumentation  \citep{Vigan_2021}. RV studies nicely complement the searches for planets at shorter separations.

In this paper, we describe this \harps \ Sco-Cen   survey and present out results. We describe the sample in \cref{survey_description} and describe the detection of GP, BD, or stellar companions in \cref{comp}.
We finally combine the \harps \ and \sophie \ YNS surveys mentioned above with the present one for a global statistical analysis to derive GP and BD occurrence rates (\cref{HS}).

\section{Description of the Sco-Cen RV survey}
\label{survey_description}

\subsection{Sample}
\label{ScoCen_sample}
Our initial sample included $107$ star members of the Sco-Cen association with a distance lower than $ \SI{150}{\pc}$ according to their \hipp\footnote{Precision PARallax COllecting Satellite\hfill} \ parallaxes \citep{hipp2}. Most of our targets are also part of the \sphere\footnote{Spectro-Polarimetric High-contrast Exoplanet REsearch\hfill} \ GTO\footnote{Guaranteed Time Observations} SHINE\footnote{SpHere INfrared survey for Exoplanets} survey sample \citep{Chauvin_shine,Desidera_shine}.  We chose to limit our sample below $\SI{150}{\pc}$ to ensure the best detection limits with \sphere \ at separations larger than typically $\SI{15}{\au}$. The Gaia mission revised the individual distances of the targets, which now range from $86$ to  $\SI{177}{\pc}$ \citep{EDR3}. 
We then excluded the stars that were identified as binaries with a separation smaller than $\SI{2}{\arcsecond}$ to avoid contamination by the second component of the system in the $\SI{1}{\arcsecond}$ \citep{HARPS}) fiber. In addition, five targets\footnote{HD 143637, HD144667, HD 145132, HIP 75367, GSC 06214-00210} were not observed because telescope time was limited. 

We measured the projected rotational velocity  (\vsini) of the remaining $88$ stars  \footnote{\emph{cf.} \Cref{observable}}. The  \vsini \ of 19 stars is too high to allow computing the  RV  ($v\sin{i}>\SI{300}{\kilo\meter\per\second}$) or computing the bisector  ($v\sin{i}>\SI{150}{\kilo\meter\per\second}$)  \footnote{ HD 103589, HD 103599, HD 105613, HD 106218, HD 107821, HD 107920, HD 110058, HD 110634, HD 118878, HD 121835, HD 122259, HD 123798, HD 132723, HD 139883, HD 141190, HD 143567, HD 145468, HD 151721, and V* V853 Cen.}. They were therefore excluded from our analysis. Our final sample (hereafter referred to as the Sco-Cen final sample) accordingly includes $69$ targets. This sample has $4$ targets in common with the  \harps \ YNS survey: HD 95086, HD 102458, HD 106906, and HD 131399. 

\begin{figure*}[ht!]
  \centering
\begin{subfigure}[t]{0.23\textwidth}
\includegraphics[width=1\hsize]{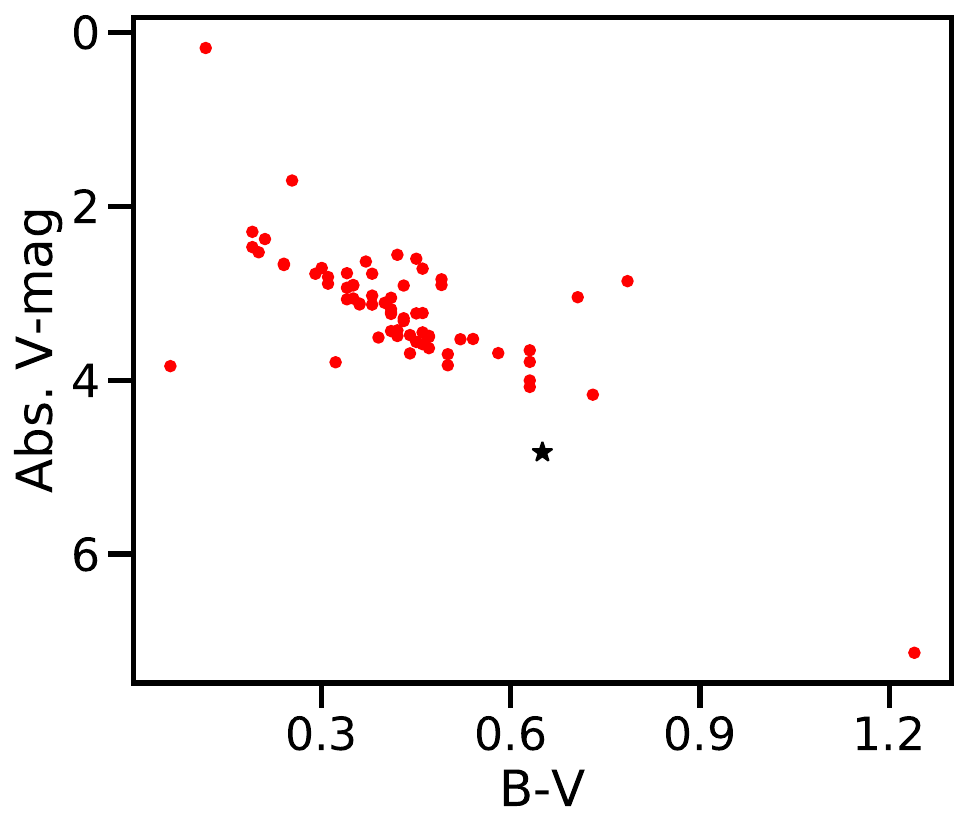}
\caption{\label{HR}}
\end{subfigure}
\begin{subfigure}[t]{0.24\textwidth}
\includegraphics[width=1\hsize]{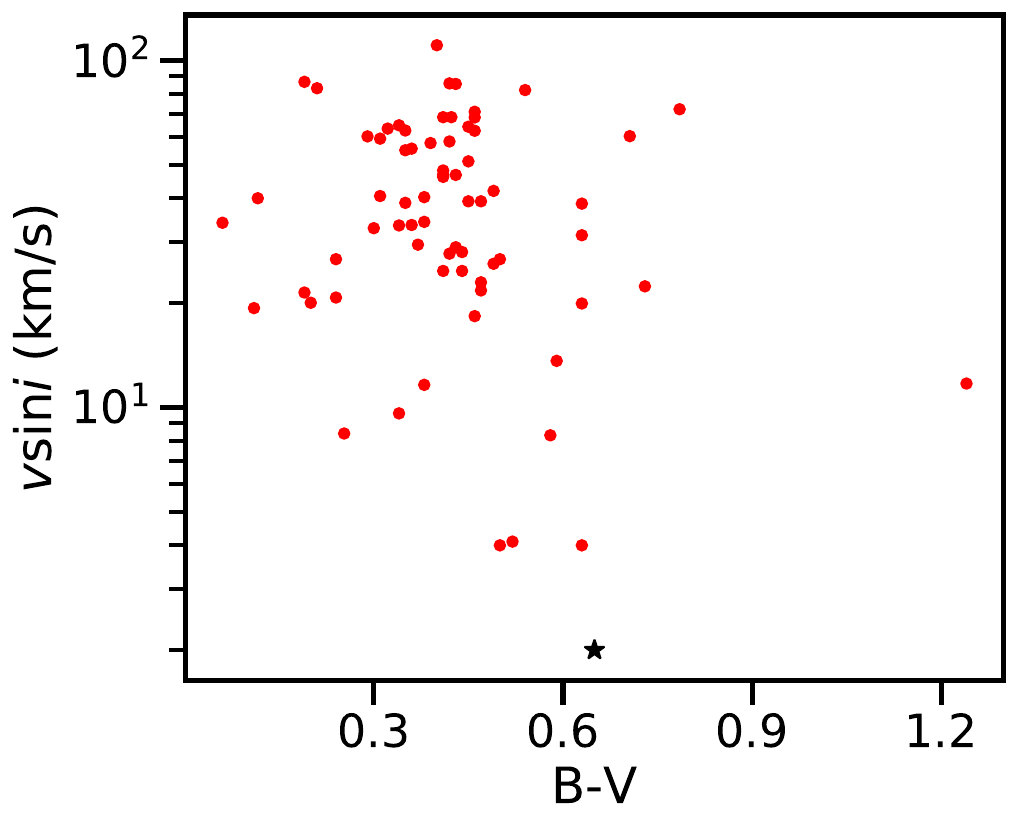}
\caption{\label{vsini}}
\end{subfigure}
\begin{subfigure}[t]{0.245\textwidth}
\includegraphics[width=1\hsize]{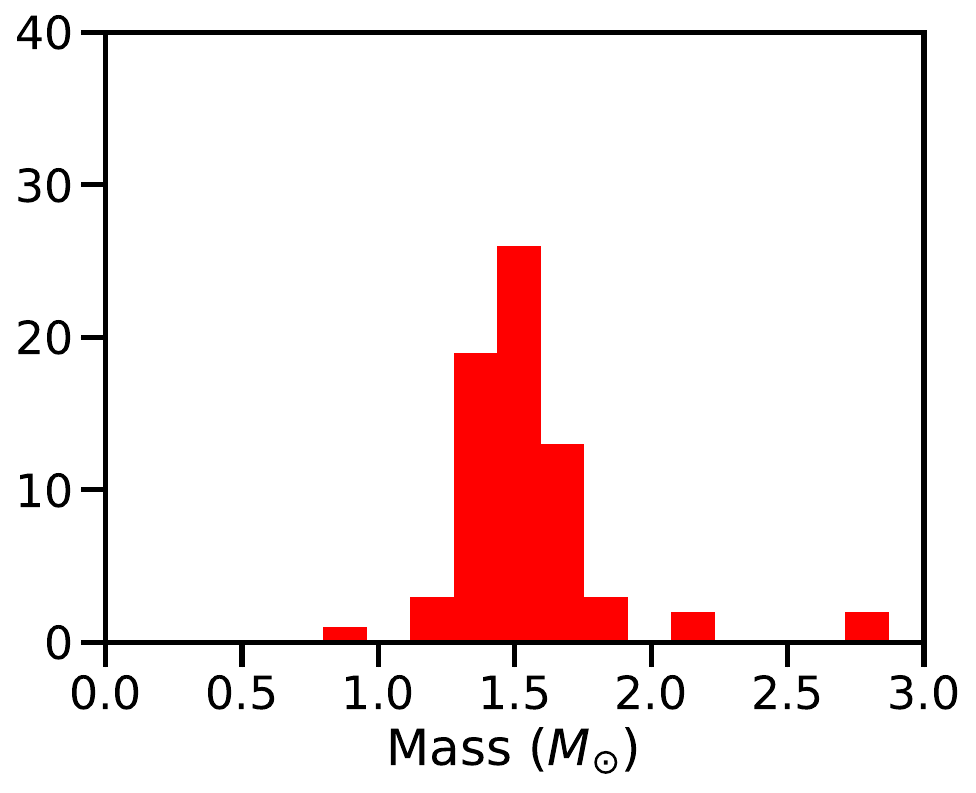}
\caption{\label{mass}}
\end{subfigure}
\begin{subfigure}[t]{0.24\textwidth}
\includegraphics[width=1\hsize]{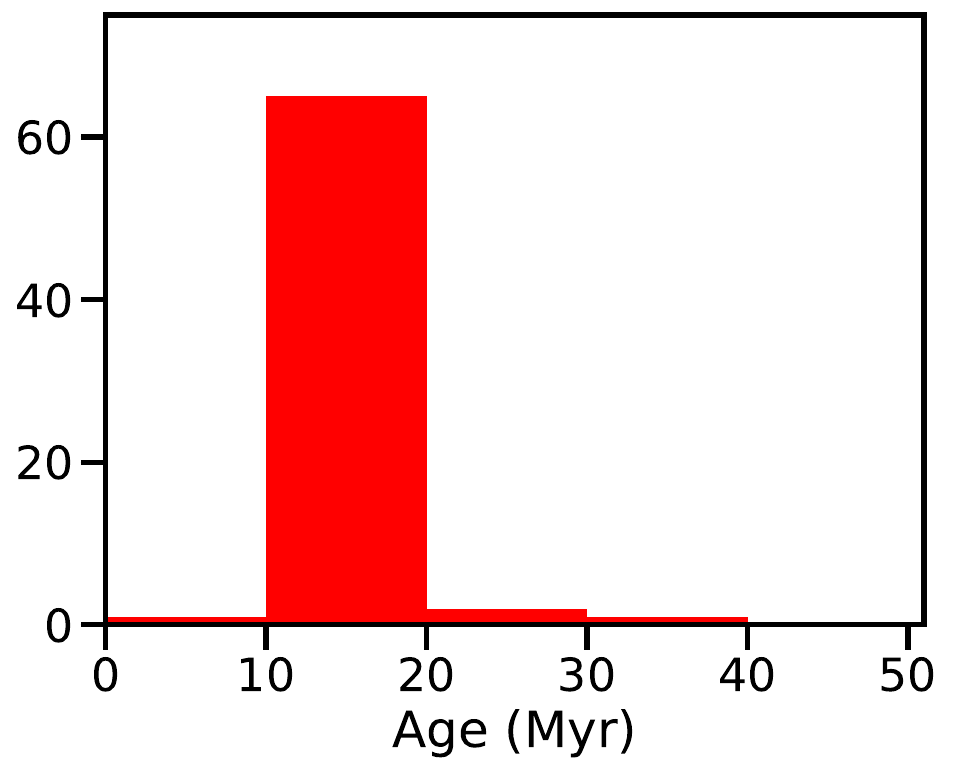}
\caption{\label{age}}
\end{subfigure}
\caption{Main physical properties of our \harps \ Sco-Cen sample (red dots).
 \subref{HR})  Absolute $V$-magnitude vs. \bv. 
 The solar parameters are displayed (black star) for comparison. 
\subref{vsini}) \vsini~vs. \bv~distribution.
\subref{mass})  Mass histogram (in \Msun).
\subref{age}) Age histogram.}
       \label{survey_carac_1_sco}
\end{figure*}

The Sco-Cen final sample (\emph{cf.} \Cref{survey_carac_1_sco}) includes $58$  A0-F5 stars ($B-V$ in the range $[-0.05:0.52[$) and $11$ F6-K5 stars ($B-V$ in the range $[0.52:1.33[$). The main characteristics of the targets are summarized in \Cref{survey_carac_1_sco} and \Cref{tab_carac_sco}. Their ages range from $10$ to $\SI{30}{\mega\year}$, with a median of $\SI{16}{\mega\year}$ (\emph{cf.} \Cref{age_mass} for the age determination).  The typical uncertainty on the ages is smaller than $\SI{8}{\mega\year}$. This sample is therefore compact in terms of age. The stellar V-band magnitudes range between  $5.7$ and $12.2$, with a median of $8.6$, and their masses are between $0.81$ and $\SI{2.87}{\msun}$, with a median of  $\SI{1.5}{\msun}$ (\emph{cf.} \Cref{age_mass} for the mass determination). 
The stellar \vsini \ range from $4$ to $\SI{110.4}{\kilo\meter\per\second}$, with a median of  $\SI{38.8}{\kilo\meter\per\second}$. We note that the median of these projected rotational velocities  is higher than that of our previous \harps \ YNS  \citep{Grandjean_HARPS} and \sophie \ YNS surveys because most of the stars are of early spectral types and are younger  \citep{Stauffer_2016, Rebull_2016, Gallet_2015}.

\begin{figure*}[ht!]
  \centering
\begin{subfigure}[t]{0.32\textwidth}
\includegraphics[width=1\hsize]{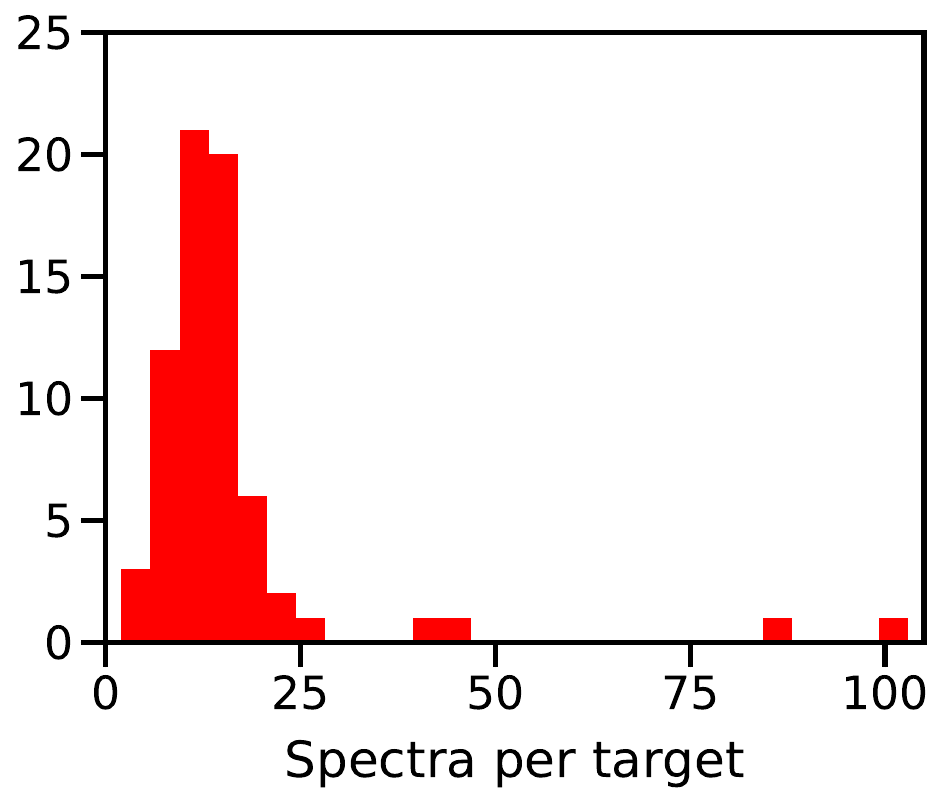}
\caption{\label{Nmes}}
\end{subfigure}
\begin{subfigure}[t]{0.32\textwidth}
\includegraphics[width=1\hsize]{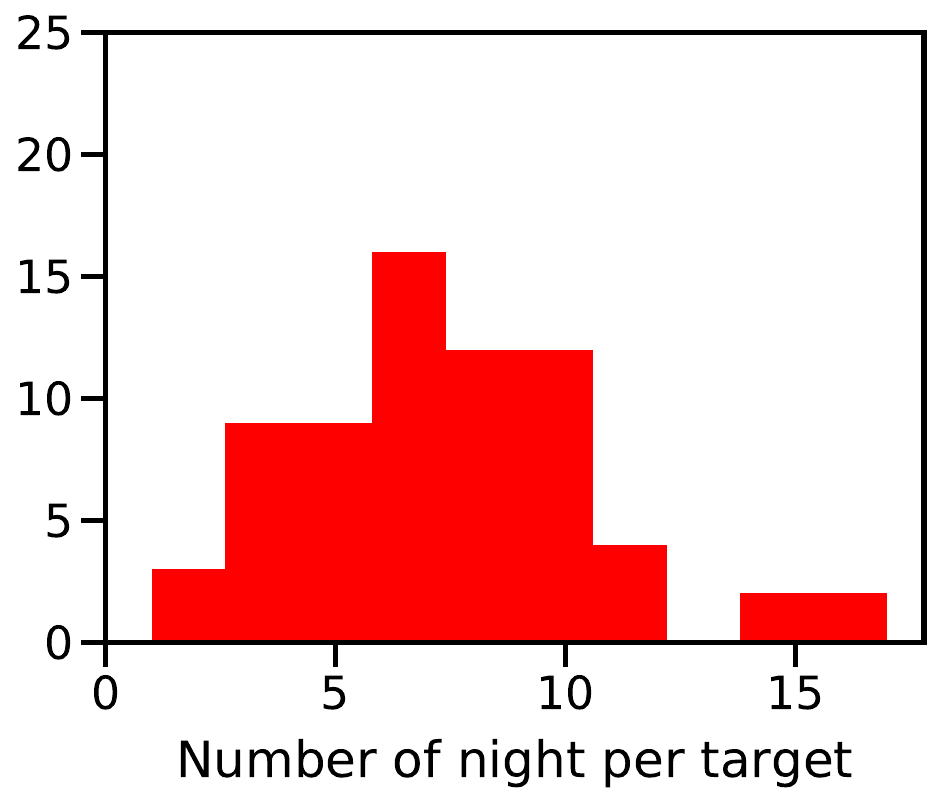}
\caption{\label{Nb_day}}
\end{subfigure}
\begin{subfigure}[t]{0.34\textwidth}
\includegraphics[width=1\hsize]{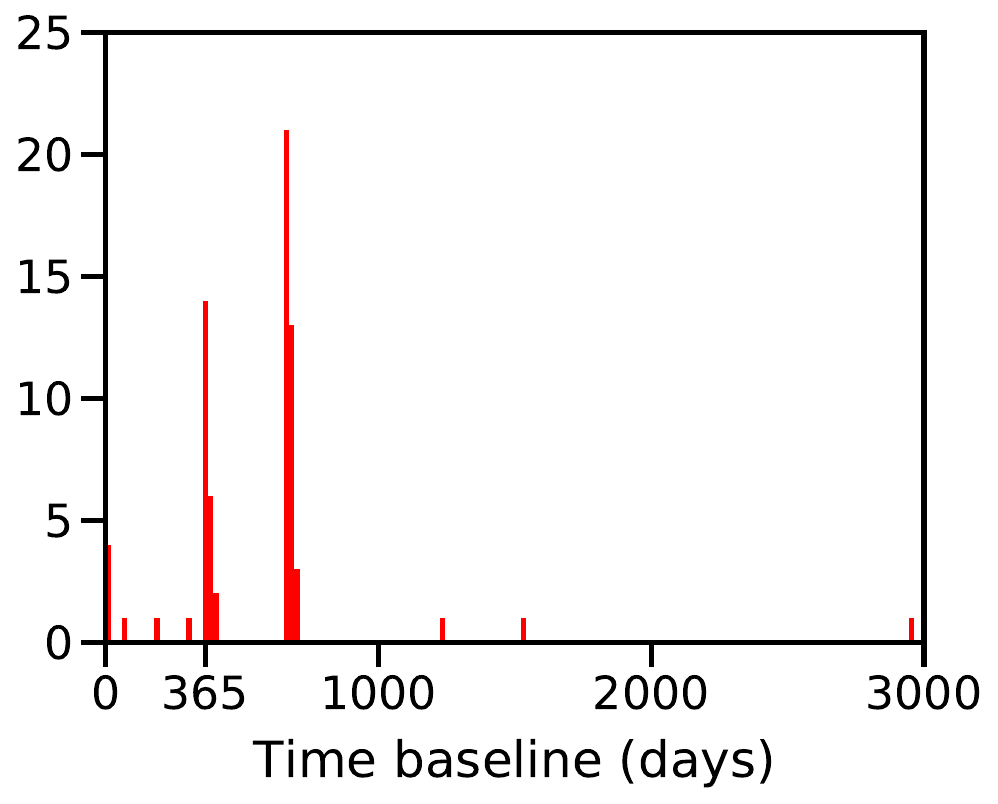}
\caption{\label{time_bsl}}
\end{subfigure}
\caption{Observation summary.
 \subref{Nmes}) Histogram of the number of spectra per target. 
 \subref{Nb_day}) Histogram of the  number of nights per target. 
\subref{time_bsl}) Histogram of the time baselines.} 
       \label{survey_carac_2_sco}
\end{figure*}

\subsection{Observations}

\label{observation}

The  observations of these $69$ stars were performed between 2018 and 2020. Some stars were previously observed as part of a previous \harps \ YNS survey \citep{Grandjean_HARPS} and as part of previous surveys made by \cite{Simon_VIII,Simon_IX,Simon_X}. The time baselines for these stars is then longer than \SI{2}{\year}, and in some cases, it is up to \SI{8}{\year}. The median time baseline  is \SI{665}{\day} (mean of  \SI{572}{\day}), the  median number of spectra per target is $13$ (mean of $16$), and the median number of nights per target is $7$  (mean of $7$, \Cref{survey_carac_2_sco}).
Details can be found in \cref{tab_result_sco}.

We adopted the observational strategy presented in \cite{Simon_VIII}, which consists of recording two spectra per visit and of observing each target on several consecutive nights. This strategy allows  sampling the short-term variations of early-type stars. Because telescope time was limited, we were unable to acquire long sequences ($>\SI{1.5}{\hour}$) for early-type stars to sample their pulsations, as was done in \cite{Simon_VIII}.

\subsection{Observables}

\label{observable}

We used the software called "spectroscopic data via analysis of the Fourier interspectrum radial velocities" (\safir; \cite{SAFIR}) to  derive the RVs and, whenever possible, the cross-correlation function (CCF), the bisector velocity span (hereafter BVS), the star \vsini \ (from  the full width at half maximum of the CCF), and the \rhk \ from the \harps \ spectra. 
\safir \ builds a reference spectrum from the median of all spectra available for a given star and computes the  RVs relative to the reference spectrum in Fourier space. The efficiency of this method was demonstrated in the search of low-mass companions  around AF-type MS stars \citep{Galland_GP}. To select the spectra that were used to build the reference, we used the quality selection criteria that were used for the  \harps \ YNS survey \citep{Grandjean_HARPS}: $S/N_{\SI{550}{\nano\meter}} \in [80;380]$\footnote{Signal-to-noise ratio per pixel between $554.753$ and $\SI{555,209}{\nano\meter}$, estimated with a sampling of one pixel every $\SI{2e-12}{\meter}$.}, $sec \ z < 3$ and $\chi^2 < 10$. However, not enough good-quality spectra  were available for some stars (HD 100282, HD 106473, HD 109832, HD113524, HD 113556, HD 113901, HD 114082, HD 115361, HD 119511, HD 120326, HD 121189, HD 125912, HD 126488, HD 129590, and HD 137057). For these stars, we adopted a relaxed selection threshold of the signal-to-noise ratio (S/N) at $$\SI{550}{\nano\meter}$$ down to $30$. We verified that the additional spectra did not substantially degrade the quality of the reference spectra. In the case of the faint ($V\sim12.2$) PDS 70, we decreased the S/N threshold to $14$ (the S/N at $\SI{550}{\nano\meter}$ of the spectra  range between $15$ and $40$). 

We did not correct the RVs for the drift induced by the secular acceleration as its amplitude ($\sim \SI{0.003}{\meter\per\second\per\year}$) is negligible with respect to the RV uncertainties for our stars.
The main sources of RV variability (magnetic activity, pulsations, or companions) were identified using either the possible correlation between BVS and RV or  the shape of their set of bisectors, as was done in \cite{Lagrange_2009,Lagrange} and \cite{Simon_IX}.
Briefly, stars with magnetic activity present a correlation between BVS and RV when their \vsini \ is high enough and the stellar lines  are resolved (\cite{Desort}). This is the case for our targets, \emph{cf.} \Cref{tab_carac_sco}. Pulsating stars rather present a (BVS, RV) diagram with an infinite slope (hereafter referred to as vertically spread (BVS, RV) diagram). Companion signals can be identified by bisectors that are parallel to each other, and constant (within noise) BVS values; hence the (BVS, RV) diagram shows a  slope of $0$ (hereafter referred to as horizontally spread (BVS, RV) diagram).
\label{safir}

\section{Detected companions in the \harps \ Sco-Cen\ survey}
\label{comp}

In addition to HD 106906 and HD 131399, which were already known to be spectroscopic binaries \citep{Lagrange_106b,Lagrange_131}, $13$ stars present evidence of a companion in their RV time series.

When enough data were available, we constrained  the orbital elements and the $m_c\sin{i}$ of the companions  (SB1, BD, GP) using  \emph{yorbit} \citep{Segransan}.

When long-term trends were observed, we used the toy model of  \cite{Grandjean_SOPHIE} to estimate the minimum mass needed for a companion to produce the observed trend. This minimum mass corresponds to the mass below which a companion cannot explain the observed RV amplitude. 

\subsection{Binaries without orbital solutions}

We present below the  single-line spectroscopic binaries (SB1) for which the available RV data are insufficient to perform orbital fitting, as well as the double-line spectroscopic binaries (SB2). Associated figures are shown in \Cref{binary_noncarac} and \ref{sco_ccf}.

\subsubsection{HD 111102} 

HD 111102 is an F0III-type star with a  $\SI{1.5}{\msun}$ companion imaged at $\SI{35}{\milli\arcsecond}$ ($\SI{4}{\au}$ projected separation; \cite{Bonavita_sphere}). 
We observe an amplitude of  $\SI{10}{\kilo\meter\per\second}$ in the HD 111102 RVs (\emph{cf.} \Cref{binary_noncarac}). The bisectors of HD 111102 are parallel to each other (\emph{cf.} Figure \ref{111102_bis}), which indicates that the high-amplitude variations are  due to a stellar companion. A correlation between BVS and RV  ($R=-0.96$, $p_{value}< \num{5e-6}\%$, \emph{cf.} \Cref{binary_noncarac}), in addition to the  $<$\rhk$>$ \  ($-4.222$, with a standard deviation of $0.006$), indicates that the RVs are also affected by the magnetic activity of the star.

\subsubsection{HD 114319}  
\label{sec_114319}
HD 114319 (F0) has a visual companion at a separation of $\sim\ang{;;2.3}$ ($\sim \SI{235}{\au}$ projected separation; \cite{Janson_2013, Bonavita_sphere}).
The RV time series of HD 114319  show variations of $\SI{1}{\kilo\meter\per\second}$ over $\SI{680}{\day}$ ($\SI{1.86}{\year}$) with a curvature, in addition to  short-term variations of about $\SI{100}{\meter\per\second}$ (\emph{cf.} \Cref{binary_noncarac}). The (BVS, RV) diagram shows that the RV variations are due to a companion plus stellar variability. The imaged companion cannot induce the observed trend given its large separation. HD114319 is thus a triple system.

\subsubsection{HD 121176} 

HD 121176 (F8V)  shows RV variations with an amplitude of $\SI{1}{\kilo\meter\per\second}$ over one day (\emph{cf.} \Cref{binary_noncarac}). The (BVS, RV) diagram is spread horizontally. This indicates that  the RV variations are due to a stellar companion.

\subsubsection{HD 126488} 

HD 126488 (F2) shows RV variations with an amplitude of $\SI{10}{\kilo\meter\per\second}$. The shape of the (BVS, VR) diagram ($R=-0.04$, $p_{value}< 92\%$; \emph{cf.} \Cref{binary_noncarac}) indicates that these variations are due to a  companion.

\subsubsection{HD 129590}

HD 129590 is a G3V type star with a resolved debris disk \citep{LiemanSifry,Matthews_17} that may consist of two rings \citep{Cotten,Matthews_18}. 
The variations in CCF (\emph{cf.} Figure \ref{129590_ccf}), with strong changes over time in the line width, depth, and with a variable width at the continuum, suggest a possible binary of type SB2.

\subsubsection{HD 137057}

HD 137057 is an F3V type star, known to be the second component of a $\SI{120.55}{\arcsecond}$ separation  binary system, together with HD 137015 \citep{Andrews}. HD 137057 also presents an IR excess \citep{McDonald,Mittal} that is attributed to a debris disk located at $\SI{16.6}{\au}$. 

\begin{figure}[h]
\centering
\includegraphics[width=1\hsize]{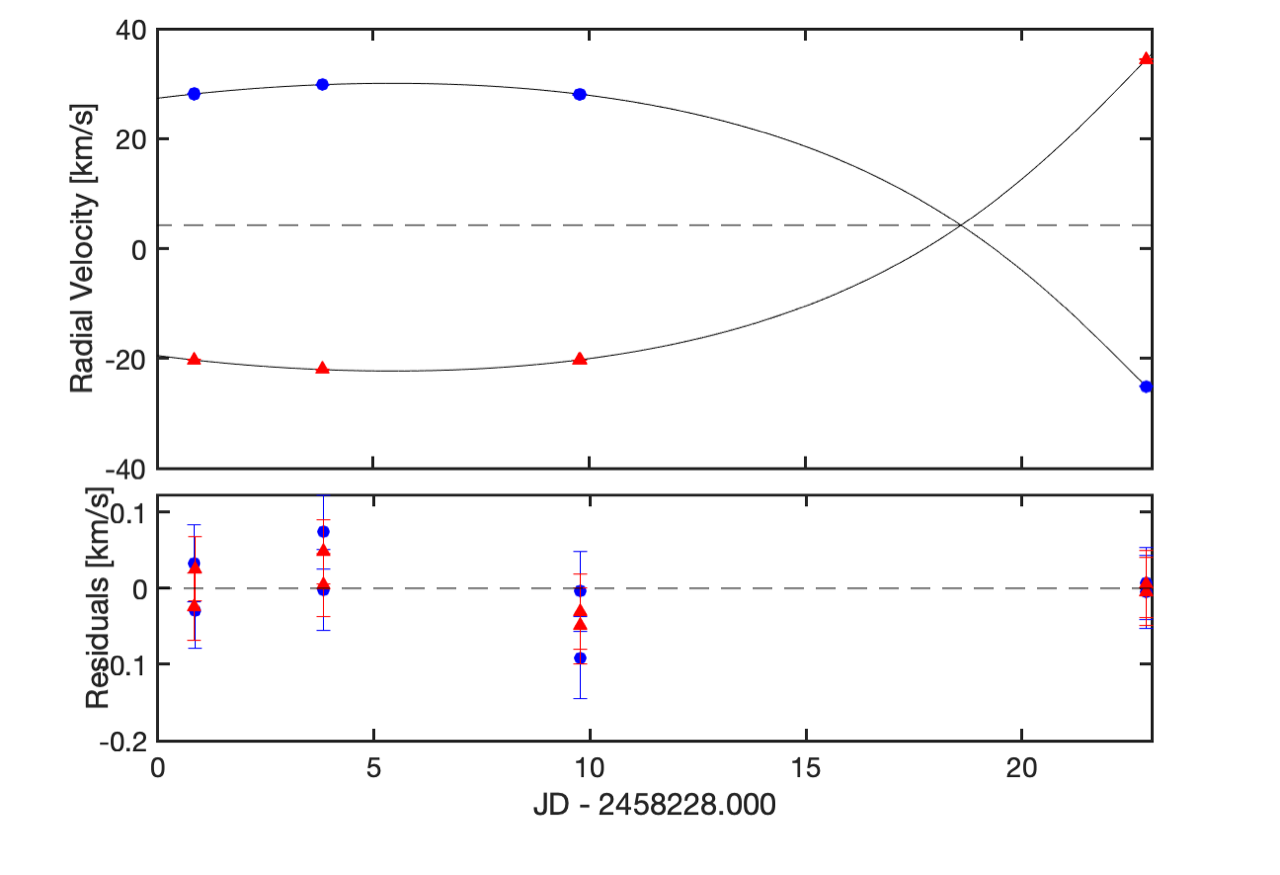}
\caption{Keplerian fits of the RV time series of the two components of the HD 137057 AB system, together with the associated residuals.\label{137057}}
\end{figure}

The CCF shows the double bump that is characteristic of SB2 binary stars (\emph{cf.} Figure \ref{137057_ccf}). Each peak represents one component of the system.
We used TODMOR \citep{Zucker_2003,Zucker_2004}, which is a 2D CCF technique that can derive the RVs of the two SB2 components at once from our HARPS spectra (thorough details of the methods and practical cases are described in \cite{Kiefer_2016,Kiefer_2018,Halbwachs}). The two spectral components were first matched with PHOENIX model spectra around the CaI lines ($\SI{6120}{\angstrom}$) at four different epochs with the largest CCF-peak separation, leading to $Teff_A= \SI{6700}{\kelvin}$, $\log g_A=4.4$ cgs, $V \sin{i_A}=\SI{23}{\kilo\meter\per\second}$,  $Teff_B=\SI{6600}{\kelvin}$, $\log g_B=4.4$ cgs, $V \sin{i_B}=\SI{20}{\kilo\meter\per\second}$, $[Fe/H]=-0.3\ dex$, and a flux ratio in the optical of $F_B/F_A\sim0.85\pm0.05$, with typical uncertainties on $\sigma_{Teff}$$\sim$$200$ K, $\sigma_{\log g}$$\sim$$0.1$ cgs, $\sigma_{Fe/H}$$\sim$$0.1 \ dex$, and $\sigma_{v\sin{ i}}$$\sim$$\SI{1}{\kilo\meter\per\second}$ \citep{Kiefer_2016,Kiefer_2018}. Then these templates were used to calculate a multi-order 2D cross-correlation with the observed spectra using TODMOR. This led to a time series of RVs for each component. Keplerians were simultaneously fit on these RVs, and the common orbital period, time of periastron passage, periastron angle, eccentricity, systemic velocity, and two different RV semi-amplitudes, one for each component (\emph{cf.} \Cref{137057}) were varied.  From this fit, we estimate a period of $48\pm\SI{13}{\day}$ and a mass ratio of $0.975\pm0.001$ for the HD 137057 AB system.
The HD 137015-137057 AB is consequently a triple hierarchical system.

\subsubsection{HD 143811}

HD 143811 is an F5V type star, presenting an IR excess \citep{Chen_11,Ballering}) that is attributed to a debris disk located at $\SI{8.5}{\au}$ \citep{Cotten}. The CCF shows a double component (\emph{cf.} Figure \ref{143811_ccf}) that might be indicative of an SB2 type binary.

\subsection{Single-line binaries with orbital solutions or constraints, and targets with RV long-term trends}
\label{trend_bin}

We present hereafter the SB1 binaries for which we were able to constrain the companion  properties (\emph{cf.}  \Cref{binary}). We also present the stars with a long-term trend for which a linear regression was performed (\emph{cf.} \Cref{trend}). 

\subsubsection{HD 108857} 

\label{HD 108857}

\begin{figure}[h]
\centering
\includegraphics[width=0.99\hsize]{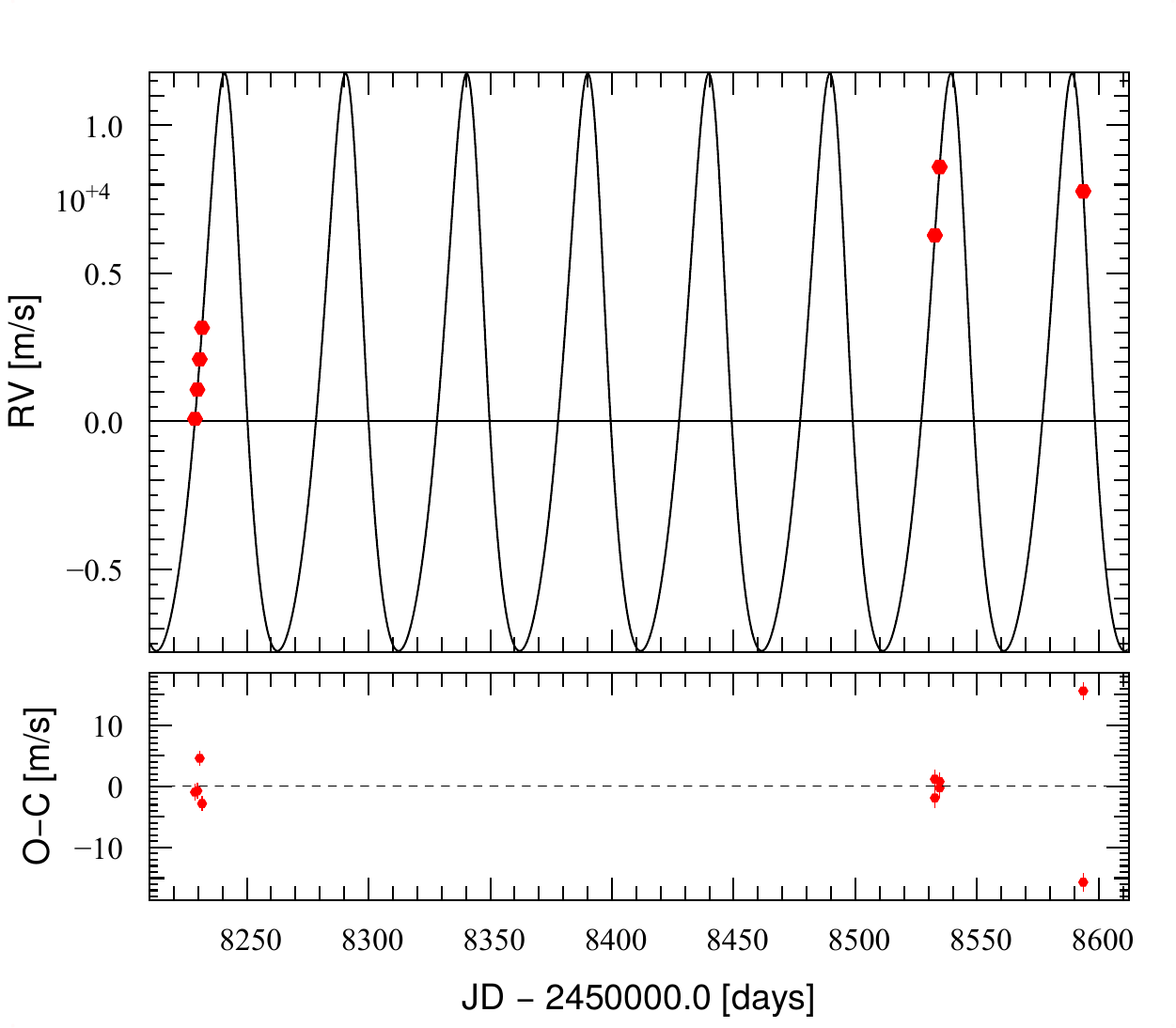}
\caption{HD 108857 Keplerian fit and associated residuals.\label{108857}}
\end{figure}

HD 108857 is an F7V type star with a mass of $\SI{1.4}{\msun}$ \citep{Mittal}. 
It shows an IR excess \citep{Chen_11,McDonald} that is attributed to two debris belts located at $2.6$ and $\SI{5.3}{\au}$. Direct-imaging observations did not reveal companions  \citep{Mamajek_99,Janson_2013,Nielsen_2019_b}. Finally, HD 108857 presents a proper motion anomaly that can be attributed to a companion \citep{Kervella_HIPGAIA}.
The RV time series of HD 108857 presents variations with an amplitude of $\SI{10}{\kilo\meter\per\second}$ that we attribute to a companion.
We used \emph{yorbit} to fit the RVs with one Keplerian model  to constrain the companion parameters  (\emph{cf.} \Cref{108857}). The solution of the fit is not unique because only a few spectra were available and the phase coverage is poor.  The less eccentric solution we obtained has a period of  $49.735\pm   0.007 \ \si{\day}$, an eccentricity of  $0.23 \pm 0.02$,  and an $m_c\sin{i}$ of $0.21\pm\SI{0.02}{\msun}$  ($215\pm\SI{21}{\MJ}$).

\subsubsection{HD 108904} 

\begin{figure*}[ht!]
  \centering
\begin{subfigure}[t,valign=t]{0.49\textwidth}
\includegraphics[width=0.99\hsize]{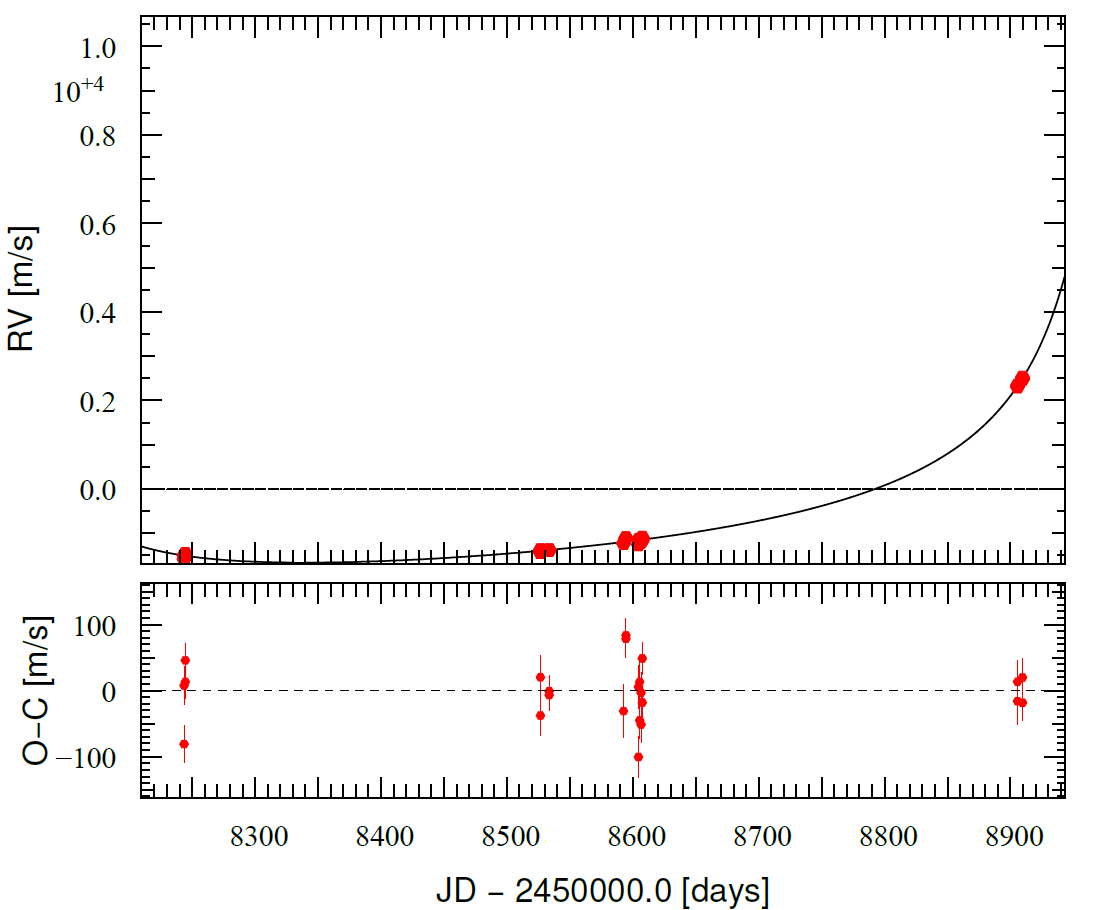}
\caption{\label{108904_sol}}
\end{subfigure}
\begin{subfigure}[t,valign=t]{0.49\textwidth}
\includegraphics[width=0.99\hsize]{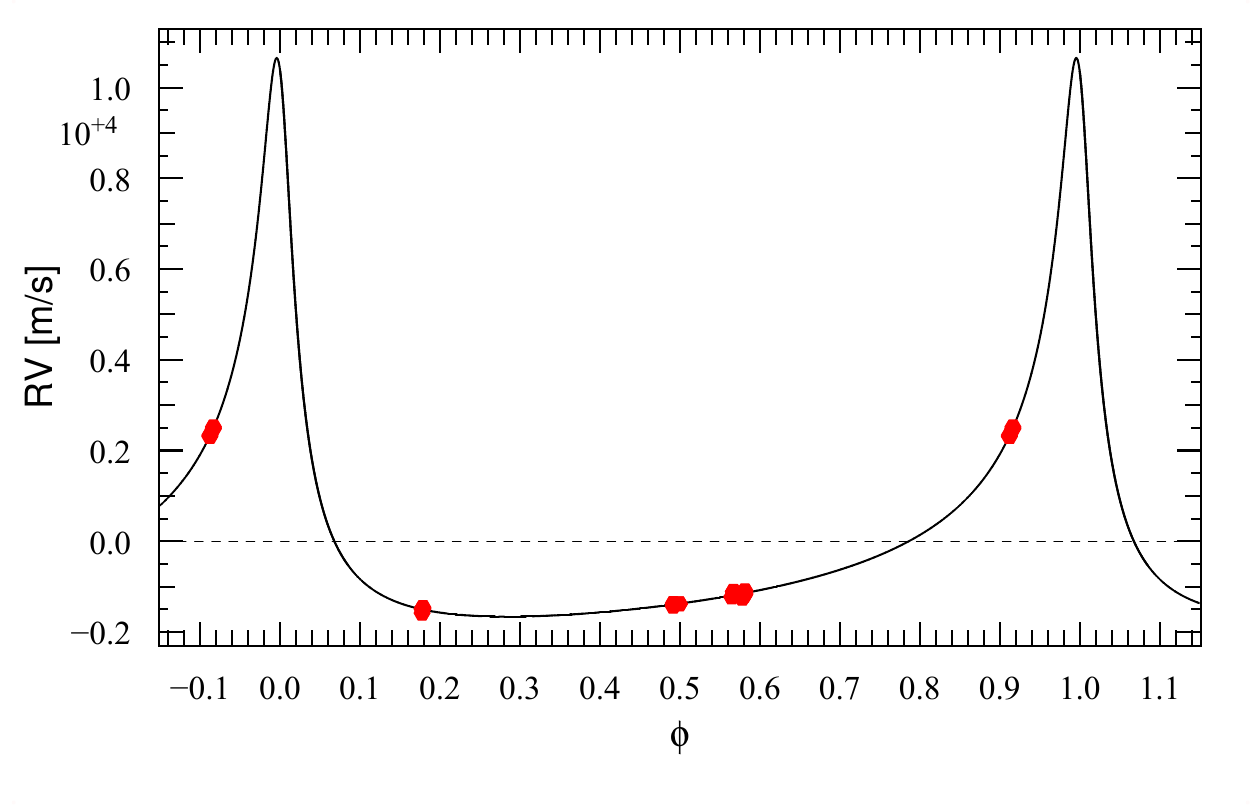}
\vfill
\caption{\label{108904_sol_phase}}
\end{subfigure}
\caption{HD 108904 companion anlysis. \subref{108904_sol}) Best fit of HD 108904's RV with one Keplerian and its associated residuals. \subref{108904_sol_phase})  Phase-folded plot of the fit.}
       \label{108904}
\end{figure*}

HD 108904 is an F6V type star with a mass of $\SI{1.5}{\msun}$ \citep{Mittal}, known to present an IR excess \citep{McDonald}. 
\cite{Bonavita_sphere} imaged a companion at  $\SI{54}{\milli\arcsecond}$ ($\sim\SI{6}{\au}$ projected separation) with \sphere \ and  estimated its mass from evolutionary models  at $0.18\pm0.02 \si{\msun}$ ($190^{+20}_{-20} \si{\MJ}$). HD 108904 also presents a proper motion anomaly \citep{Kervella_HIPGAIA} that may have been induced by the imaged companion.
We observe RV variations of $\SI{4}{\kilo\meter\per\second}$ over $\SI{666}{\day}$ with a curvature, in addition to a $\SI{100}{\meter\per\second}$ short-term jitter (\emph{cf.} \Cref{binary}). The (BVS, RV) diagram is spread horizontally ($R=-0.19$, $p_{value}< 40\%$, \emph{cf.} \Cref{binary}), which indicates that the long-term signal is due to a companion. A Keplerian fit of the RVs alone with \emph{yorbit}  (\emph{cf.} \Cref{108904}) gives a period of $900\pm  220 \ \si{\day}$ ($2.1^{+0.3}_{-0.4} \si{\au}$; $19\pm\SI{3}{\milli\arcsecond}$), an eccentricity of $0.76 \pm 0.11$, and a mass of $0.24^{+0.07}_{-0.08} \si{\msun}$ ($247^{+75}_{-86} \si{\MJ}$). The sma we found is inconsistent with the projected separation of the imaged companion.  This is expected because our RV data do not properly sample the companion orbit. Clearly, more RV and more imaging data are needed to sample the companion orbit properly.

\begin{figure}[h]
\centering
\includegraphics[width=0.99\hsize]{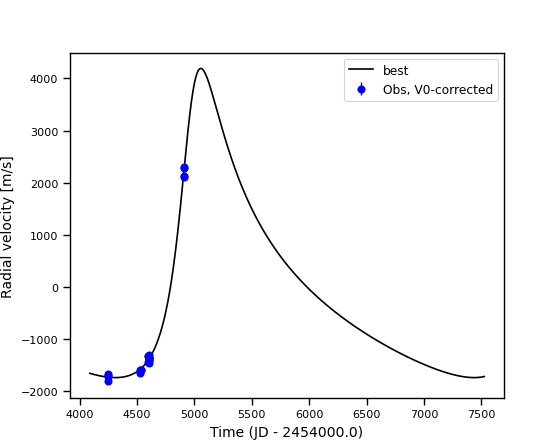}
\caption{HD 108904 best fit of the MCMC applied on RV and relative astrometry simulatenously.\label{108094_best_fit}}
\end{figure}

We then used a Markov chain Monte Carlo (MCMC), described in \cite{Lagrange_2020}), to fit the RVs and the available relative astrometry \citep{Bonavita_sphere} of the imaged companion simulatenously. A correct fit of the RVs is obtained with an sma of about $\SI{4}{\au}$ ($P \sim \SI{2400}{\day}$), a high eccentricity ($>0.4$), and a mass compatible with that deduced from evolution models: $\sim \SI{0.2}{\msun}$ (\emph{cf.} \Cref{108094_best_fit} and \Cref{corner_108904}). We stress that the data are sparse, so that these results should be regarded as indicative.

\subsubsection{HD 126838}

HD 126838 is a $\SI{1.78}{\msun}$ F3V star that is known to present an IR excess \citep{McDonald}. It is also known as a visual binary, with a projected separation of $\SI{2.93}{arcsec}$ ($\sim \SI{354}{\au}$ projected separation; \cite{Mason_WDS,Chen_11,Kervella_HIPGAIA, Bonavita_sphere}).
Our RVs show a trend with a slope of $\SI{7.1}{\kilo\meter\per\second\per\year}$ observed on  $\SI{398}{days}$. The amplitude of the observed trend is too high to be attributed to a possible magnetic cycle. Short-term variations are present in addition to the trend, with an amplitude of $\SI{100}{\meter\per\second}$ (\emph{Cf.} \Cref{trend}).
 Moreover, the (BVS, VR) diagram is spread horizontally ($R=0.29$, $p_{value}< 48\%$; \emph{Cf.} \Cref{trend}) and the bisectors are parallel to each other. The imaged companion cannot induce the observed trend due to its large separation. We  therefore attribute the long-term drift to an additional companion, with a minimum mass of $\SI{0.25}{\msun}$ ($\SI{260}{\MJ}$), and a period longer than one year. 
The residuals of a linear regression on the RVs present a standard deviation of $\SI{110}{\meter\per\second}$ , and they are not significantly correlated with BVS ($R=-0.29$, $p_{value}< 26\%$; \emph{cf.} \Cref{trend}).

The system was identified as a TESS object of interest (TOI 1946). The transiting candidate has a period of $\SI{10.845}{\day}$ and a radius of $30.4\pm\SI{0.9}{\rearth}$ \citep{Bonavita_sphere}. \cite{Bonavita_sphere} discussed the probable system architecture, arguing that the companion that causes the transit signature is a star and orbits HD 126838 B. If confirmed, this would make the system a hierarchical quadruple system.

\subsubsection{HD 133574} 

HD 133574 is  a $\SI{1.6}{\msun}$ \citep{Pecaut_2012} A9-type star, located at $156.3013\pm1.7125 \si{\pc}$ \citep{DR2A1}, which presents an IR excess \citep{McDonald}. 
\cite{Bonavita_sphere} imaged a low-mass star companion at $\SI{88.9}{\milli\arcsecond}$ ($\SI{13.9}{au}$ projected separation) and  estimated its mass from evolutionary models to be $\SI{0.59}{\msun}$ ($\SI{620}{\MJ}$). 
HD 133574 also presents a proper motion anomaly \citep{Kervella_HIPGAIA} that might be induced by the imaged companion.
We observe a drift in HD 133574 RVs with a slope of $\SI{320}{\meter\per\second\per\year}$ over  $\SI{681}{\day}$. In addition, the RVs present short-term variations with an amplitude of $\sim\SI{50}{\meter\per\second}$ (\emph{Cf.} \Cref{trend}), but these variations are poorly sampled.
The HD 133574  bisectors are parallel to each other, and its BVS are not significantly correlated with the RVs ($R=-0.36$, $p_{value}< 12\%$). This indicates that the trend is due to a companion. Assuming a semimajor axis of $\SI{13.9}{\au}$, the imaged companion would induce an RV trend with a mean slope of $\sim\SI{360}{\meter\per\second\per\year}$. We therefore attribute  the RV trend to this companion.

\subsection{Interesting substellar companions}

\subsubsection{Probable BD around HD 149790}

\begin{figure*}[t]
  \centering
\includegraphics[width=0.8\hsize]{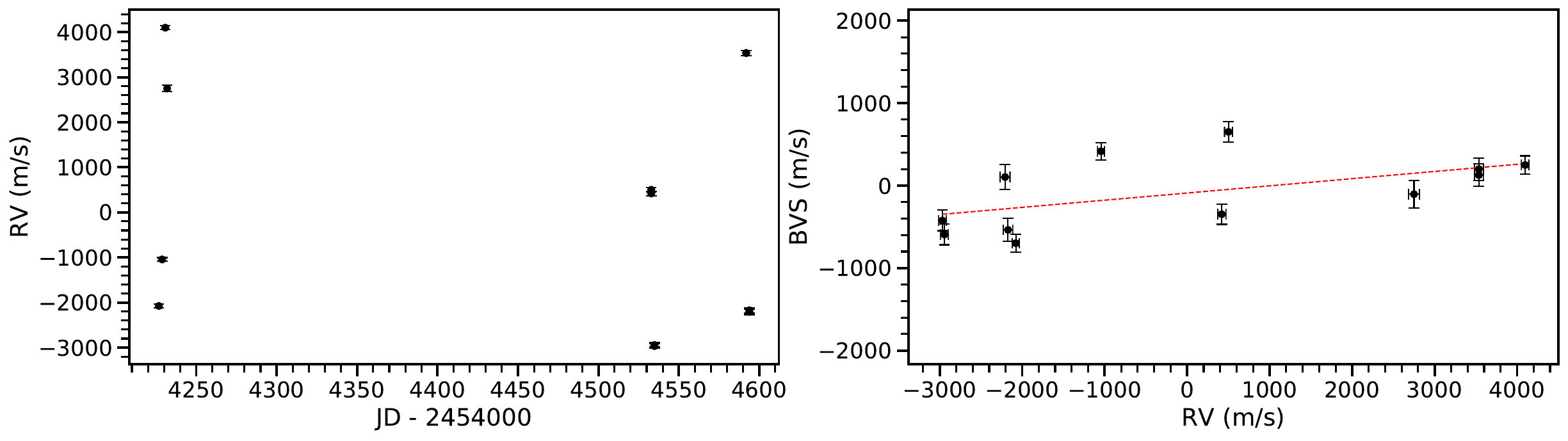}
\caption{HD 149790 data. \emph{Left:} RV time variations. \emph{Right:} BVS vs. RV diagram (black) and its best linear model (dashed red line).}
       \label{HD149790_data}
\end{figure*}

HD 149790 is an F3V-type star with a mass of $\SI{1.4}{\msun}$ \citep{Pecaut_2012}. It shows an IR  excess \citep{McDonald}, attributed to a debris disk located at $\SI{1.2}{\au}$ \citep{Cotten}.  
Our RV data show variations with amplitude  of about  $\SI{7}{\kilo\meter \per\second}$ (\emph{cf.} \Cref{HD149790_data}). The  bisectors are roughly parallel to each other (\emph{cf.} Figure \ref{149790_bis}) and are spread horizontally (BVS vs. VR diagram: $R=0.55$, $p_{value}< 7\%$ \emph{cf.} \Cref{binary}), which indicates the presence of a companion.

\begin{figure}[h]
\centering
\begin{subfigure}[t]{0.49\textwidth}
\includegraphics[width=1\hsize,valign=m]{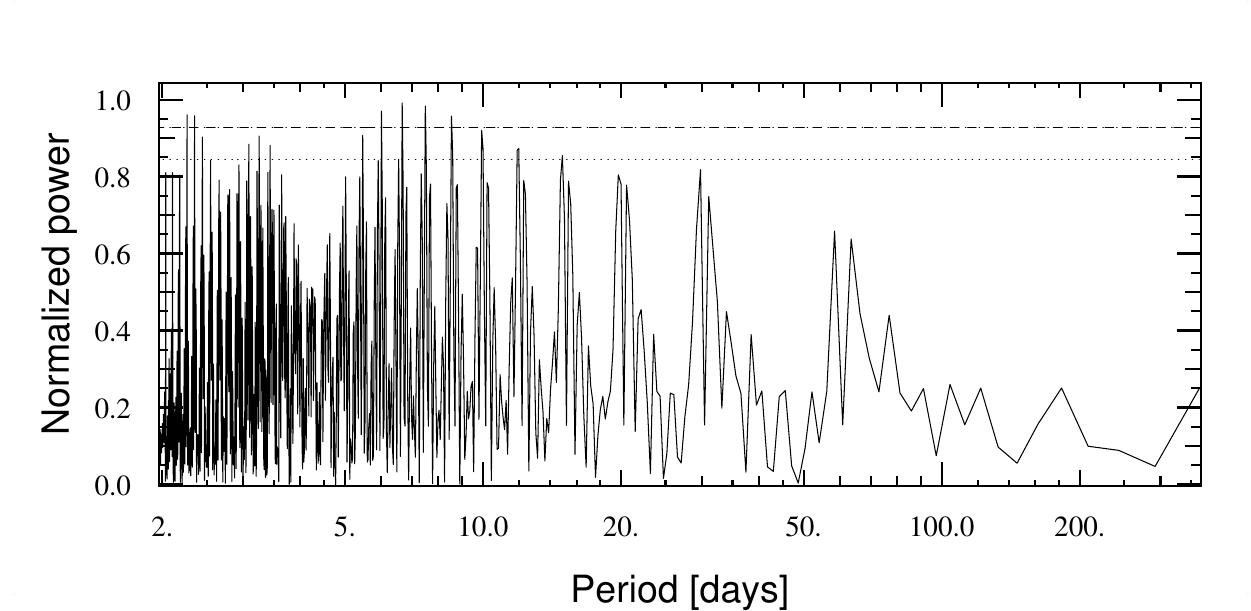}
\caption{\label{periodo_147_rv}}
\end{subfigure}

\begin{subfigure}[t]{0.49\textwidth}
\includegraphics[width=1\hsize,valign=m]{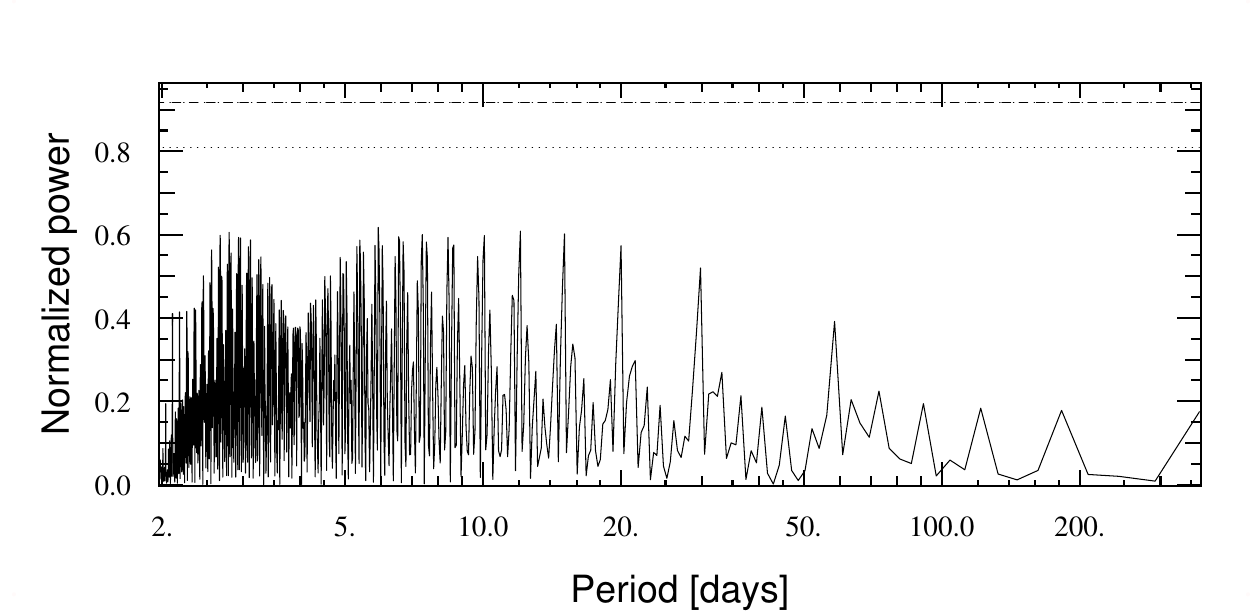}
\caption{\label{periodo_147_span}}
\end{subfigure}

\begin{subfigure}[t]{0.49\textwidth}
\includegraphics[width=1\hsize,valign=m]{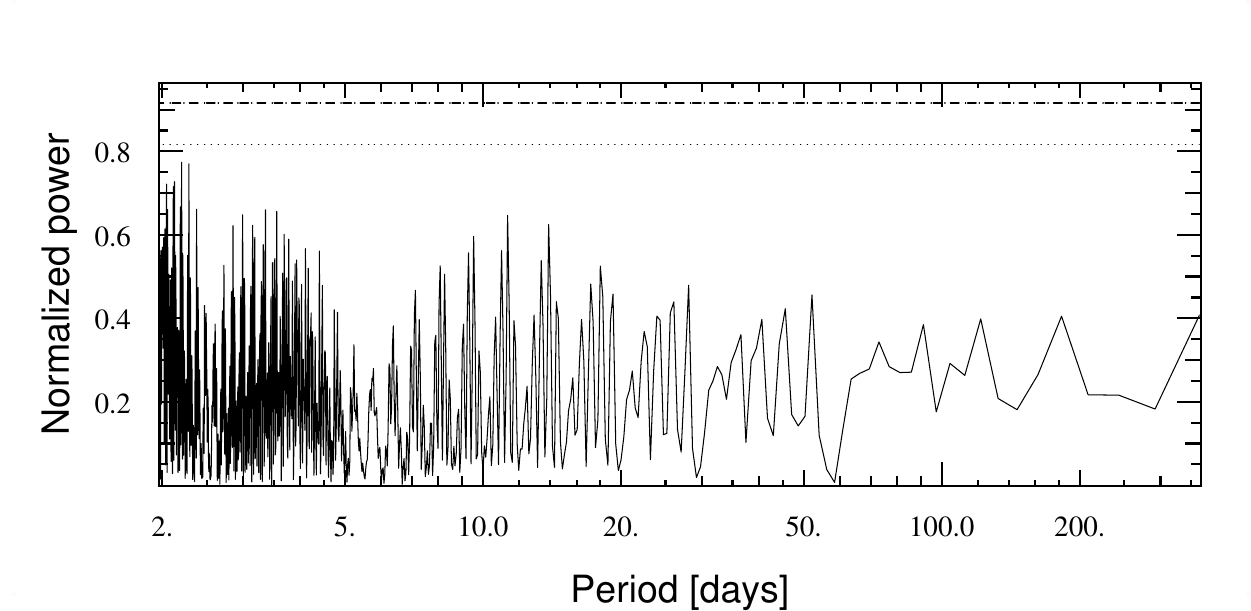}
\caption{\label{periodo_147_omc}}
\end{subfigure}

\caption{HD 149790 anylysis. \protect\subref{periodo_147_rv}) Periodogram of the RV time series.  
\protect\subref{periodo_147_span}) Periodogram of the BVS time series. \protect\subref{periodo_147_omc})  Periodogram of the residuals after subtracting the best Keplerian fit. The   $1 \%$ and $10 \%$ FAP are presented as dashed and dotted lines, respectively. \label{periodo_147}}
\end{figure}

 The periodogram of the RVs presents peaks above the $1\%$  false-alarm probability (FAP) at $2.26$ and $\SI{2.35}{\day}$, as well as at $6$, $6.67$, $7.5$, and $\SI{8.5}{\day}$ (\emph{cf.} \Cref{periodo_147_rv}).
 The BVS periodogram does not show peaks above the $10\%$ FAP (\emph{cf.} \Cref{periodo_147_span}).

\begin{figure*}[ht!]
  \centering
\begin{subfigure}[t,valign=t]{0.49\textwidth}
\includegraphics[width=0.99\hsize]{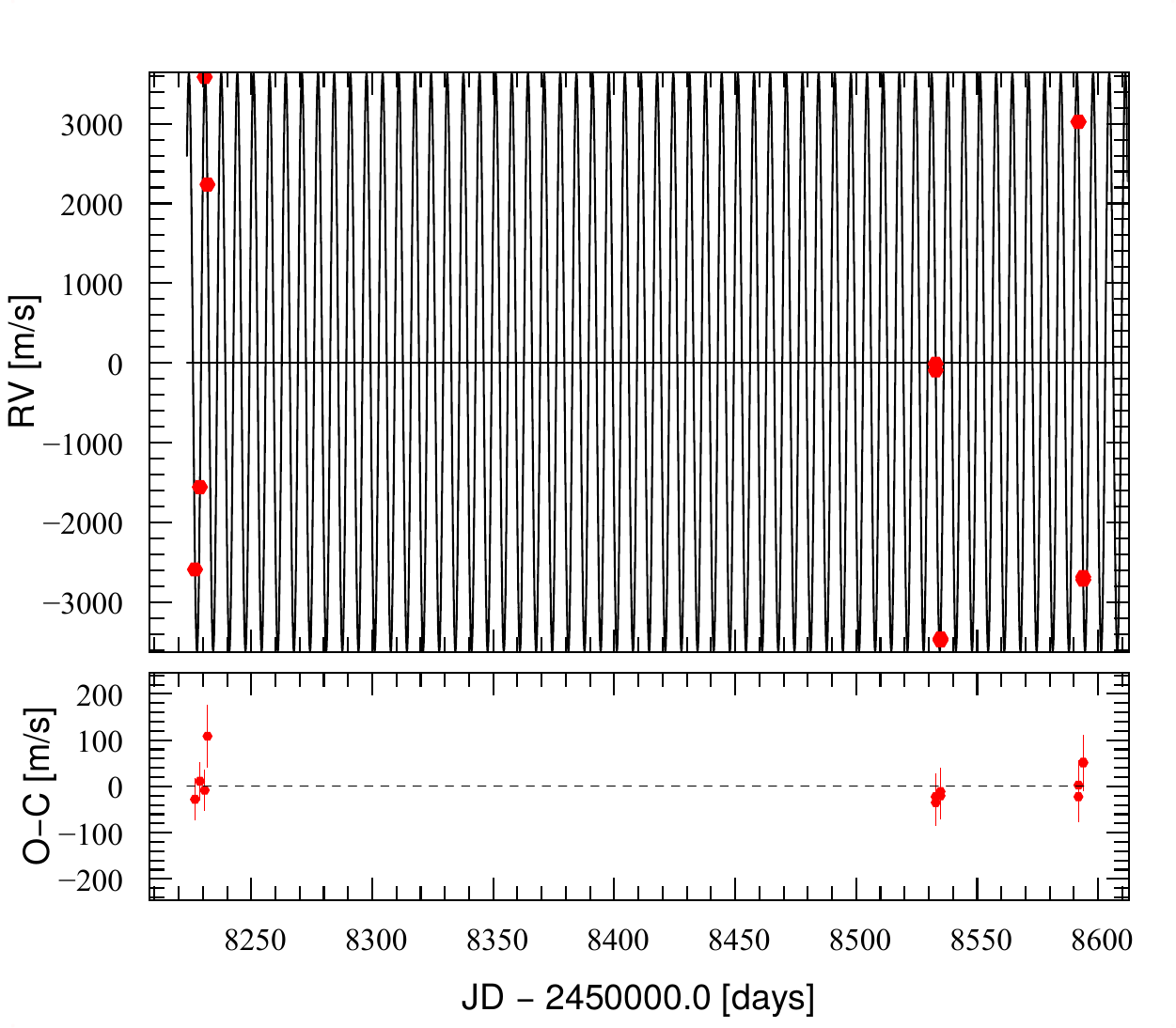}
\caption{\label{HD 149790_sol}}
\end{subfigure}
\begin{subfigure}[t,valign=t]{0.49\textwidth}
\includegraphics[width=0.99\hsize]{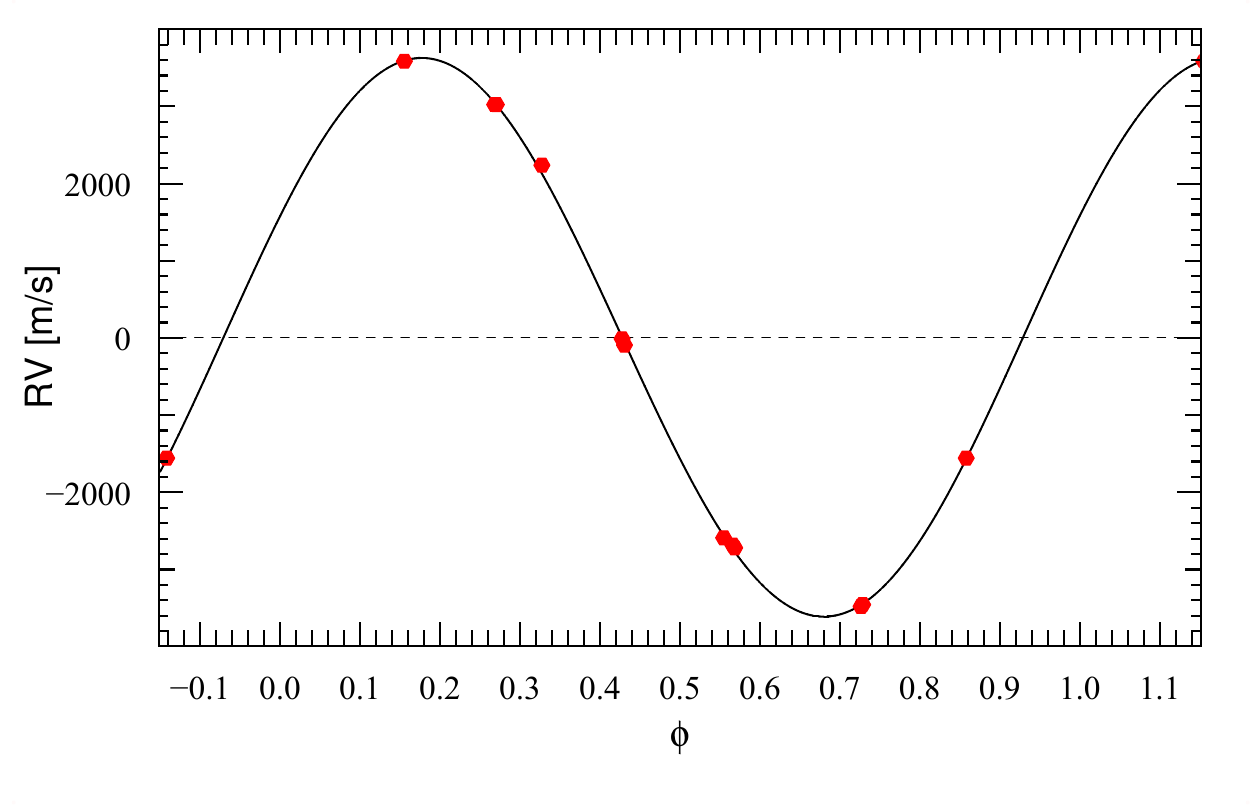}
\vfill
\caption{\label{HD 149790_sol_phase}}
\end{subfigure}
\caption{HD 149790 companion analysis. \subref{HD 149790_sol}) HD 149790 best Keplerian fit (one planet) and its residuals. \subref{HD 149790_sol_phase})  Phase-folded plot of the fit.}
       \label{149790}
\end{figure*}

To constrain the period and the minimum mass of the companion candidate, we fit the RVs with a Keplerian model (\emph{cf.} \Cref{149790} and Figure \ref{periodo_147_omc}) using  \emph{yorbit}. We obtain a period of $6.673\pm 0.003 \si{\day}$, a low eccentricity ($e=0.005 \pm 0.006$), and a $m_c\sin{i}$ of $42\pm0.3\si{\MJ}$. This companion candidate would therefore be a BD. Additional RV observations are mandatory to definitely confirm the orbital parameters and the $m_c\sin{i}$ of this companion. If confirmed, it will be the first young ($<\SI{100}{\mega\year}$) short-period ($<\SI{10}{\day}$) BD ($\sin{i}$-wise)  ever detected with the RV technique, and the second of its kind ever detected, together with NGTS-7A b \citep{Jackman}.

The residuals of the fit have an amplitude of $\SI{50}{\meter\per\second}$, and the (BVS, RV residuals) diagram  is spread vertically (\emph{Cf.} \Cref{binary}), which indicates that the variations of the residuals can be due to pulsations. This would explain why the (BVS, VR) diagram of HD 149790 is not completely flat, as it is in fact the combination of a vertically spread diagram due to pulsations and a horizontally spread diagram due to the companion.

\subsubsection{GP candidate: HD 145467 b}

\label{sec_145467}

\begin{figure*}[h!]
\centering
\begin{subfigure}[t]{0.8\textwidth}
\centering
\includegraphics[width=1\hsize,valign=m]{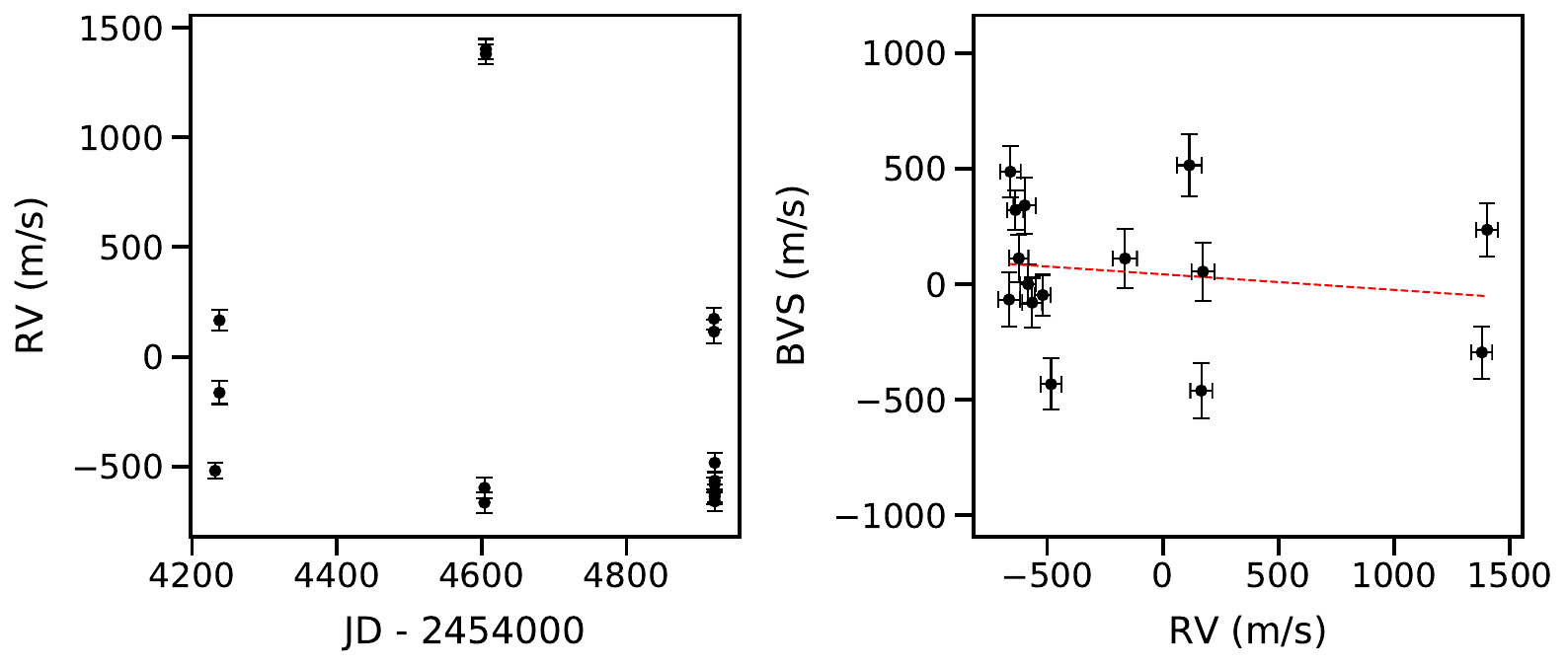}
\caption{\label{HD145_rv}}
\end{subfigure}
\begin{subfigure}[t]{0.8\textwidth}
\centering
\includegraphics[width=1\hsize,valign=m]{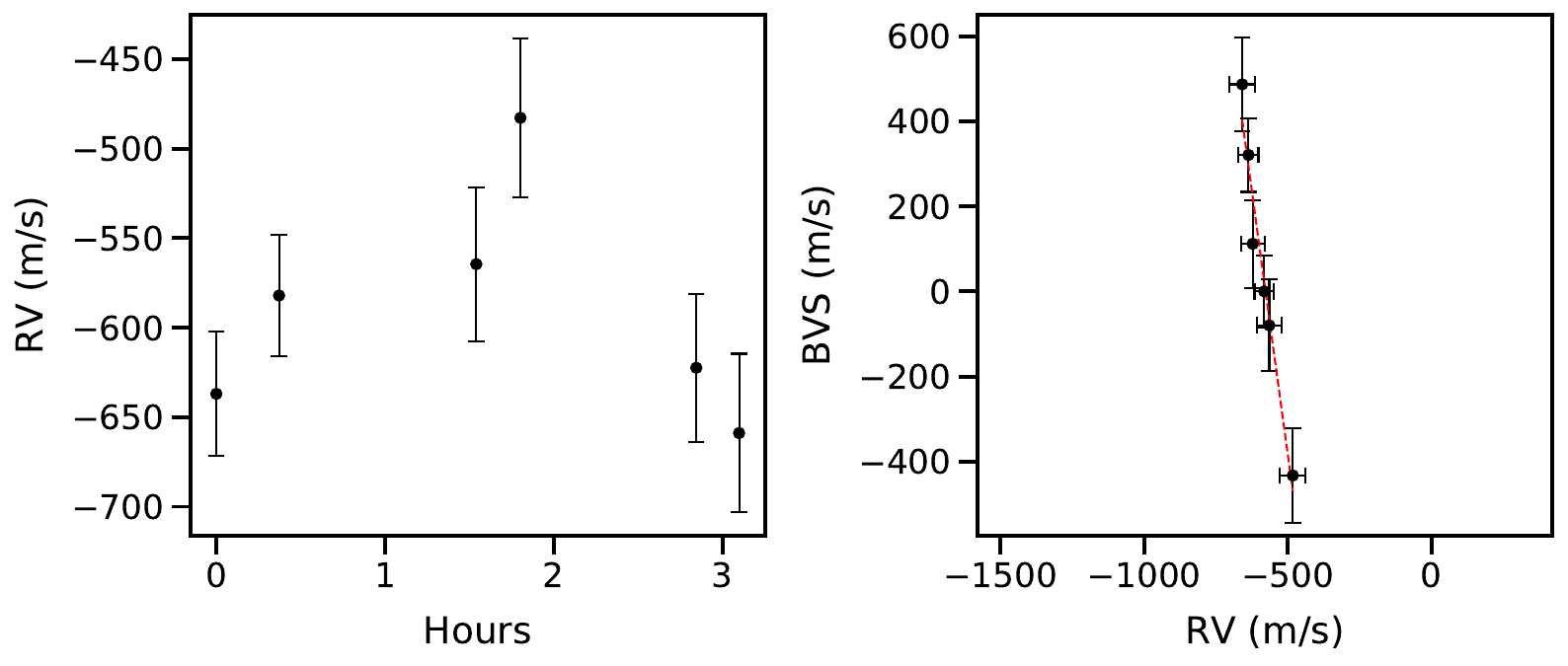}
\caption{\label{HD145_rv_zoom}}
\end{subfigure}
\caption{HD 145467 analysis. \protect\subref{HD145_rv}) RV variations  (\emph{left}) and corresponding (BVS, VR) diagram (\emph{right}).  
\protect\subref{HD145_rv_zoom})  Zoom on a 3-hour sequence (\emph{left}) and  corresponding  (BVS, VR) diagram (\emph{right}).  \label{HD145467_rv}}
\end{figure*}

HD 145467 is a $\SI{1.56}{\msun}$ F0V-type star.
We obtained $24$ spectra between April 2018 and March 2020 on this star. However, $8$ spectra (from the nights of March 12 and 13, 2020) were taken with a lunar angular separation below $\SI{9}{\degree}$. To avoid any contamination from the moon, we excluded these spectra from our analysis.
The  RV variations present a total amplitude of $\SI{2.5}{\kilo\meter\per\second}$ , and the (BVS, VR) diagram is spread both vertically (over $\SI{1}{\kilo\meter\per\second}$) and  horizontally  (over $\SI{2}{\kilo\meter\per\second}$; $R=-0.16$, $p_{value}< 58\%$, \emph{cf.} Figure \ref{HD145_rv}). This indicates that the RV variations are mainly due to a companion that produces the horizontal spreading, and to stellar pulsations, which produce the vertical spreading. A sequence of six continuous spectra  obtained on the night of March $12$, 2020, shows RV variations of more than $\SI{100}{\meter\per\second}$ over  $3$ hours  (\emph{cf.} Figure \ref{HD145_rv_zoom}), associated with BVS variations over $\SI{1}{\kilo\meter\per\second}$. The  (BVS, VR) diagram of this sequence is spread vertically (\emph{cf.} Figure \ref{HD145_rv_zoom}), which confirms that the short-term variations are due to pulsations. The RV variations of HD 145467 are then the sum of a companion signal with an amplitude of about $\SI{2}{\kilo\meter\per\second}$ and of pulsations with an amplitude of about $\SI{100}{\meter\per\second}$.

\begin{figure*}[h!]
\centering
\begin{subfigure}[t]{0.49\textwidth}
\centering
\includegraphics[width=1\hsize,valign=m]{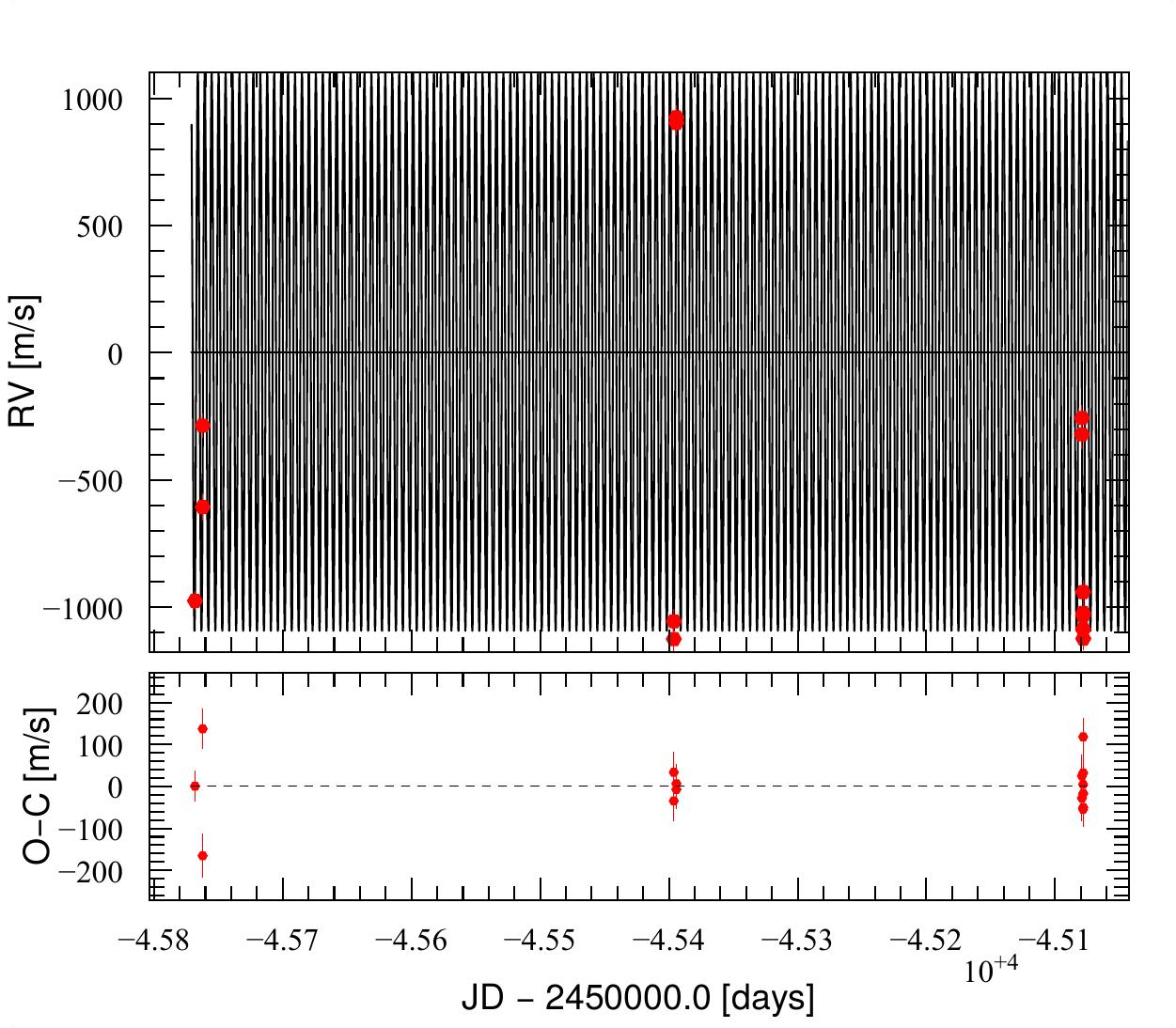}
\caption{\label{HD145_e0_fit}}
\end{subfigure}
\begin{subfigure}[t]{0.49\textwidth}
\centering
\includegraphics[width=1\hsize,valign=m]{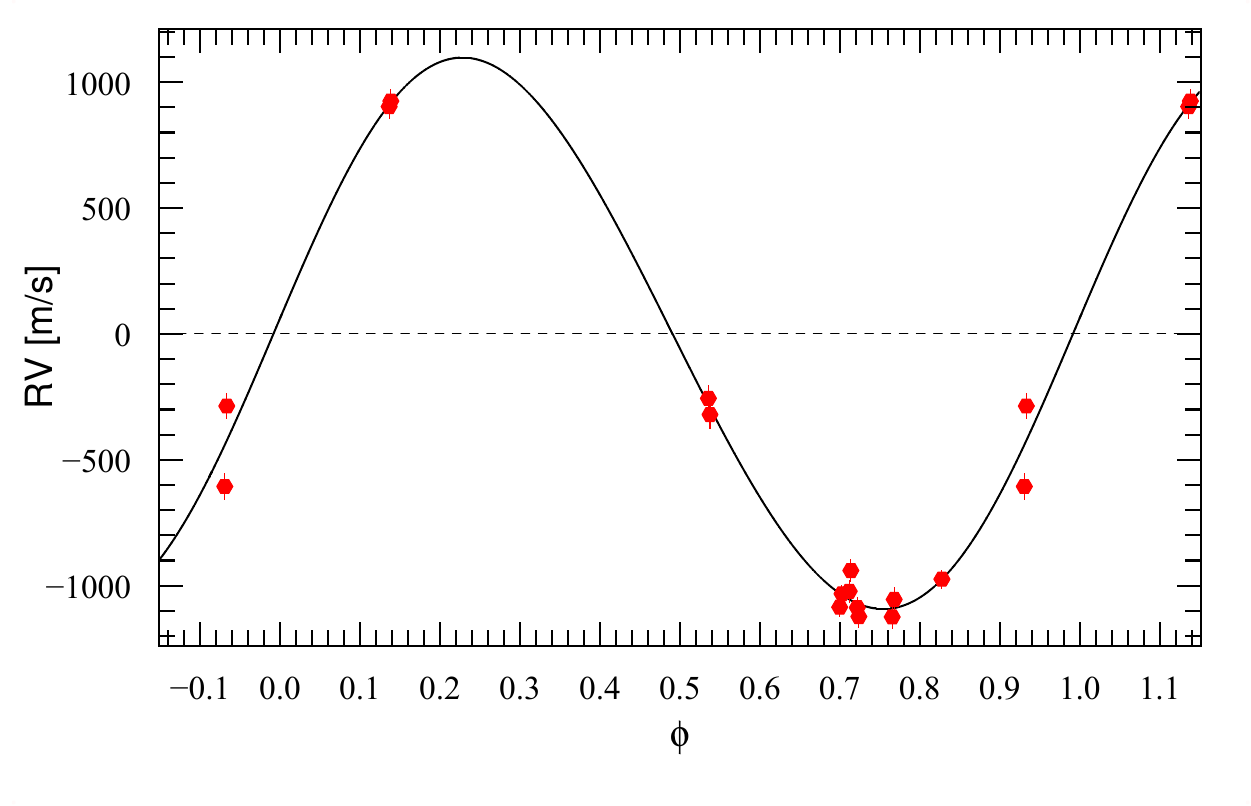}
\caption{\label{HD145_e0_phase}}
\end{subfigure}

\caption{HD 145467 companion analysis. \subref{HD145_e0_fit})  Best one-planet Keplerian fit of HD 145457's RV with $e<0.2$ and its associated residuals. \subref{HD145_e0_phase}) Phase-folded plot of the fit. \label{HD145467_yorbit}}
\end{figure*}

\begin{figure*}[h!]
\centering
\begin{subfigure}[t]{0.8\textwidth}
\centering
\includegraphics[width=1\hsize,valign=m]{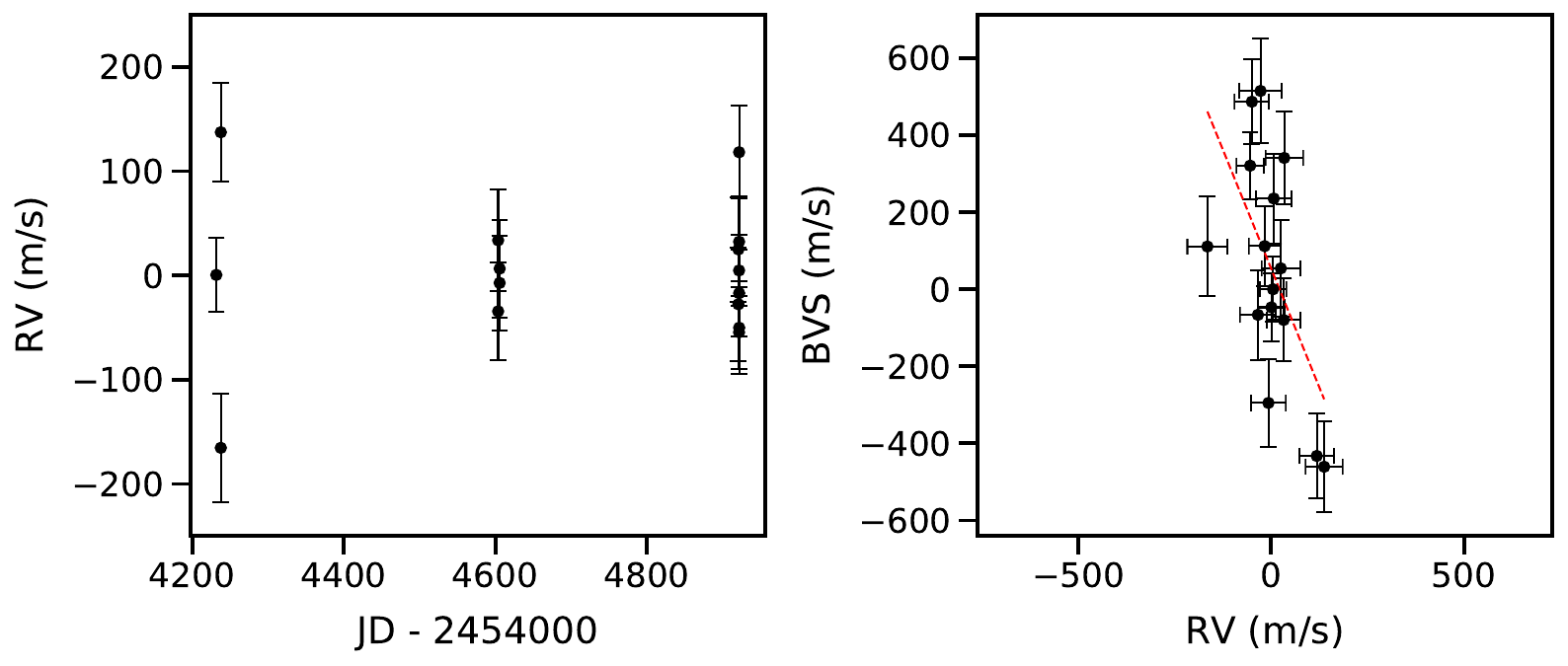}
\end{subfigure}

\caption{HD 145467 companion analysis. \emph{Left:} RV residuals of HD 145467 best one-planet Keplerian fit with $e<0.2$. \emph{Right:} (BVS, RV residuals)  diagram and its best linear model (dashed red line). \label{HD145_fit}}
\end{figure*}

We used \emph{yorbit} to constrain the period and $m_c\sin{i}$ of the companion. The number of data points is small, however, and the data are clearly impacted by pulsations. For simplicity, we therefore assume an eccentricity lower than $0.2$.
We obtain a period of  $\sim5.4 \ \si{\day}$, and the $m_c\sin{i}$ is $\sim\SI{13}{\MJ}$  (\emph{cf.} Figure \ref{HD145_e0_fit} and \ref{HD145_e0_phase}). The residuals of the fit present variations with a standard deviation of  $\SI{70}{\meter\per\second}$ , and their (BVS, RV residuals) diagram is spread vertically (\emph{cf.} Figure \ref{HD145_fit}). This indicates that the residuals are due to  pulsations.

We conclude that HD 145467 has a companion with a minimum mass  of $\sim\SI{13}{\MJ}$. However, due to a  lack of RV data, its period and eccentricity are not precisely constrained. Additional data are then needed to confirm this HJ or short-period BD companion candidate. If confirmed, this companion would be of particular interest, as only a few confirmed HJs of that age ($< \SI{20}{\mega\year}$) are known to date \citep{Cameron,Deleuil, Tanimoto,Mann, Alsubai,David,Rizzuto}.

\subsection{Summary of the detected companions}

$11$ spectroscopic binaries ($8$ SB1 and $3$ SB2) were identified in our analysis of the Sco-Cen final sample; two additional were known. This sample therefore includes $13$ spectroscopic binaries ($10$ SB1 and $3$ SB2). 
In addition, one short-period ($P < \SI{10}{\day}$) BD companion candidate was identified around HD 149790, and one tentative short-period ($P < \SI{10}{\day}$) companion candidate, at the limit between BD and planets, was identified around HD 145467.

\section{\harps \  Sco-Cen, \sophie \ YNS, and \harps \ YNS combined survey analysis}
\label{HS}

To improve the statistics of GPs around young stars, we combine below the \harps \  Sco-Cen survey described above with the  \sophie \ \citep{Grandjean_SOPHIE} and the  \harps \  \citep{Grandjean_HARPS} \ YNS surveys.

\subsection{Combined sample}

\label{descritpion}

The \harps \ and \sophie \ YNS combined sample included $143$ targets \citep{Grandjean_SOPHIE}. Four of them are in common with our\  Sco-Cen sample: HD 95086, HD 102458, HD 106906, and HD 131399 (which are part of the \harps \ YNS survey). 

We excluded the targets whose data were not suitable for the  statistical analysis. From the \harps \ and \sophie \ YNS combined sample, we excluded the targets that were excluded in Section 2.1 and 4.1 of \cite{Grandjean_SOPHIE}: stars whose \vsini \ is too high ($>\SI{300}{\kilo\meter\per\second}$) to allow RV measurements, stars that are too old, and binary stars for which the companion signal could not be fitted. From the Sco-Cen survey, we excluded the stars excluded in \cref{ScoCen_sample} of this paper. In addition, we excluded the stars with companions whose RV signal could not be fitted: HD 111102, HD 121176, HD 126488, HD 129590, HD 137057, and HD 143811. Finally, we excluded HD 114319 and HD 145467 because their companions still need confirmation.

\begin{figure}[ht!]
  \centering
\begin{subfigure}[t]{0.32\textwidth}
\includegraphics[width=1\hsize,valign=m]{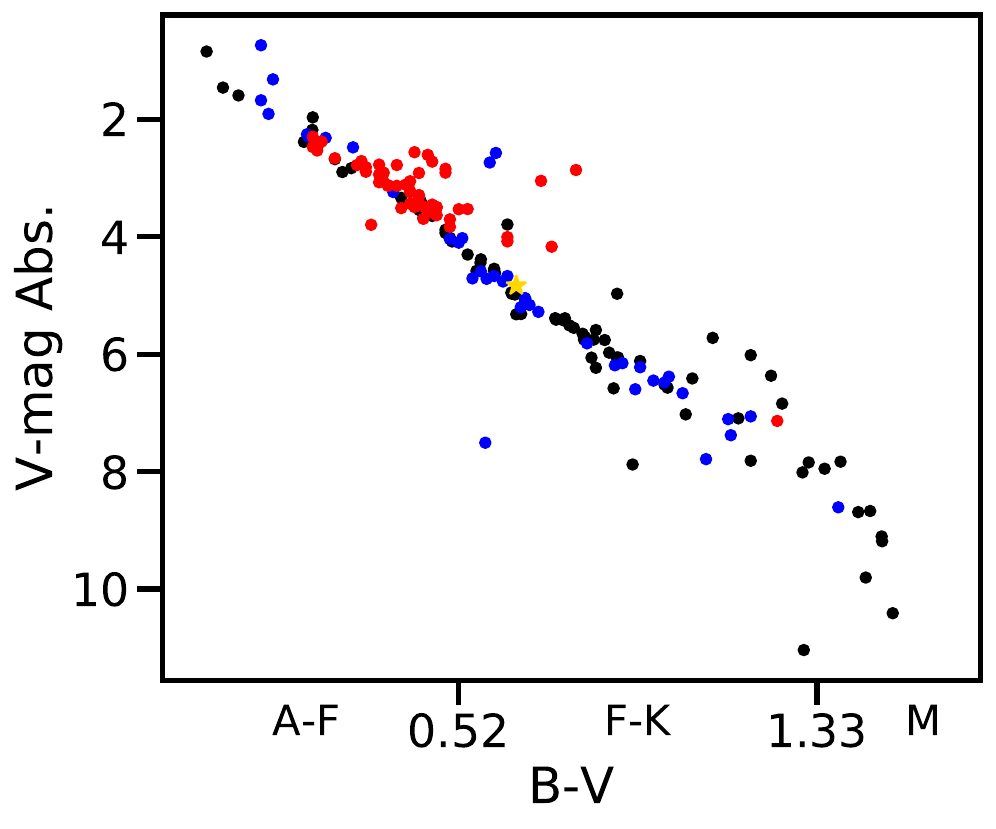}
\caption{\label{HSSCO_HR}}
\end{subfigure}
\begin{subfigure}[t]{0.32\textwidth}
\includegraphics[width=1\hsize,valign=m]{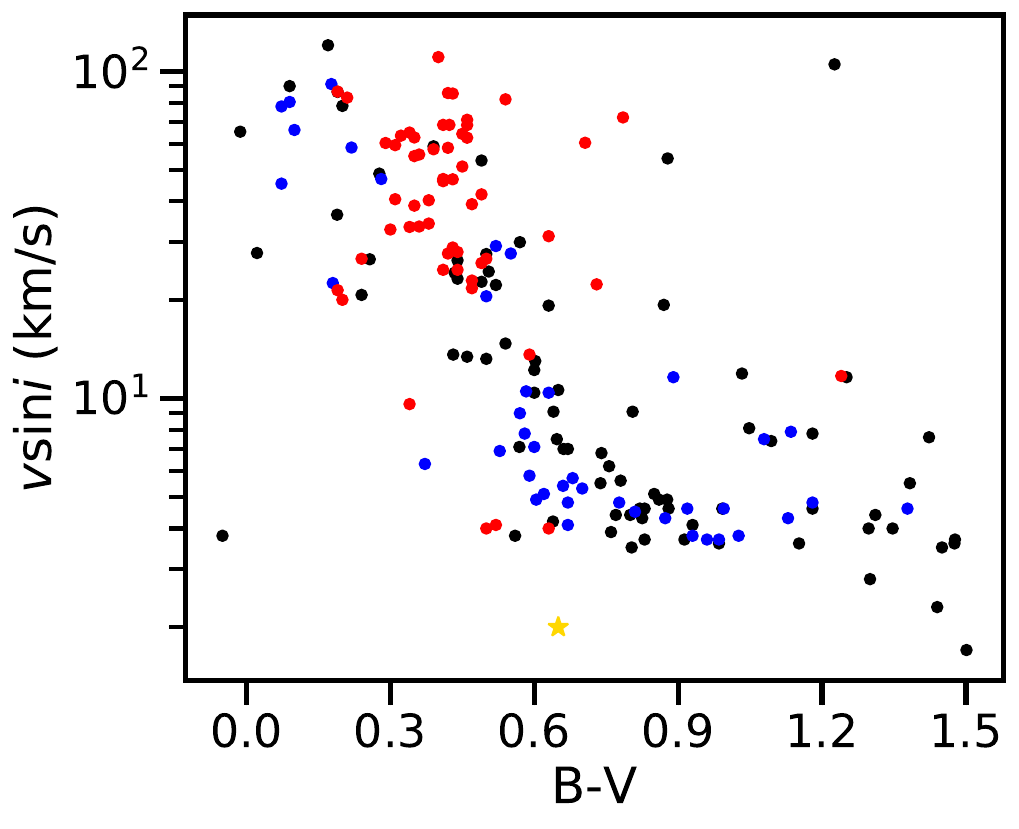}
\caption{\label{HSSCO_vsini}}
\end{subfigure}

\caption{Main physical properties of our final combined sample. Black shows \harps \ YNS targets, blue shows \sophie \ YNS targets, and red shows \harps \ Sco-Cen targets. 
 \subref{HSSCO_HR})  Absolute $V$-magnitude vs. \bv. 
 The solar properties are displayed (yellow stars) for comparison. 
\subref{HSSCO_vsini}) \vsini~vs. \bv~distribution.}
       \label{survey_carac_1_hssco}
\end{figure}

\begin{figure}[ht!]
  \centering
\begin{subfigure}[t]{0.32\textwidth}
\includegraphics[width=1\hsize]{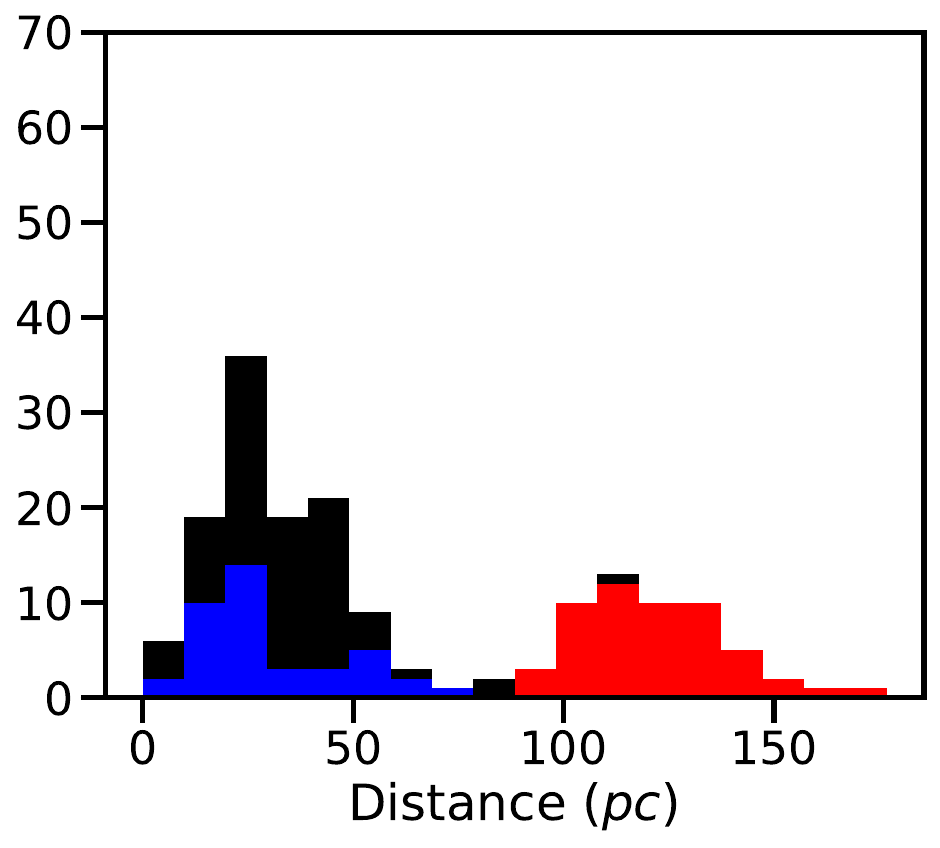}
\caption{\label{HSsco_dist}}
\end{subfigure}
\caption{Gaia DR2 \citep{DR2A1} distance histogram of our final  combined  sample. The  \harps \ YNS (black) histogram, the \sophie \ YNS (blue) histogram, and the  \harps \ Sco-Cen (red) histogram  are stacked.}
       \label{survey_dist_hssco}
\end{figure}

\begin{figure}[ht!]
  \centering
\begin{subfigure}[t]{0.32\textwidth}
\includegraphics[width=1\hsize]{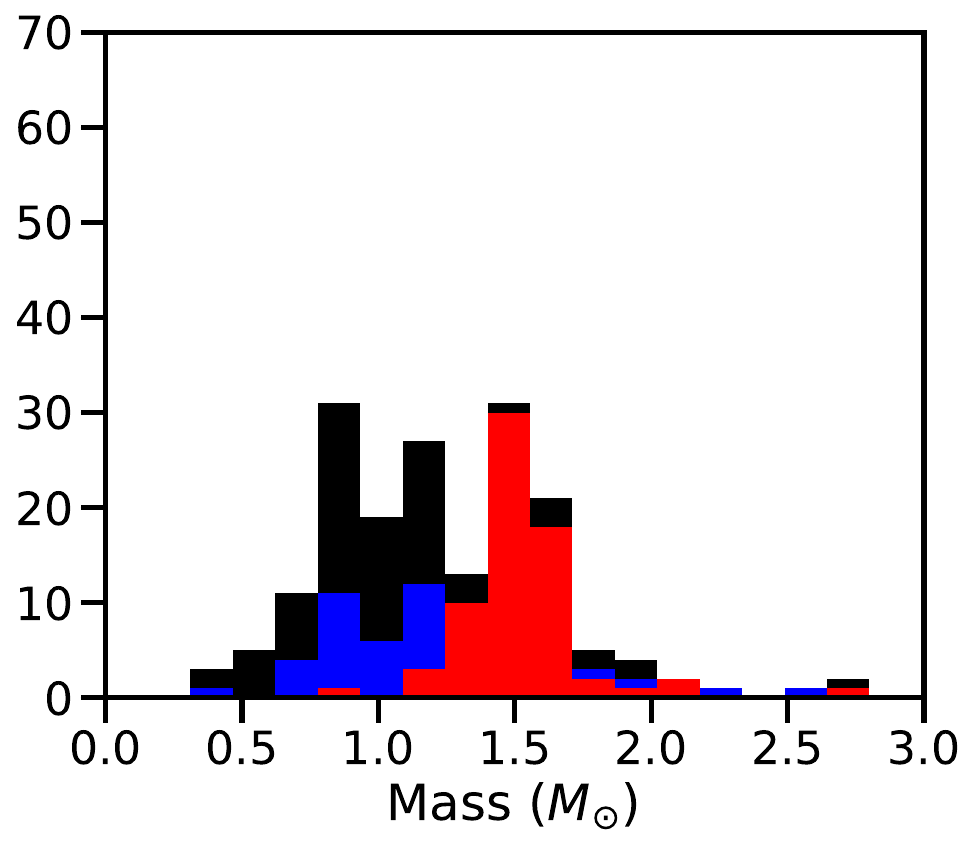}

\caption{\label{HSSCO_mass}}
\end{subfigure}
\begin{subfigure}[t]{0.32\textwidth}
\includegraphics[width=1\hsize]{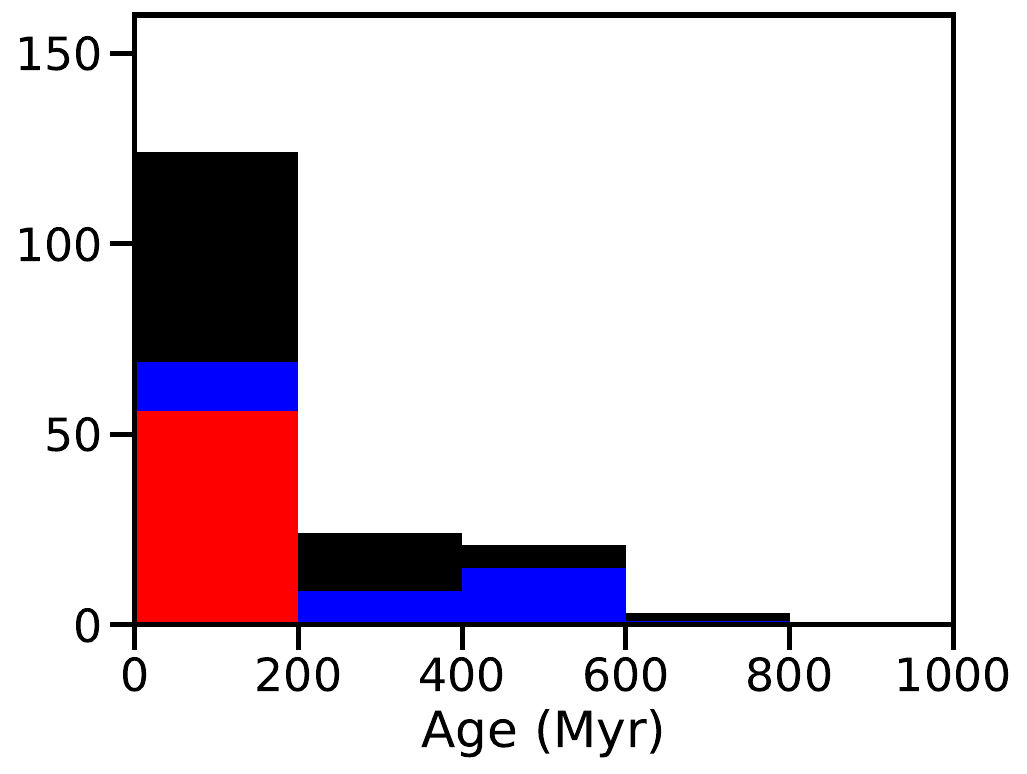}
\caption{\label{HSSCO_age}}
\end{subfigure}

\caption{Main physical properties of our final combined sample.
\subref{HSSCO_mass}) Histogram of the stellar masses (in \Msun).
\subref{HSSCO_age}) Age histogram. The bin size has been chosen to be greater than the standard deviation of the ages of the sample.
The  \harps \ YNS (black) histogram, the \sophie \ YNS (blue) histogram, and the  \harps \ Sco-Cen (red) histogram  are stacked.}
       \label{survey_age_hssco}
\end{figure}

\begin{figure}[ht!]
  \centering
\begin{subfigure}[t]{0.32\textwidth}
\includegraphics[width=1\hsize]{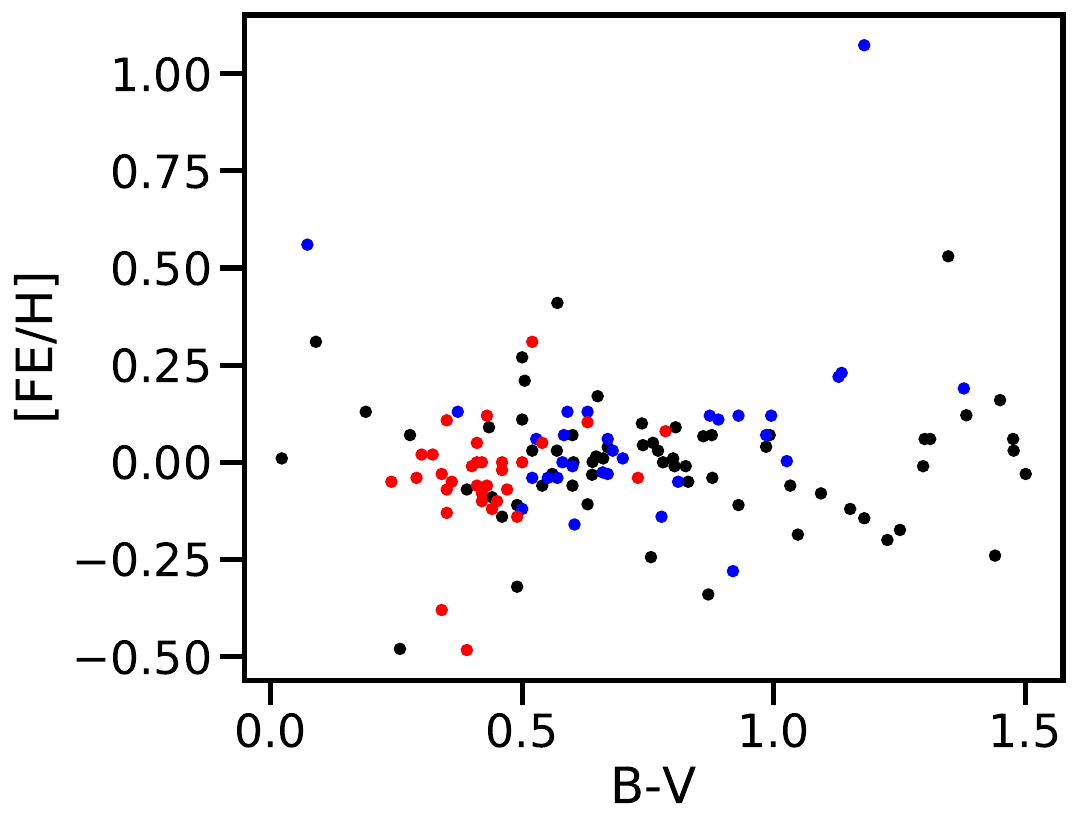}
\caption{\label{FEH_BV}}
\end{subfigure}
\begin{subfigure}[t]{0.32\textwidth}
\includegraphics[width=1\hsize]{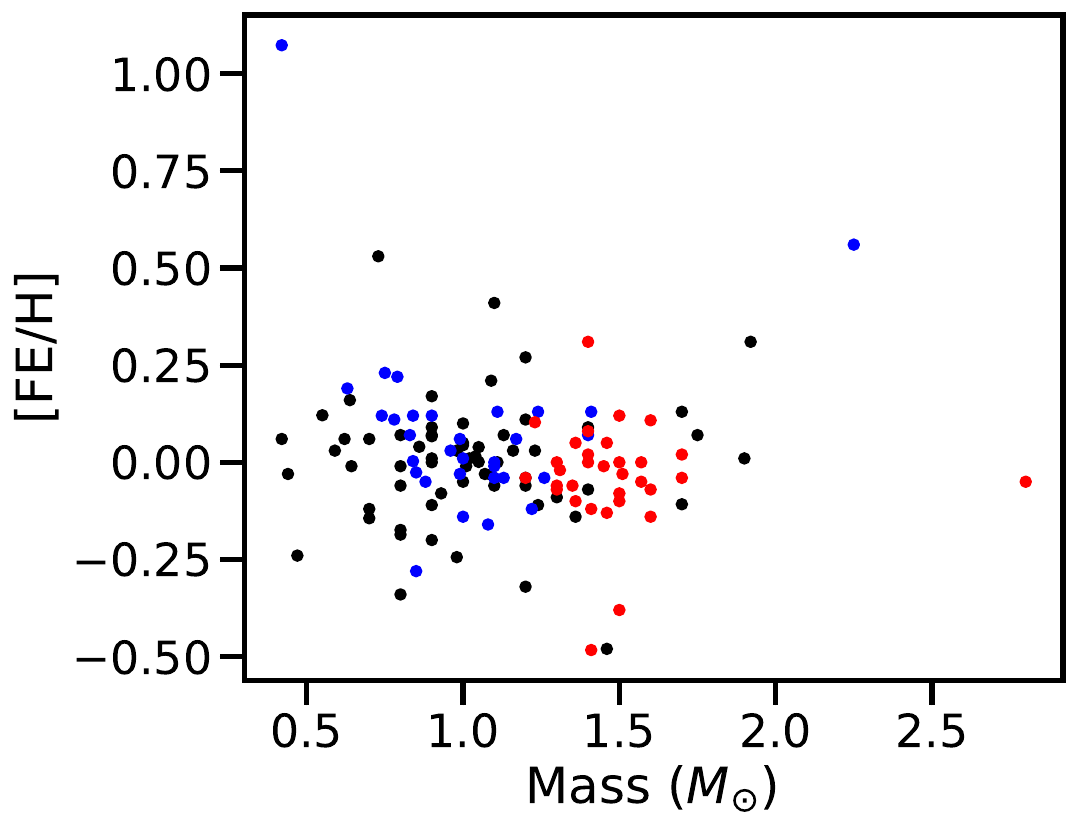}
\caption{\label{FEH_M}}
\end{subfigure}
\caption{Metallicity of our final combined sample against \bv \ (\subref{FEH_BV}) and against stellar mass (\subref{FEH_M}).  Black shows \harps \ YNS targets, blue shows \sophie \ YNS targets, and red shows \harps \ Sco-Cen targets. }
       \label{FEH}
\end{figure}

\begin{figure}[h!]
  \centering

\includegraphics[width=0.75\hsize]{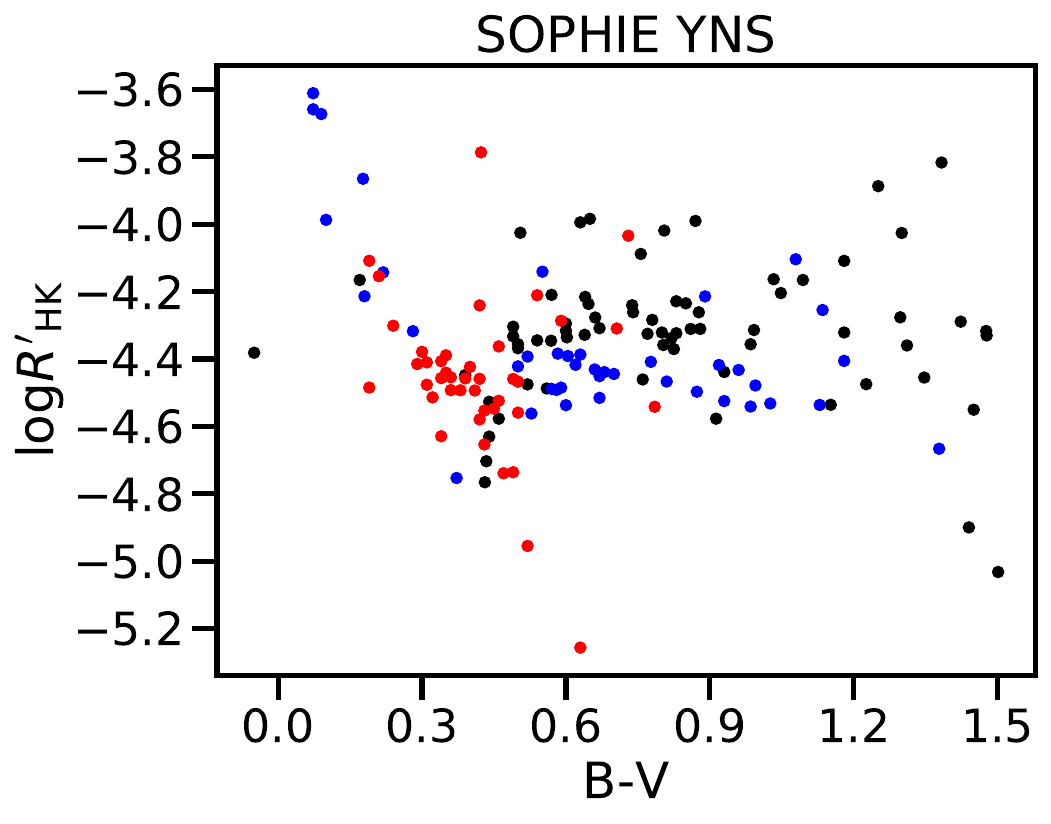}

\caption{Final combined sample $<$\rhk$>$ \ vs.  \bv.  Black shows \harps \ YNS targets, blue shows \sophie \ YNS targets, and red shows \harps \ Sco-Cen targets.} 
       \label{stell_rhk_hssco}
\end{figure}

\begin{figure*}[ht!]
  \centering
\begin{subfigure}[t]{0.32\textwidth}
\includegraphics[width=1\hsize]{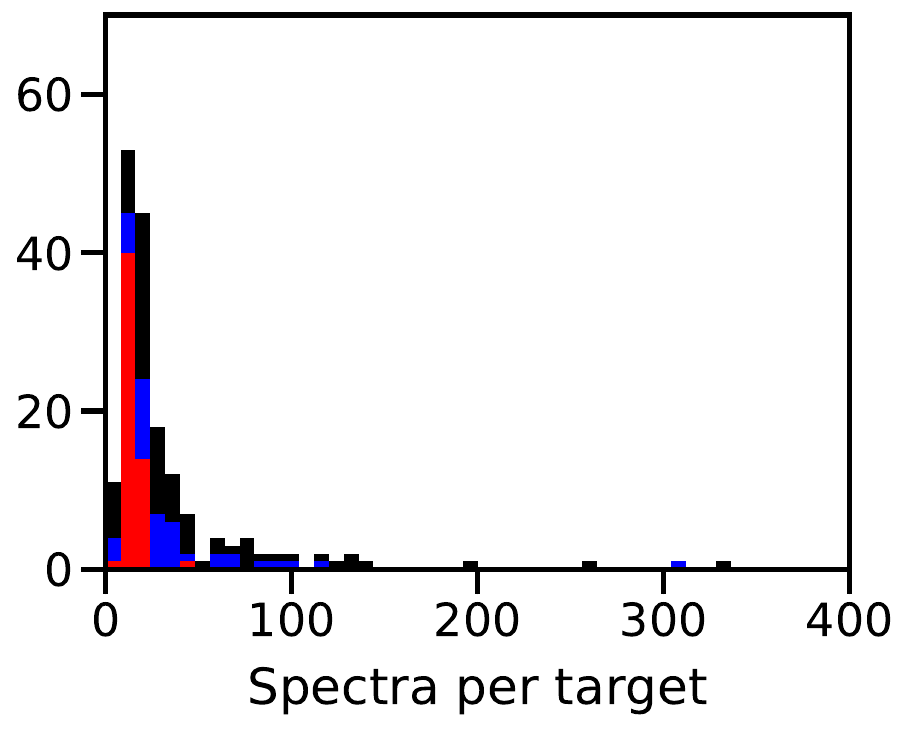}
\caption{\label{Nmes}}
\end{subfigure}
\begin{subfigure}[t]{0.32\textwidth}
\includegraphics[width=1\hsize]{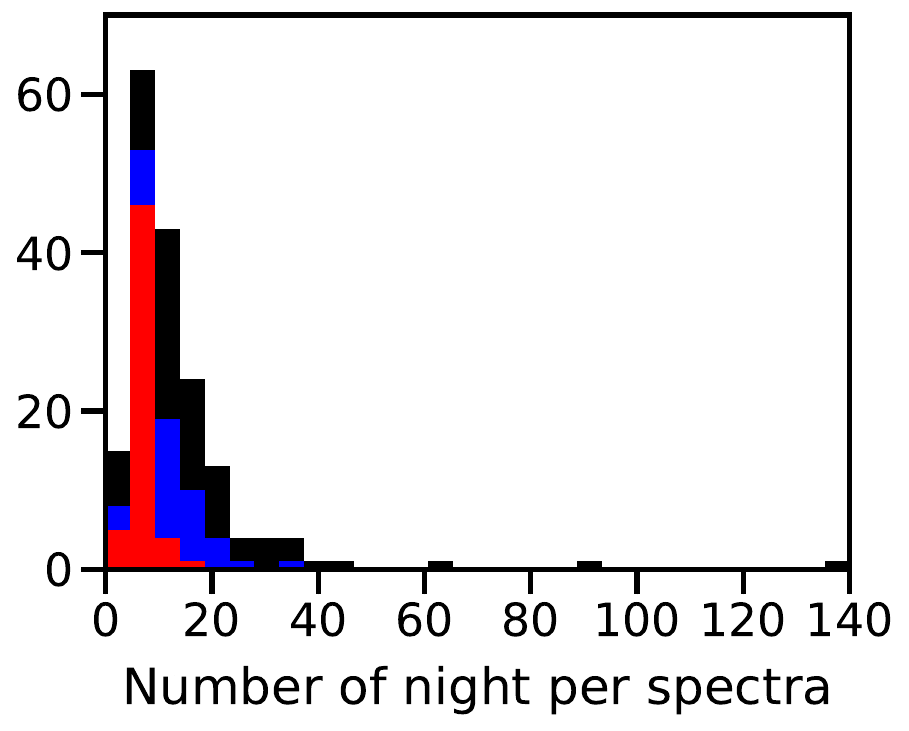}
\caption{\label{Nb_day}}
\end{subfigure}
\begin{subfigure}[t]{0.32\textwidth}
\includegraphics[width=1\hsize]{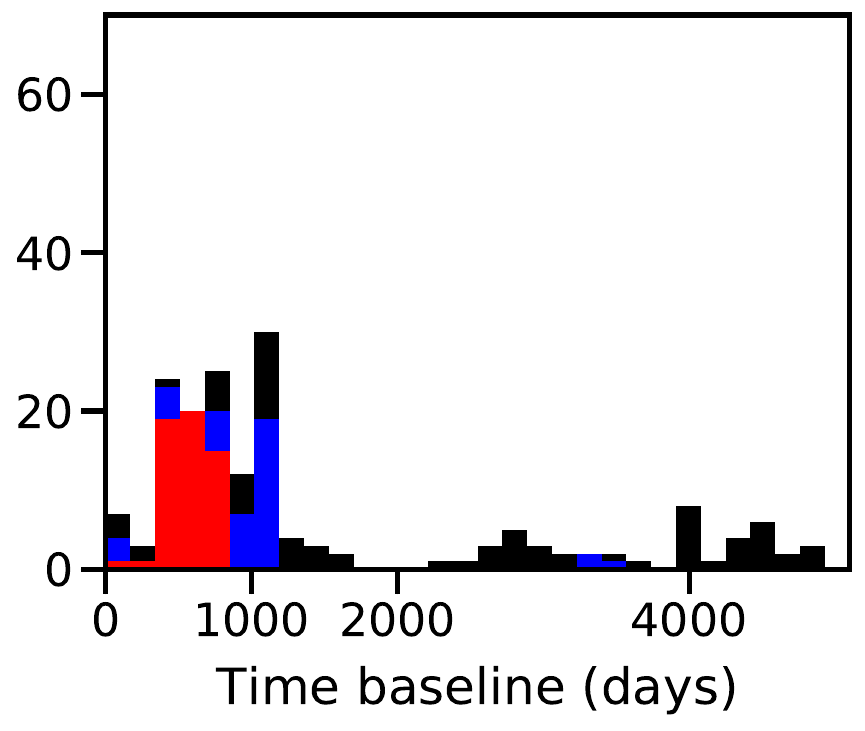}
\caption{\label{time_bsl}}
\end{subfigure}
\caption{Observation summary of the final combined sample.
 \subref{Nmes}) Histogram of the number of spectra per target, HD 216956 (Fomalhaut, $834$ spectra) and HD 039060 ($\beta$ Pic, $5108$ spectra) are not displayed.
 \subref{Nb_day}) Histogram of the  number of nights per target. 
\subref{time_bsl}) Histogram of the time baselines.
The  \harps \ YNS (black) histogram, the \sophie \ YNS (blue) histogram, and the  \harps \ Sco-Cen (red) histogram  are stacked.} 
       \label{survey_carac_2_hssco}
\end{figure*}

The final combined sample gathers $176$ A0V to M5V targets (\emph{cf.} \Cref{survey_carac_1_hssco}) that are suitable for our statistical analysis.
Their main properties are listed hereafter:
\begin{itemize}
\item[-] $80$ targets have a spectral type between  A0 and F5V ($B-V \in [-0.05:0.52[$; hereafter AF subsample), $87$  have a spectral type between F6 and K5 ($B-V \in [0.52:1.33[$; hereafter FK subsample), and $9$ have a spectral type between K6 and M5 ($B-V \geq 1.33$; hereafter M subsample);
\item[-] Their distances range between $3$ and $\SI{177}{\pc}$, with a median of $\SI{43}{\pc}$ (\emph{cf.} \Cref{survey_dist_hssco}, \cite{DR2A1}. 
\item[-] Their projected rotational velocities (\vsini) range from $1.7$ to $\SI{120}{\kilo\meter\per\second}$, with a median of  $\SI{13.5}{\kilo\meter\per\second}$;
\item[-] Their V-band relative magnitude ranges between  $1.16$ and $12.18$, with a median of $8.11$;
\item[-] Their masses range between $0.42$ and $\SI{2.8}{\msun}$, with a median of  $\SI{1.2}{\msun}$ (\emph{cf.} \Cref{HSSCO_mass}; \emph{cf.} \Cref{age_mass} for the determination of the masses). The AF (FK and M) subsample has a median mass of $1.5$ ($0.98$ and $0.6$) solar mass, with a standard deviation of $0.31$ ($0.22$, and $0.08$) solar mass;
\item[-] Their median age (\emph{cf.} histogram in \Cref{HSSCO_age}) is $\SI{45}{\mega\year}$ (\emph{cf.} \Cref{age_mass} for the determination of the ages). $97\%$ are younger than  $\SI{600}{\mega\year}$;
\item[-] Their metallicities are close to the solar metallicity (\emph{cf.} \Cref{FEH}), with a median of  $0.0 \ dex$ (mean of $0.01 \ dex$) and a standard deviation of $0.18 \ dex$ (metallicity measurements are only available for $127$ of our targets in the Strasbourg astronomical data center (hereafter CDS database));
Similarly to the \harps \ and \sophie \ YNS surveys \citep{Grandjean_SOPHIE}, we observe no statistically significant correlation between the metallicity and the \bv, nor between the metallicity and the stellar mass in our final combined sample; 
\item[-] $166$ stars out of $176$ present Ca II emission in their spectra.  \Cref{stell_rhk_hssco} shows the $<$\rhk$>$ \ vs. \bv. The median $<$\rhk$>$ is $-4.4$, with a standard deviation of $0.3$. Eight targets present signs of low activity ($<$\rhk$>$ $<-4.75$), $132$ are active ($-4.75<$ $<$\rhk$>$ $< -4.2$), and $26$ present signs of high activity ($<$\rhk$>$ $-4.2$);
\item[-] Finally, the median time baseline is $\SI{906}{\day}$ (mean time baseline of  $\SI{1546}{\day}$), and the median number of spectra per target is $19$ (mean of $67$), spaced on a median number of $\SI{11}{\night}$  (mean of $14$, \Cref{survey_carac_2_hssco}).
\end{itemize}

Details can be found in the Tables \ref{tab_carac_sco},  \ref{tab_result_sco}, \ref{tab_carac_s}, \ref{tab_result_s}, \ref{tab_carac_h},  and \ref{tab_result_h} (Tables \ref{tab_carac_s} \ref{tab_result_s}, \ref{tab_carac_h}, and \ref{tab_result_h}   are updated versions of the Tables A.1, A.2, A.3, and A.4  presented in \citep{Grandjean_SOPHIE}, respectively).

\subsection{Stellar intrinsic variability}

\begin{figure*}[t!]
  \centering
\begin{subfigure}[m]{0.32\textwidth}
\includegraphics[width=1\hsize]{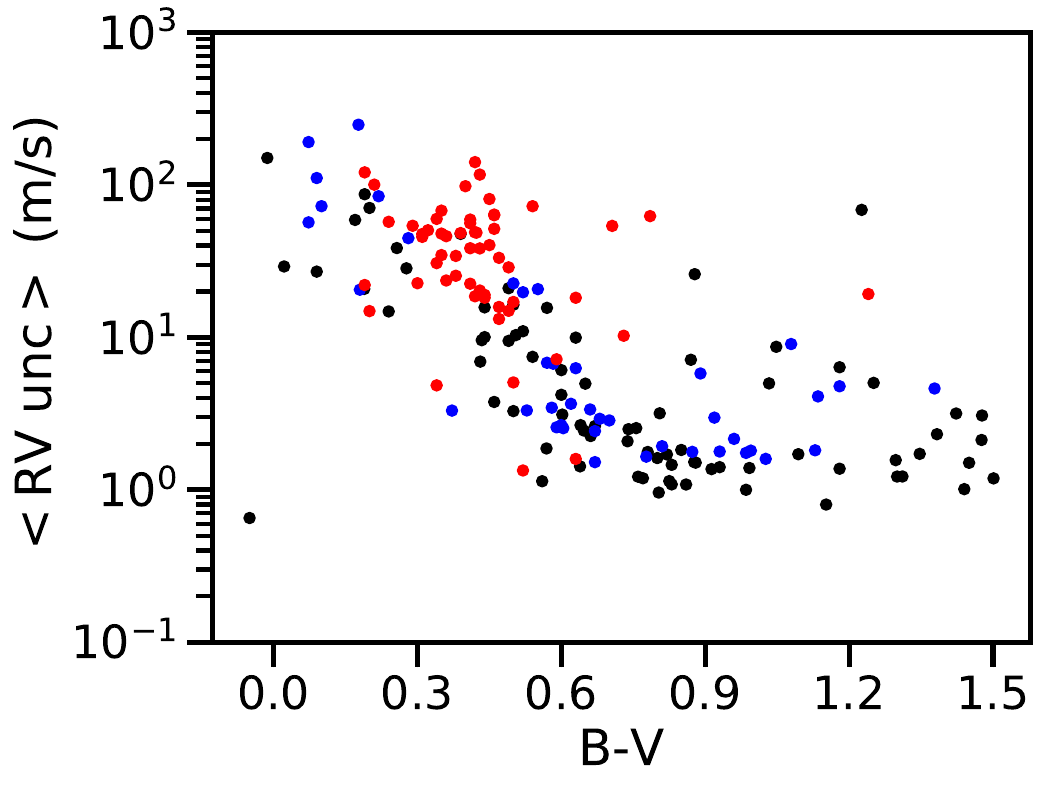}
\caption{\label{HSSCO_unc_BV}}
\end{subfigure}
\begin{subfigure}[m]{0.32\textwidth}
\includegraphics[width=1\hsize]{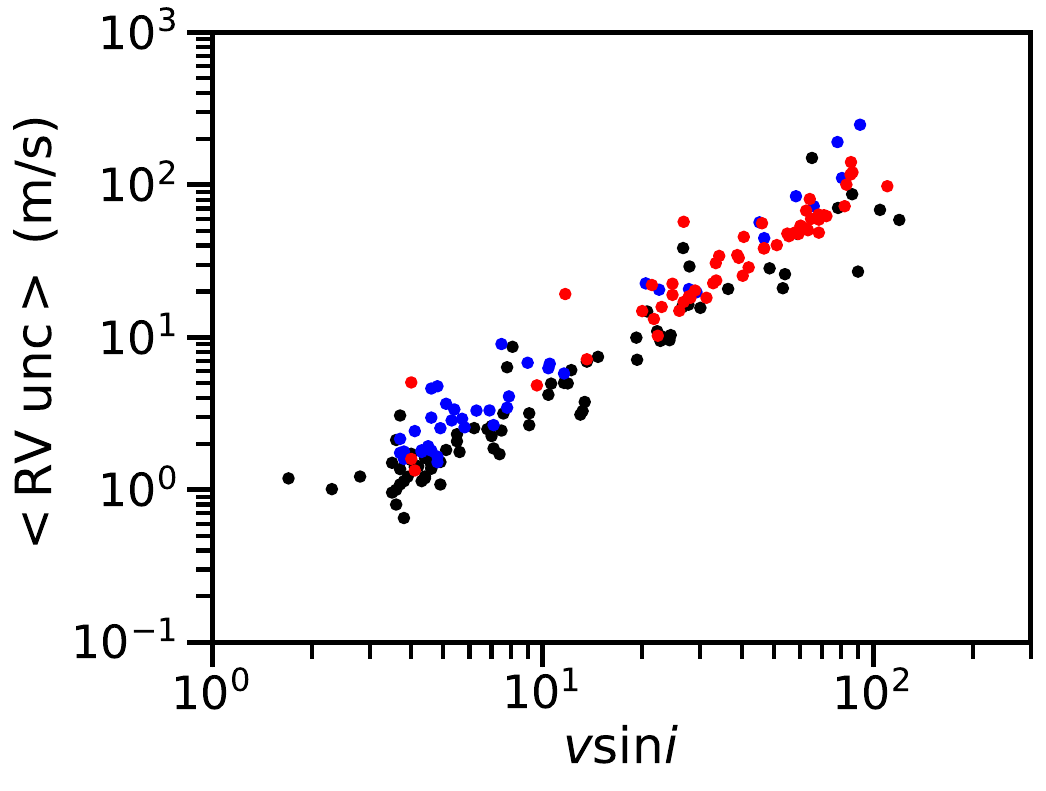}
\caption{\label{HSSCO_unc_vsini}}
\end{subfigure}
\begin{subfigure}[m]{0.32\textwidth}
\includegraphics[width=1\hsize]{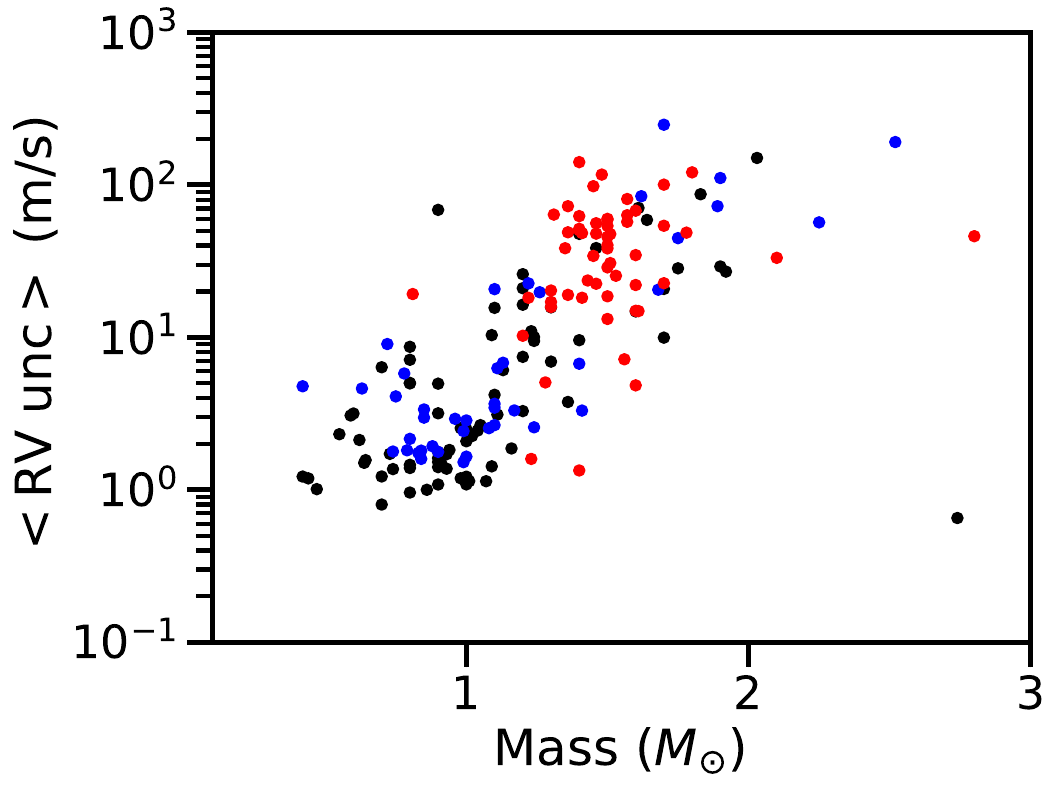}
\caption{\label{HSSCO_unc_M}}
\end{subfigure}
\caption{Summary of the RV uncertainties of the  final combined sample.  Mean RV uncertainty (accounting for the photon noise only) vs. \bv (\subref{HSSCO_unc_BV}), vs. \vsini \ (\subref{HSSCO_unc_vsini}) and vs. \Mstar \ (in \Msun, \subref{HSSCO_unc_M}).  Black shows \harps \ YNS targets, blue shows \sophie \ YNS targets, and red shows \harps \ Sco-Cen targets.}
       \label{stell_var_hssco_1}
\end{figure*}

\begin{figure*}[t!]
  \centering
\begin{subfigure}[m]{0.32\textwidth}
\includegraphics[width=1\hsize]{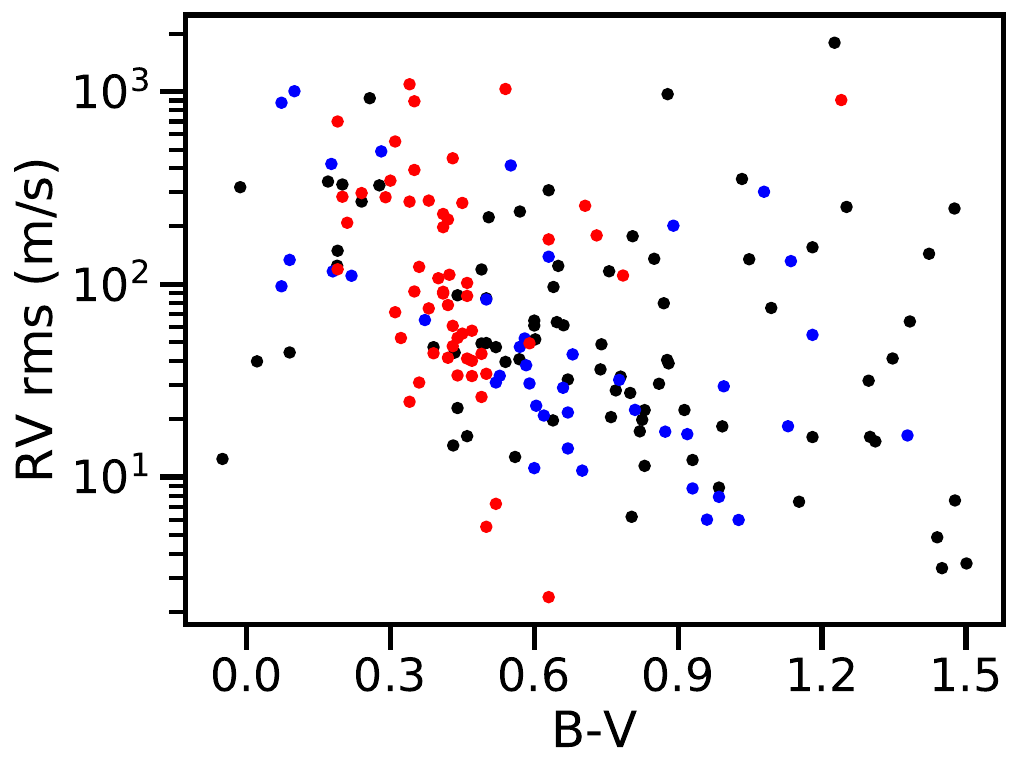}
\caption{\label{HSSCO_rms_BV}}
\end{subfigure}
\begin{subfigure}[m]{0.32\textwidth}
\includegraphics[width=1\hsize]{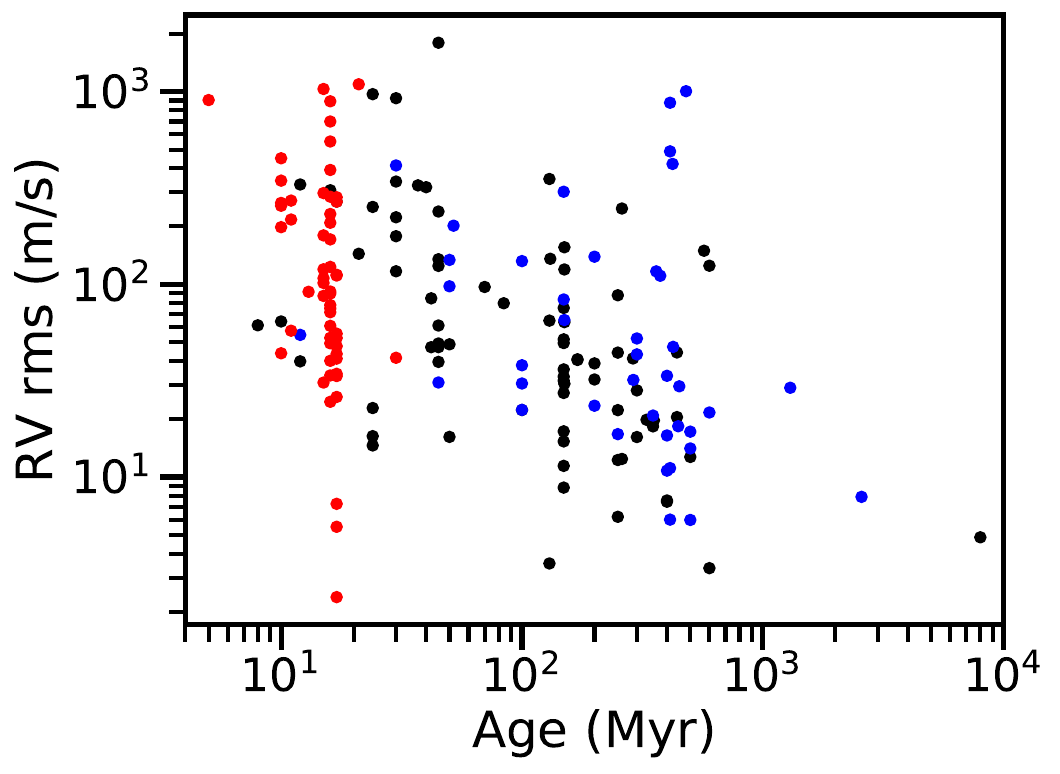}
\caption{\label{HSSCO_rms_A}}
\end{subfigure}
\caption{Summary of the RV rms of the final combined sample. RV rms vs. \bv \ (\subref{HSSCO_rms_BV}), and vs. age  (\subref{HSSCO_rms_A}).  Black shows \harps \ YNS targets, blue shows \sophie \ YNS targets, and red shows \harps \ Sco-Cen targets.}
       \label{stell_var_hssco_2}
\end{figure*}

The stars in our final combined sample present strong jitter in their RV time series, which is consistent with the stars being active (solar- to late-type stars) or pulsating (early-type stars).
Pulsating stars stars show a vertically spread (BVS, RV) diagram, while stars with magnetic activity present a correlation between BVS and RV\footnote{This applies when the stellar lines are resolved  \citep{Desort}, which is the case for our targets (\emph{cf.} \Cref{tab_carac_sco,tab_carac_s,tab_carac_h})}  \citep{Lagrange_2009}. 
According to these criteria, we report the main  origin of the RV variations in Tables \ref{tab_result_sco}, \ref{tab_result_s} and  \ref{tab_result_h}.  \newline
After removing the signal due to a companion when relevant (\emph{cf.} \Cref{comp}), and additionally removing the drift induced by the secular acceleration for HD 217987 \citep{Grandjean_HARPS}, the median of the RV rms of the final combined sample is $\SI{54}{\meter\per\second}$ (mean of $\SI{153}{\meter\per\second}$) and the ratio of the RV  rms and the mean RV uncertainty varies between $1$ and $120$, with a median at $9$.
We display in \Cref{stell_var_hssco_1} the mean RV uncertainty versus  the \bv, versus the \vsini, and versus the $M_*$. We also display the RV rms versus \bv and versus age in \Cref{stell_var_hssco_2}.
The mean RV uncertainty is strongly correlated with the \vsini \ ($R=0.84$, $p_{value}< \num{4e-47}\%$). This correlation arises because the spectral lines become broader and shallower with increasing \vsini , which decreases our precision on RV computation. In addition, we observe a weak negative correlation between  mean RV uncertainty and \bv \ ($R=-0.52$, $p_{value}< \num{2e-13}\%$). This correlation arises because early-type stars present fewer spectral lines and higher \vsini \ than late-type stars, which decreases our precision on the RV computation for these stars.

\subsection{RV correction for activity and companions}

To obtain the best detection limits, we need to correct the targets RVs for stellar activity and from any known companion signal whenever possible. This has been done in the analyses of the \harps \ and \sophie \ YNS surveys  \citep{Grandjean_HARPS,Grandjean_SOPHIE}. Here, we proceeded as in \cite{Grandjean_HARPS}: 
\begin{itemize} 
\item[-]  When the RV variations are mainly due to spots (marked A in the \cref{tab_result_sco}), we corrected the RVs for this stellar activity as proposed in \cite{Melo_BVS_RV_corr}: A linear regression between the RVs and the BVS was  made and the resulting RV linear function was subtracted from the observed RVs; 
\item[-] When a trend due to a companion was present (\emph{cf.} \Cref{trend_bin}), we applied a linear regression on their RVs. If the corresponding residuals present a correlation between the BVS and the RVs, we corrected them for this correlation using the \cite{Melo_BVS_RV_corr} method (\emph{Cf.} point above); 
\item[-] When a companion was present and its properties are well characterized (\emph{cf.} \cref{trend_bin}), we used the residuals of the fit. If these residuals present a correlation between the BVS and the RVs, we corrected them for this correlation again using  \cite{Melo_BVS_RV_corr} method.
\end{itemize}

The detection limits were then computed using the RV residuals of these corrections.

\subsection{Detection limits}
\label{detlim}

We used the local power analysis (LPA; \cite{Meunier,Simon_IX}) to compute the $m_c\sin{i}$ detection limits for periods between $1$ and \SI{1000}{days} in the GP domain (between $1$ and $\SI{13}{\MJ}$) and in the BD domain (between $13$ and  $\SI{80}{\MJ}$). The LPA method  determines the minimum $m_c\sin{i}$  for all periods $P$
for which a companion on a circular orbit with a period $P$ would lead to a signal consistent with the data by comparing the maximum power of the periodogram of the synthetic companion  to the maximum power of the data periodogram within a small period range around the period $P$. For a given star, the detection limit is infinite for periods greater than its time baseline. We made this choice because the high jitter and the moderate number of spectra per target prevent us from obtaining a strong constraint on a companion signal  that has a period longer than the time baseline.

\begin{figure*}[t!]
  \centering
 \begin{subfigure}[m]{0.49\textwidth}
\includegraphics[width=1\hsize]{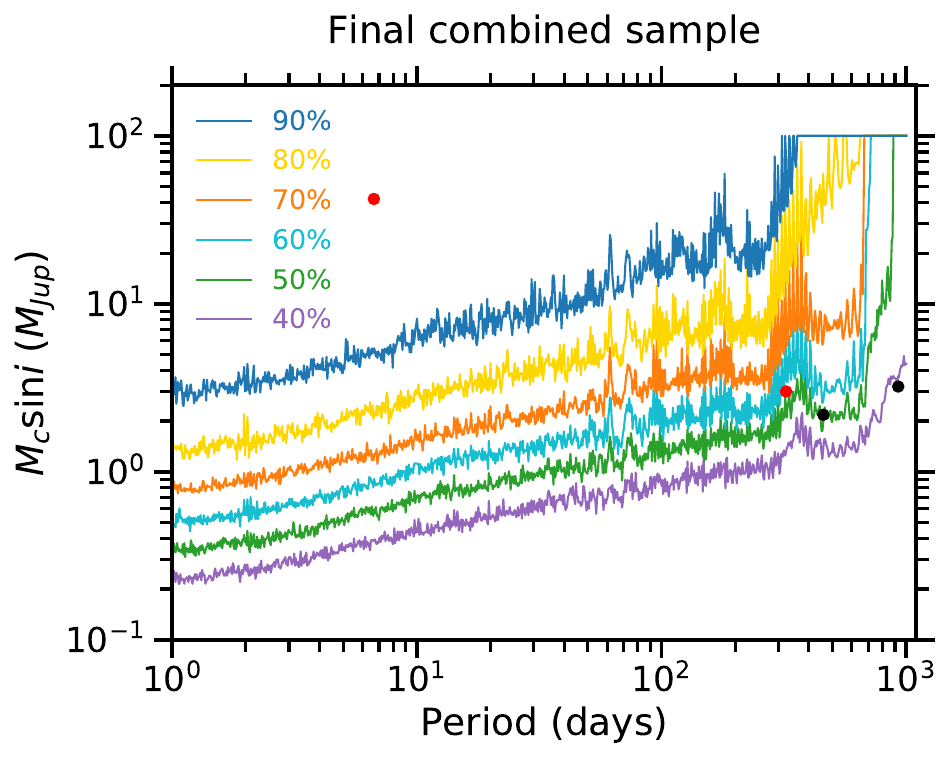}
\caption{\label{Completeness_ALL}}
\end{subfigure}
\begin{subfigure}[m]{0.49\textwidth}
\includegraphics[width=1\hsize]{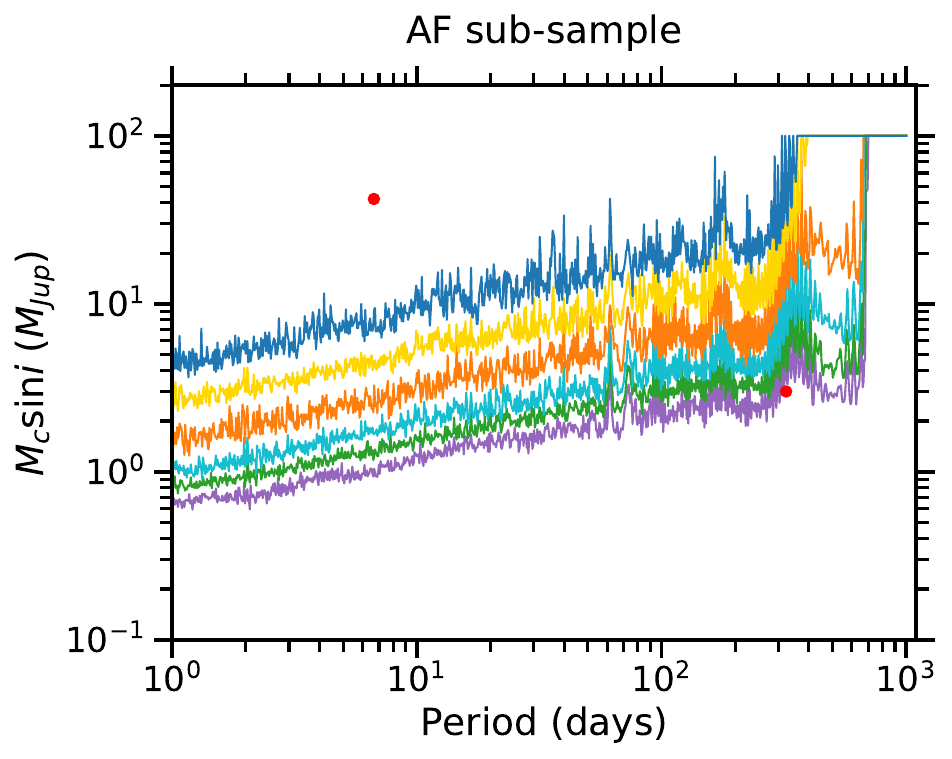}
\caption{\label{Completeness_AF}}
\end{subfigure}

\begin{subfigure}[m]{0.49\textwidth}
\includegraphics[width=1\hsize]{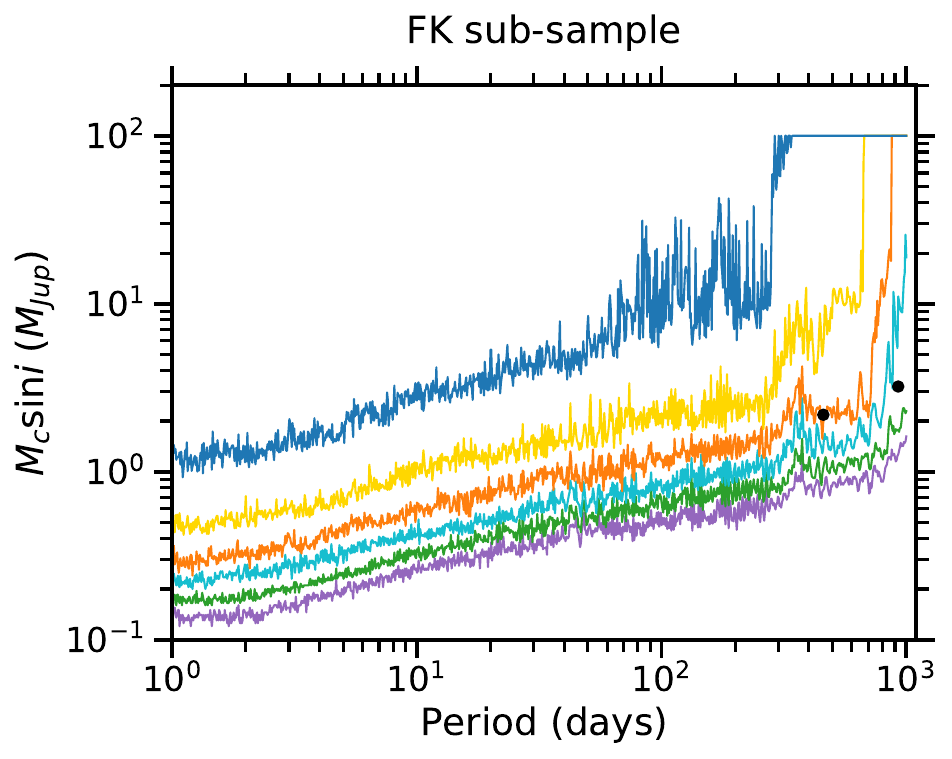}
\caption{\label{Completeness_FK}}
\end{subfigure}
\begin{subfigure}[m]{0.49\textwidth}
\includegraphics[width=1\hsize]{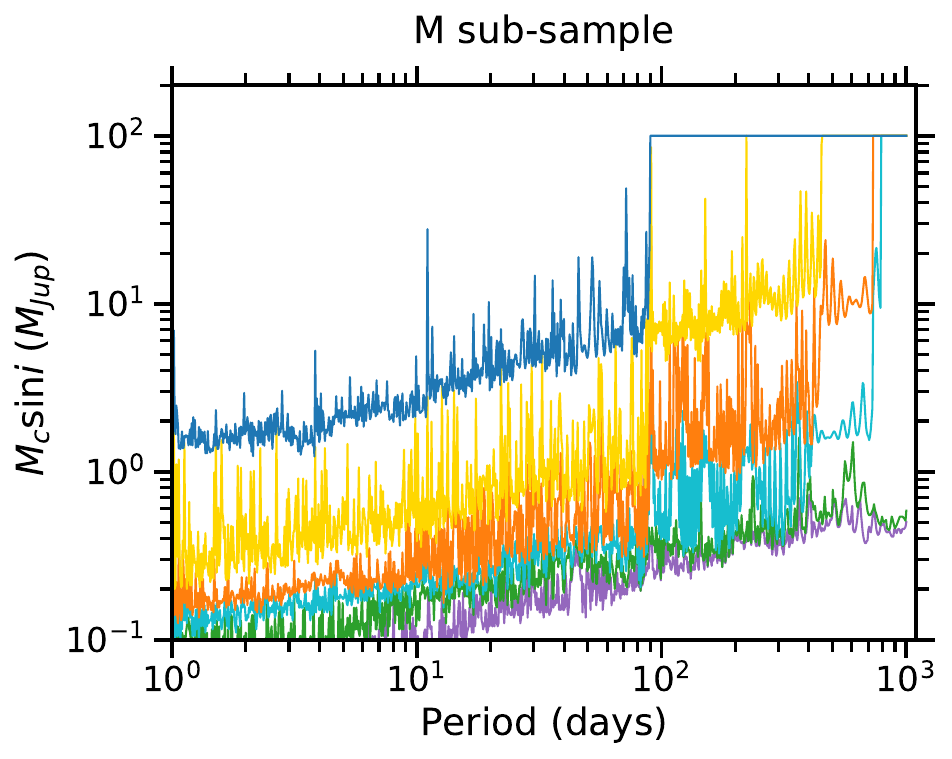}
\caption{\label{Completeness_M}}
\end{subfigure}
\caption{Search completeness of the whole final combined sample (\subref{Completeness_ALL}), and of the AF (\subref{Completeness_AF}), FK (\subref{Completeness_FK}), and M (\subref{Completeness_M})  subsamples. This corresponds to the lower $m_c \sin{i}$ for which $X\%$ of the stars in the corresponding sample have detection limits below this $m_c \sin{i}$ at a given period $P$; from bottom to top: $40\%$ to $90\%$. Our detected substellar companions HD 113337 b and HD 149790 B are shown as a red dot, while the nondetected planets of the \sophie\ survey, HD 128311 b and c, are shown as black dots (\emph{cf.} \citealt{Grandjean_SOPHIE}).}
       \label{Completeness}
\end{figure*}

Then, we computed the completeness function $C(m_c\sin{i},P)$ of the final combined sample and of the host star spectral type subsamples described in \Cref{descritpion}: AF, FK, and M. The completeness for a given pair $(m_c\sin{i},P)$ corresponds to the fraction of stars in the regarded sample for which a companion with this mass and period is excluded given the detection limits within the sample \citep{Simon_IX}. To compute the completeness, we excluded the targets for which at least one substellar companion with a period between $1$ and $$\SI{1000}{\day}$$ was discovered during the \harps \ Sco-Cen survey, the \sophie \ YNS survey, or the \harps \ YNS survey: HD 113337 and HD 149790. We did not take the  HJ companion candidate of HD 145467 in this analysis into account because we consider its discovery not robust enough.
The $40$ - $90\%$ search completeness of the final  combined sample and of the AF, FK, and M subsamples is presented in \Cref{Completeness}.
We note that $60\%$ of the stars have detection limits below $\SI{1}{\MJ}$ for periods shorter than $\SI{10}{\day}$ and $40\%$ have detection limits below $\SI{5}{M_{Jup}}$ for periods shorter than $\SI{1000}{\day}$ (\emph{cf.} \Cref{Completeness_ALL}). Finally, we also computed the search completeness function $C_D$ \citep{Simon_IX} of the final combined sample in this period and mass range (\emph{cf.} \cref{tab_occur}). It is greater than $70 \ \%$ for the AF and FK subsamples of the final  combined sample.

\subsection{GP and BD occurrence rates} 

\begin{figure}[h!]
  \centering
\includegraphics[width=1\hsize]{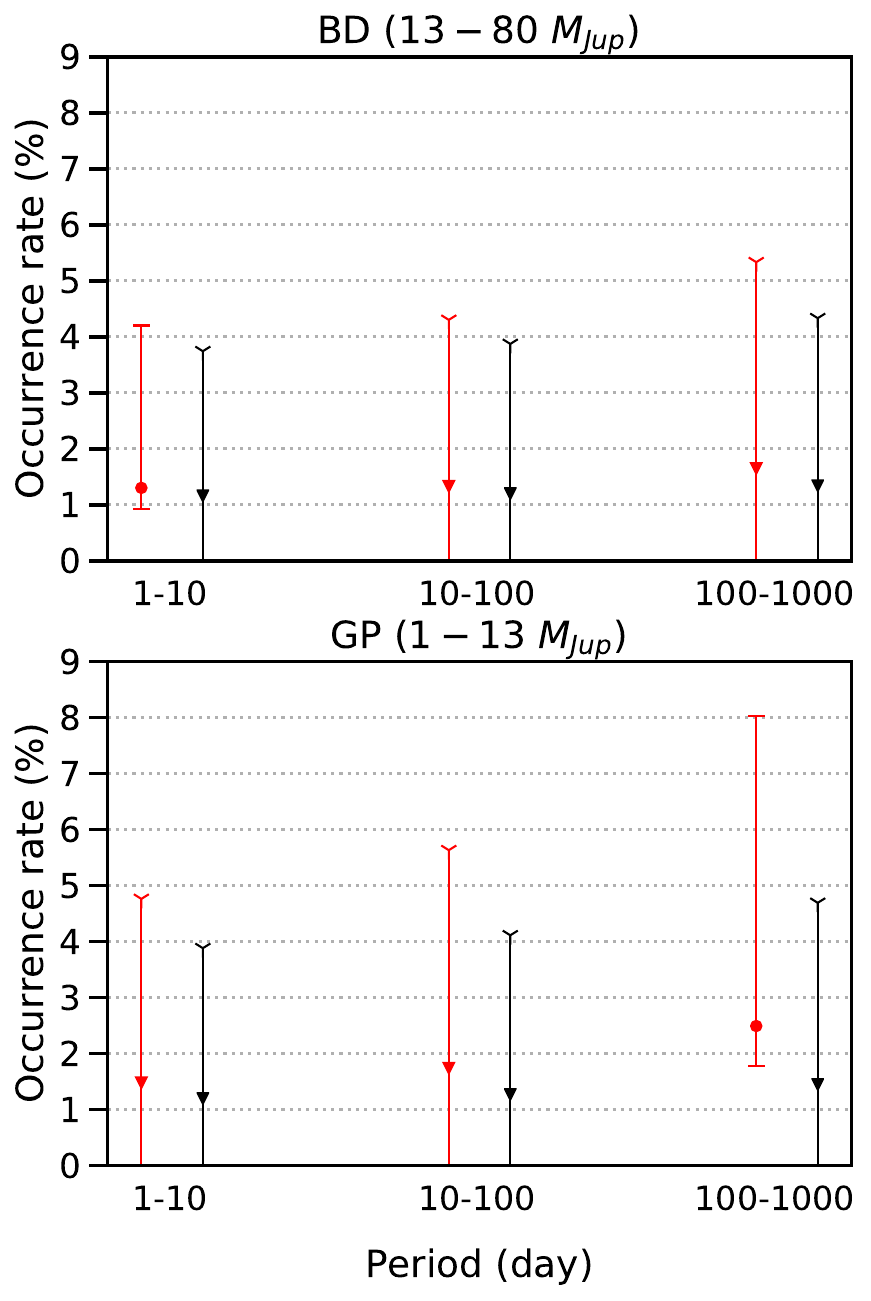}
\caption{Occurrence rates (dots) and upper limits on the occurrence rates (triangles)  and their respective $1 \sigma$ ranges  for the period ranges of $1$-$10$, $10$-$100$, and $100$-$$\SI{1000}{\day}$$ in the GP domain ($1-13$ \Mjup,  \emph{top}) and BD domain ($13-80$ \Mjup,  \emph{bottom}) for the AF subsample (\emph{red}) and the FK subsample (\emph{black}).} 
       \label{sample}
\end{figure}

We computed the occurrence rates of GPs ($1$ to $\SI{13}{\MJ}$) and BDs ($13$ to $\SI{80}{\MJ}$) for the whole final combined sample\footnote{176 stars, $B-V \geq -0.05$, $M\in[0.42;2.8]\si{\msun}\quad\quad\quad\quad\quad\quad$} and for different ranges of host star spectral types: AF\footnote{80 stars, $B-V \in [-0.05:0.52[$,  $M\sim 1.5\pm\SI{0.31}{M_{\odot}}\quad\quad\quad\quad$}, FK\footnote{87 stars, $B-V \in [0.52:1.33[$,  $M\sim 0.98\pm\SI{0.22}{\msun}\quad\quad\quad\quad$}, and M\footnote{9 stars, $B-V \geq 1.33$,  $M\sim 0.6\pm\SI{0.08}{\msun}$}, and  for different ranges of periods: $1$-$10$, $10$-$100$, $100$-$1000$, and $1$-$\SI{1000}{\day}$. To do this, we used the method described in \cite{Simon_IX}. For the range without detected companions in the survey, only the upper limits of the occurrence rates can be estimated. The results are presented in \cref{tab_occur} and in \Cref{sample}. 
We find an occurrence rate of $0.7_{-0.2}^{+1.6} \ \%$ for GPs with periods between $1$ and $ \SI{1000}{\day}$ and an occurrence rate of  $0.6_{-0.2}^{+1.4} \ \%$ for BDs in the same period range. These values are compatible within $1\sigma$ with those estimated from the \harps \ and \sophie \ YNS  combined sample \citep{Grandjean_SOPHIE}. They provide a better estimate, however: the size of the $1\sigma$ and $2\sigma$ intervals of the occurrence rates is reduced by $25\%$ for GPs. Moreover, through the discovery of HD 149790 B, an  occurrence rate for BDs is computed instead of an upper limit.

\renewcommand{\arraystretch}{1.25}
\begin{table*}[t!]
\caption{GPs ($m_c\sin{i} \in [1;13] \ \si{\MJ}$) and BDs  ($m_c\sin{i} \in [13;80] \ \si{\MJ}$) occurrence rates around young stars, derived from our \harps \ Sco-Cen, \sophie \ YNS, and \harps \ YNS combined sample. The parameters are displayed in normal, bold, italic or bold  italic fonts when considering the full final combined sample, the  AF sub-sample, the FK sub-sample or the M sub-sample, respectively.}
\label{tab_occur}
\begin{center}
\resizebox{1\textwidth}{!}{ 
\begin{tabular}{l c c c c c c l l}\\
\hline
\hline
\msini    & Orbital period  & \bv  & Search       & Detected    &  Missed       & Companion    & \multicolumn{2}{c}{Confidence intervals}\\
interval  & interval        && completeness & companion  & companion systems    &               occurrence rate  & $1\sigma$ & $2\sigma$        \\
(\Mjup)   & (day)           & & $C_D$  (\%)   &    systems          &     upper limit          &                (\%)         & (\%)      &  (\%)            \\

\hline 
1-13 & 1-10 & all & 92 & 0 & 0.1 & <0.6 & 0-2.0 & 0-3.5 \\ 
(GP) &  & $[-0.05:0.52[$ & {\bf 87} & {\bf 0} & {\bf 0.2} & {\bf <1.5} & {\bf 0-4.8} & {\bf 0-8.1} \\ 
 &  & $[0.52:1.33[$ & {\it 96} & {\it 0} & {\it 0.0} & {\it <1.2} & {\it 0-3.9} & {\it 0-6.6} \\ 
 &  & $\geq 1.33$ &  \textbf{\textit{99}} &  \textbf{\textit{0}} & \textbf{\textit{0.0}} & \textbf{\textit{<11.2}} & \textbf{\textit{0-29.7}} & \textbf{\textit{0-45.6}} \\ 
\hline 
1-13 & 1-100 & all & 87 & 0 & 0.1 & <0.7 & 0-2.1 & 0-3.7 \\ 
&  & $[-0.05:0.52[$ & {\bf 80} & {\bf 0} & {\bf 0.2} & {\bf <1.6} & {\bf 0-5.2} & {\bf 0-8.7} \\ 
 &  & $[0.52:1.33[$ & {\it 93} & {\it 0} & {\it 0.1} & {\it <1.2} & {\it 0-4.0} & {\it 0-6.8} \\ 
 &  & $\geq 1.33$ &  \textbf{\textit{97}} &  \textbf{\textit{0}} & \textbf{\textit{0.0}} & \textbf{\textit{<11.5}} & \textbf{\textit{0-30.4}} & \textbf{\textit{0-46.6}} \\ 
\hline 
1-13 & 1-1000 & all & 80 & 1 & 0.2 & 0.7 & 0.5-2.3 & 0.2-4.0 \\ 
 &  & $[-0.05:0.52[$ & {\bf 71} & {\bf 1} & {\bf 0.4} & {\bf 1.8} & {\bf 1.3-5.9} & {\bf 0.4-9.9} \\ 
 &  & $[0.52:1.33[$ & {\it 88} & {\it 0} & {\it 0.1} & {\it <1.3} & {\it 0-4.2} & {\it 0-7.1} \\ 
 &  & $\geq 1.33$ &  \textbf{\textit{89}} &  \textbf{\textit{0}} & \textbf{\textit{0.1}} & \textbf{\textit{<12.5}} & \textbf{\textit{0-33.0}} & \textbf{\textit{0-50.7}} \\ 
\hline 
\hline 
13-80 & 1-10 & all & 99 & 1 & 0.0 & 0.6 & 0.4-1.9 & 0.1-3.2 \\ 
(BD) &  & $[-0.05:0.52[$ & {\bf 147} & {\bf 1} & {\bf 0.0} & {\bf 1.3} & {\bf 0.9-4.2} & {\bf 0.3-7.1} \\ 
 &  & $[0.52:1.33[$ & {\it 96} & {\it 0} & {\it 0.0} & {\it <1.2} & {\it 0-3.9} & {\it 0-6.6} \\ 
 &  & $\geq 1.33$ &  \textbf{\textit{100}} &  \textbf{\textit{0}} & \textbf{\textit{0.0}} & \textbf{\textit{<11.1}} & \textbf{\textit{0-29.4}} & \textbf{\textit{0-45.2}} \\ 
\hline 
13-80 & 1-100 & all & 97 & 1 & 0.0 & 0.6 & 0.4-1.9 & 0.1-3.3 \\ 
 &  & $[-0.05:0.52[$ & {\bf 97} & {\bf 1} & {\bf 0.0} & {\bf 1.3} & {\bf 0.9-4.2} & {\bf 0.3-7.2} \\ 
 &  & $[0.52:1.33[$ & {\it 97} & {\it 0} & {\it 0.0} & {\it <1.2} & {\it 0-3.8} & {\it 0-6.5} \\ 
 &  & $\geq 1.33$ &  \textbf{\textit{100}} &  \textbf{\textit{0}} & \textbf{\textit{0.0}} & \textbf{\textit{<11.2}} & \textbf{\textit{0-29.5}} & \textbf{\textit{0-45.3}} \\ 
\hline 
13-80 & 1-1000 & all & 92 & 1 & 0.1 & 0.6 & 0.4-2.0 & 0.1-3.5 \\ 
 &  & $[-0.05:0.52[$ & {\bf 91} & {\bf 1} & {\bf 0.1} & {\bf 1.4} & {\bf 1.0-4.6} & {\bf 0.3-7.7} \\ 
 &  & $[0.52:1.33[$ & {\it 94} & {\it 0} & {\it 0.1} & {\it <1.2} & {\it 0-4.0} & {\it 0-6.7} \\ 
 &  & $\geq 1.33$ &  \textbf{\textit{94}} &  \textbf{\textit{0}} & \textbf{\textit{0.1}} & \textbf{\textit{<11.8}} & \textbf{\textit{0-31.4}} & \textbf{\textit{0-48.1}} \\ 
\hline 
\hline

\end{tabular}}
\end{center}
\end{table*}
\renewcommand{\arraystretch}{1.}

\subsection{Comparison of the occurrence rates for main-sequence stars} 

In our final  combined sample, two systems have substellar companions with periods shorter than $\SI{1000}{\day}$: HD 113337 \cite{Simon_VIII,Simon_X} and HD 149790. Both belong to the AF subsample.  However, we may have missed some GPs with low masses and long periods, as only $40\%$ of the stars in our final  combined sample have detection limits lower than $\SI{5}{\MJ}$ between $100$ and $\SI{1000}{\day}$ (\emph{cf.} \Cref{Completeness}).

We adopted two statistical tests to compare our results for young stars with those of the surveys carried out on older stars: the $p_{value}$ test, and the pooled version of the two proportion Z-test (which are both described in Appendix \ref{test_stat}). The null hypothesis that we adopted for these tests is that the occurrence rates of companions in a specific (P, $m_c\sin{i}$) box are identical for the two surveys that are compared. If these tests are passed, then the null hypothesis can be rejected with a confidence level of $1-\alpha$, indicating a statistical difference between the populations being compared. 

\subsubsection{Giant planets}

In our final combined sample, one GP was detected and confirmed with a period between $1$ and $\SI{1000}{\day}$, HD 113337 b \citep{Simon_VIII,Simon_X}. In addition, no HJ is firmly detected, and it is unlikely that we missed these objects because $60\%$ of the stars in our final combined sample have detection limits below $\SI{1}{\MJ}$ between $1$ and \SI{10}{\day}, and $80\%$ have detection limits below $\SI{3}{\MJ}$ in the same period interval.

For A-F stars, an occurrence rate of $1.8_{-0.5}^{+4.1} \%$ for GPs with periods between $1$ and $\SI{1000}{\day}$ is found. This is lower than but compatible at $1\sigma$ with the occurrence rate of $3.7_{-1.1}^{+2.8}. \%$ estimated by \cite{Simon_X} around older A-F stars in the same period interval.  The \emph{p-value} is $15_{-12}^{+11}\%$, and the Z-test is only validated with a confidence level of $59\%$ (and only the relaxed criterion of this test, $n>30$, is validated). We cannot conclude that there is a deficit of GPs with short separations around young A-F stars. Moreover, it should be noted that the survey of \cite{Simon_X} has stars in common with the \sophie \ YNS survey; the ages of the two surveys therefore overlap. 

For F-K stars,
 \cite{Grandjean_SOPHIE} noted a possible lack of young GPs with periods  between $1$ and $\SI{1000}{\day}$ in comparison to their older counterparts. However, the confidence level of this observation was only $90\%$.
Our Sco-Cen survey only provides a limited number of additional FK-type targets, making it difficult to confirm this observation with a higher level of confidence. 
In our study, we find an upper limit on the occurrence rate of these planets of $1.3^{+2.9}_{-1.3}\%$, comparable to the $1.4^{+3.1}_{-1.4}\%$ found by \cite{Grandjean_SOPHIE}. This is lower than but still compatible at $1\sigma$ with the occurrence rate  estimated around older stars of $4.3\pm 1\%$ \citep{cumming}. The \emph{p-value} is $2_{-1}^{+3}\%$. 
Thus, there might be a deficit of small-separation GPs around young F-K stars in comparison to their older counterparts, but the confidence level of at least $95\%$ is rather low. A statistical analysis based on a larger number of F-K  young  targets is needed to determine whether the GPs occurrence rate is significantly lower for young F-K stars than for MS F-K stars.

The  occurrence rate of HJs with $m_c\in[1;13]\si{\MJ}$ and $P\in[1;10]\si{\day}$ around F-K stars is found to be lower than  $1.2_{-1.2}^{+2.7} \%$, which is similar to the one found for their older counterparts ($1.2\pm0.2 \%$ \cite{Marcy};  $0.46^{+0.3}_{-0.3} \ \%$ \cite{cumming}). No massive HJ  has been discovered for $87$ F-K stars in  our final combined sample, the corresponding \emph{p-value} is $35^{+7}_{-5}\%$ and $67^{+20}_{-15}\%$, respectively. It is possible that the  occurrence rate of these planets is identical for young and old stars.

Finally, an occurrence rate of $1.8_{-0.5}^{+4.1} \%$ (against $4.3_{-1.0}^{+9.1} \%$ previously; \cite{Grandjean_SOPHIE}) for   A-F stars and an upper limit of the occurrence rate of $1.3^{+2.9}_{-1.3}\%$ (against $1.4_{-1.4}^{+3.1} \%$ previously) for F-K stars is estimated  for GPs with periods between $1$ and $\SI{1000}{\day}$.  
When identical occurrence rates are assumed for young A-F and F-K stars, the \emph{p-value} of the non-detection of these objects around the $87$ stars of our F-K subsample is $21_{-21}^{+11}\%$ (against $25_{-24}^{+13}\%$ previously). There is thus no evidence that the compared populations are different.

\subsubsection{Brown dwarfs}

A BD with a period between $1$ and  $\SI{1000}{\day}$ was found in our final combined sample (HD 149790 B), and we compute an  occurrence rate of  $0.6_{-0.2}^{+1.4}. \%$ for close BDs ($m_c\in[13;80]\si{\MJ}$, $P\in[1;1000]\si{\day}$). This rate agrees with the results from surveys carried out for older stars, which highlighted that BDs with small separations are rare around the latter \citep{Grether,Sahlmann,Grieves,Jones_BD,Simon_X,Kiefer}.

We find an upper limit on the occurrence rate of close BDs of $1.2_{-1.2}^{+2.8} \ \%$ for young F-K stars. This agrees with the upper limits on the occurrence rate of these objects derived around older stars of similar spectral types, which was estimated at $1\%$ ($m_c\in[13;80]\si{\MJ}$, $P<\SI{5}{\year}$; \citealt{Grether}), $0.6\%$ ($m_c\in[13;80]\si{\MJ}$, $P<\SI{12}{\year}$; \citealt{Sahlmann}), $0.56\%$ ($m_c\in[13;80]\si{\MJ}$, $P<\SI{300}{\day}$; \citealt{Grieves}), and $2\%$ ($m_c\in[12.5;90]\si{\MJ}$, $P<\SI{10000}{\day}$; \citealt{Kiefer}).    
The occurrence rate of close BDs  for A-F stars is found to be  $1.4_{-0.4}^{+3.2} \%$, which is higher than but still compatible at $1\sigma$ with the upper limit found by \cite{Simon_X}  around older stars of the same spectral types and over the same period interval, $0.5_{-0.5}^{+3.1} \%$.
The probability that \cite{Simon_X} found no BD on the $225$ stars of their sample, knowing the occurrence rate we computed, is $4^{+6}_{-4}\%$. The \emph{p-value} test is therefore validated with a confidence level of $90\%$. However, the Z-test is only validated with a confidence level of $52\%$ (and only the relaxed criterion of this test, $n>30$, is validated). It therefore cannot be concluded that there is an overabundance of close BDs around A-F young stars in comparison to older stars of similar spectral types. 

\section{Conclusions}
\label{conc}

We have carried out a two-year survey of $88$ stars of the Sco-Cen association with  \harps  in the search for close GPs and BDs companions.
Among the stars investigated, one HJ candidate that is yet to be confirmed was found around HD 145467, and one short-period ($P<\SI{10}{\day}$) BD candidate was discovered around HD 149790.
In addition, $11$ binaries (eight SB1 and three SB2) were identified. 

We then combined this survey with the \sophie \  and  \harps \ YNS surveys presented in \cite{Grandjean_SOPHIE} and in \cite{Grandjean_HARPS}, respectively, leading to a statistical analysis of $176$ young stars.
We obtain an occurrence rate of  $0.7_{-0.2}^{+1.6} \ \%$ for GPs with periods between $1$ and  $\SI{1000}{\day}$ and a BD occurrence rate  of $0.6_{-0.2}^{+1.4} \ \%$ in the same period range.

We observe a more significant difference in the GP occurrence rates between young and MS FK type stars than was found previously. The associated confidence level  is now at  $95\%$, against $90\%$ previously \citep{Grandjean_SOPHIE}.
However, an analysis of a larger number of young stars is still needed to determine whether this difference is really significant.

\begin{acknowledgements}
   We acknowledge support from the French CNRS and from the Agence Nationale de la Recherche (ANR grant GIPSE ANR-14-CE33-0018). 
This work has been supported by a grant from Labex OSUG@2020 (Investissements d’avenir – ANR10 LABX56).
These results have made use of the SIMBAD database, operated at the CDS, Strasbourg, France.  ESO
SD acknowledges the support by INAF/Frontiera through the "Progetti
Premiali" funding scheme of the Italian Ministry of Education,
University, and Research.
Based on observations collected at the European Southern Observatory under ESO programme(s) 060.A-9036(A),072.C-0488(E),072.C-0636(A),072.C-0636(B),073.C-0733(A),073.C-0733(C),073.C-0733(D),073.C-0733(E),074.C-0037(A),074.C-0364(A),075.C-0202(A),075.C-0234(A),075.C-0234(B),075.C-0689(A),075.C-0689(B),076.C-0010(A),076.C-0073(A),076.C-0279(A),076.C-0279(B),076.C-0279(C),077.C-0012(A),077.C-0295(A),077.C-0295(B),077.C-0295(C),077.C-0295(D),078.C-0209(A),078.C-0209(B),078.C-0751(A),078.C-0751(B),079.C-0170(A),079.C-0170(B),079.C-0657(C),080.C-0032(A),080.C-0032(B),080.C-0664(A),080.C-0712(A),081.C-0034(B),081.C-0802(C),082.C-0308(A),082.C-0357(A),082.C-0412(A),082.C-0427(A),082.C-0427(C),082.C-0718(B),083.C-0794(A),083.C-0794(B),083.C-0794(C),083.C-0794(D),084.C-1039(A),085.C-0019(A),087.C-0831(A),089.C-0732(A),089.C-0739(A),090.C-0421(A),091.C-0034(A),094.C-0946(A),098.C-0739(A),099.C-0205(A),0104.C-0418(A),1101.C-0557(A),183.C-0437(A),183.C-0972(A),184.C-0815(A),184.C-0815(B),184.C-0815(C),184.C-0815(E),184.C-0815(F),191.C-0873(A),192.C-0224(B),192.C-0224(C),192.C-0224(G),192.C-0224(H).
Based on observations collected at the Observatoire de Haute Provence under the programme(s)
06B.PNP.CONS,07A.PNP.CONS,07B.PNP.CONS, 08A.PNP.CONS,13A.PNP.DELF,13A.PNP.LAGR,14A.PNP.LAGR.
\end{acknowledgements}


\bibliography{41235corr}

\bibliographystyle{aa}

%
%


\clearpage

\begin{appendix}

\onecolumn
\section{Combined sample}

\subsection{\harps \  Sco-Cen sample}

\setcounter{footnote}{0}
\begin{landscape}
\begingroup
\renewcommand\arraystretch{1.3}
\begin{longtable}{lccccccccccc}
\caption{Properties of the $69$ stars in the Sco-Cen final sample. The stars excluded from our statistical analysis (not enough spectra ($N_{\rm m}<3$), $v\sin{i}>\SI{150}{\kilo\meter\per\second}$, and stars for which the companion signal could not be fitted) are indicated with a $^\dagger$.
Spectral types (ST) were taken from the CDS database at the beginning of the survey and the  \bv~values are  taken from the CDS database (up to date).
The \vsini~ are computed with \safir~based on the CCF width.
The IR/D column reports if either an IR excess is reported ($y$) or not ($n$) and if a disk has been resolved ($y$) or not ($n$) in the literature.}
\label{tab_carac_sco}
 \\ \hline
Name & HIP & ST & \bv & plx & $pm_{Ra}$ & $pm_{Dec}$ & Mass & Age  &\vsini          &  IR/D\\ 
  HD/BD/CD & && & ($\si{mas}$) & ($\si{mas\,yr^{-1}}$) & ($\si{mas\,yr^{-1}}$) &(\Msun) & (\si{Myr}) &(\kms) &  & \\
\hline \hline
\endfirsthead
\caption{continued.}\\
 \\ \hline
Name & HIP & ST & \bv & plx & $pm_{Ra}$ & $pm_{Dec}$ & Mass & Age  &\vsini         &  IR/D\\ 
  HD/BD/CD & &&& ($\si{mas}$) & ($\si{mas\,yr^{-1}}$) & ($\si{mas\,yr^{-1}}$) &(\Msun) & (\si{Myr})  &(\kms)   & \\
 \\ \hline
  \hline
\endhead
\hline
\endfoot
HD95086 & 53524 &  A8III & 0.24 & $11.57\pm0.02$\footnote{\label{EDR3}\cite{EDR3}} & $-41.13\pm0.02$\footnoteref{EDR3} & $12.86\pm0.02$\footnoteref{EDR3} & 1.6\footnote{\label{Rameau_95086}\cite{Rameau_95086}} & $17_{-4}^{+4}$\footnote{\label{Meshkat_2013}\cite{Meshkat_2013}}  & 20.7 & y\footnote{\label{Rhee}\cite{Rhee}}/y\footnote{\label{Moor}\cite{Moor}} \\ 
HD100282 & 56227 &  F0III & 0.31 & $7.88\pm0.02$\footnoteref{EDR3} & $-41.36\pm0.02$\footnoteref{EDR3} & $-3.04\pm0.02$\footnoteref{EDR3} & 1.51\footnote{\label{Sylvano_mass_sco} This work (\emph{cf.} \Cref{age_mass})} & $16_{-4}^{+4}$\footnoteref{Sylvano_mass_sco}  & 59.4 & y\footnote{\label{Chen_11}\cite{Chen_11}}/- \\ 
HD100453 & 56354 &  A9e & 0.30 & $9.64\pm0.02$\footnoteref{EDR3} & $-36.96\pm0.02$\footnoteref{EDR3} & $-5.86\pm0.02$\footnoteref{EDR3} & 1.7\footnote{\label{Dominik}\cite{Dominik}} & $10_{-2}^{+2}$\footnote{\label{Collins_2009}\cite{Collins_2009}}  & 32.8 & y\footnote{\label{Vioque}\cite{Vioque}}/y\footnote{\label{Marinas}\cite{Marinas}}\footnote{\label{Benisty}\cite{Benisty}} \\ 
HD102458 & 57524 &  G4V & 0.63 & $8.79\pm0.01$\footnoteref{EDR3} & $-34.73\pm0.01$\footnoteref{EDR3} & $-9.76\pm0.01$\footnoteref{EDR3} & 1.70\footnote{\label{Reza}\cite{reza}} & $16_{-4}^{+4}$\footnoteref{Sylvano_mass_sco}  & 19.9 & y\footnote{\label{Rebull}\cite{Rebull}}/- \\ 
HD105874 & 59413 &  A6V & 0.19 & $8.16\pm0.03$\footnoteref{EDR3} & $-31.92\pm0.02$\footnoteref{EDR3} & $-9.79\pm0.02$\footnoteref{EDR3} & 1.6\footnote{\label{Kouwenhoven_05}\cite{Kouwenhoven_05}} & $15_{-3}^{+3}$\footnote{\label{Pecaut_16}\cite{Pecaut_16}}  & 21.4 & y\footnote{\label{McDonald}\cite{McDonald}}\/- \\ 
HD105994  & 59481 & F3V &  0.41 & $8.99\pm0.02$\footnoteref{EDR3} & $-34.99\pm0.02$\footnoteref{EDR3} & $-10.98\pm0.02$\footnoteref{EDR3} & 1.5\footnote{\label{Chen_14}\cite{Chen_14}} & $13$\footnote{\label{Pecaut_12}\cite{Pecaut_2012}} &  68.5 & y\footnoteref{Chen_11}/- \\
\end{longtable}
\endgroup
\end{landscape}

\setcounter{footnote}{0}
\begin{landscape}
\begingroup
\begin{longtable}{lcc|cccccccccccccc}
\caption{Main characteristics of the RV variations for the $69$ stars in the Sco-Cen final sample : time baseline (TBL), number of computed spectra $N_{\rm m}$, total  amplitude of the RVs ({\it A}), rms and mean uncertainty $<${\it U}$>$ on the RV and BVS measurements, RV-BVS correlation factor (slope of the best linear fit), mean FWHM ($<$FWHM$>$), mean \rhk \ ($<$\rhk$>$) and associated  standard deviation $\sigma_{logR'_{\rm HK}}$.
Column V indicates the origin of the stellar jitter: A stands for stellar activity (spots), and P stands for pulsations.
Column T indicates the stars with a long term trend induced by a companion.
Column B precises the nature of the companion: C stands for sub-stellar companion, SB1 stands for single-line spectroscopic binary, and SB2 stands for double-line spectroscopic binary.}
\label{tab_result_sco}\\
\\ \hline
\multicolumn{3}{c|}{Stellar characteristics} & \multicolumn{12}{c}{Survey results.}
 \\ \hline
Name & HIP & ST &  TBL & $N_{\rm m}$   & \multicolumn{3}{c}{RV}        & \multicolumn{3}{c}{BVS}      & RV-    & $<$FWHM$>$ & $<$\rhk$>$  & V   & T & B\\ 
 HD/BD/CD & &  & & & \multicolumn{3}{c}{\raisebox{.5\baselineskip}{$\overbrace{\hspace{2.7cm}}$}} & \multicolumn{3}{c}{\raisebox{.5\baselineskip}{$\overbrace{\hspace{2.7cm}}$}} & BVS & &($\sigma_{logR'_{\rm HK}}$)  & & &\\
&& &  &  & {\it A} &rms  & $<${\it U}$>$ & {\it A} & rms &$<${\it U}$>$ & corr.    &                 &  & & &  \\
&&&&    & \multicolumn{3}{c}{\raisebox{.5\baselineskip}{$\underbrace{\hspace{2.7cm}}$}} & \multicolumn{3}{c}{\raisebox{.5\baselineskip}{$\underbrace{\hspace{2.7cm}}$}} & &  & & &\\
&& &     (day)       && \multicolumn{3}{c}{($\si{\meter\per\second}$)}      & \multicolumn{3}{c}{($\si{\meter\per\second}$)}     &   &      ($\si{\kilo\meter\per\second}$)  &    & & &\\ \hline
\endfirsthead
\caption{continued.}\\
 \\ \hline
\multicolumn{3}{c|}{Stellar characteristics} & \multicolumn{12}{c}{Survey results.}
 \\ \hline
Name & HIP & ST &  TBL & $N_{\rm m}$   & \multicolumn{3}{c}{RV}        & \multicolumn{3}{c}{BVS}      & RV-    & $<$FWHM$>$ & $<$\rhk$>$ & V  & T & V \\ 
 HD/BD/CD & &  & & & \multicolumn{3}{c}{\raisebox{.5\baselineskip}{$\overbrace{\hspace{2.7cm}}$}} & \multicolumn{3}{c}{\raisebox{.5\baselineskip}{$\overbrace{\hspace{2.7cm}}$}} & BVS & &($\sigma_{logR'_{\rm HK}}$)  & & &\\
&& &  &  & {\it A} &rms  & $<${\it U}$>$ & {\it A} & rms &$<${\it U}$>$ & corr.    &                 &    & & &\\
&&&&    & \multicolumn{3}{c}{\raisebox{.5\baselineskip}{$\underbrace{\hspace{2.7cm}}$}} & \multicolumn{3}{c}{\raisebox{.5\baselineskip}{$\underbrace{\hspace{2.7cm}}$}} & & & & & \\
&& & (day)           && \multicolumn{3}{c}{($\si{\meter\per\second}$)}      & \multicolumn{3}{c}{($\si{\meter\per\second}$)}     &   &       ($\si{\kilo\meter\per\second}$) &  & & &\\ \hline
\endhead
\hline
\endfoot
HD95086 & 53524 & A8III & 1532 & 103 & 1279.0 & 268.8 & 14.8 & 2594.0 & 36.3 & 511.1 & - & 45.4 & -4.47(0.03) & P & - &- \\ 
HD100282 & 56227 & F0III & 684 & 14 & 1578.2 & 551.1 & 47.6 & 3685.9 & 106.8 & 1112.6 & - & 123.8 & -4.48(0.01) & P & - &- \\ 
HD100453 & 56354 & A9e & 683 & 20 & 1294.1 & 345.1 & 22.6 & 3203.8 & 53.8 & 751.4 & - & 71.9 & -4.38(0.01) & P & - &- \\ 
HD102458 & 57524 & G4V & 2964 & 26 & 988.5 & 307.7 & 10.0 & 1358.4 & 23.8 & 431.7 & -0.79 & 43.6 & -3.99(0.02) & A & - &- \\ 
HD105874 & 59413 & A6V & 311 & 12 & 405.5 & 119.6 & 22.0 & 1213.1 & 51.1 & 383.4 & - & 47.5 & -4.48(0.03) & P & - &- \\ 
HD105994 & 59481 & F3V & 669 & 16 & 283.6 & 91.4 & 59.2 & 2416.9 & 149.3 & 565.7 & - & 137.4 & -4.49(0.01) & P & - &- \\ 
HD106444 & 59716 & F5V & 382 & 10 & 323.8 & 111.0 & 62.4 & 1619.5 & 151.9 & 451.9 & - & 143.2 & -4.54(0.01) & P & - &- \\ 
HD106473 & 59724 & A6III & 668 & 12 & 933.8 & 283.0 & 53.9 & 1926.2 & 114.7 & 613.9 & - & 126.2 & -4.42(0.02) & P & - &- \\
\end{longtable}
\endgroup
\end{landscape}

\clearpage

\subsection{\sophie \  YNS sample}

\setcounter{footnote}{0}
\begingroup
\renewcommand\arraystretch{1.3}
\begin{longtable}{lcccccccc}
\caption{Same as \Cref{tab_carac_sco} for the $54$ stars in  the \sophie \  YNS RV survey for which the RV could be computed  ($v\sin{i} <\SI{300}{\kilo\meter\per\second}$). 
}
\label{tab_carac_s}
 \\ \hline
Name & HIP & ST & \bv & Mass & Age  &\vsini          &  IR/D\\ 
  HD/BD/CD & &&&(\Msun) & (\si{Myr}) &(\kms) &  & \\
\hline \hline
\endfirsthead
\caption{continued.}\\
 \\ \hline
Name & HIP & ST & \bv & Mass & Age  &\vsini         &  IR/D\\ 
  HD/BD/CD & &&&(\Msun) & (\si{Myr})  &(\kms)   & \\
 \\ \hline
  \hline
\endhead
\hline
\endfoot
HD377 & 682 &  G2V & 0.630 & 1.11\footnote{\label{Sylvano_mass_s} This work (\emph{cf.} \Cref{age_mass})} & $200_{-200}^{+200}$\footnote{\label{Chavero}\cite{Chavero}}  & 10.4 & y\footnote{\label{Meyer_excess}\cite{Meyer_excess}}/y\footnote{\label{Cotten_s}\cite{Cotten}} \\ 
HD2454$^\dagger$ & 2235 &  F6V & 0.430 & 1.21\footnoteref{Sylvano_mass_s} & $800_{-300}^{+300}$\footnote{\label{Gomez}\cite{Gomez}}  & 6.7 & -/- \\ 
HD7804  & 6061 & A3V &  0.073 & 2.52\footnote{\label{Stone}\cite{Stone}} & $412$\footnoteref{Stone} &  78.0 & y\footnote{\label{McDonald_s}\cite{McDonald}}/- \\ 
HD13507 & 10321 &  G0V & 0.670 & 0.99\footnoteref{Sylvano_mass_s} & $500_{-100}^{+100}$\footnoteref{Gomez}  & 4.1 & y\footnoteref{McDonald_s}/- \\ 
HD13531  & 10339 & G0V &  0.700 & 1\footnoteref{Sylvano_mass_s} & $400$\footnote{\label{Gaspar_s}\cite{Gaspar}} &  5.3 & y\footnoteref{McDonald_s}/- \\ 
HD17250  & 12925 & F8V &  0.520 & 1.26\footnoteref{Sylvano_mass_s} & $45$\footnote{\label{Vigan}\cite{Vigan}} &  29.2 & y\footnoteref{McDonald_s}/- \\ 
HD20630 & 15457 &  G5V & 0.670 & $0.99^{+0.02}_{-0.03}$\footnoteref{Gomez} & $600_{-200}^{+200}$\footnoteref{Gomez}  & 4.8 & n\footnoteref{Gaspar_s}/- \\ 
HD25457 & 18859 &  F5V & 0.500 & 1.2\footnoteref{Vigan} & $149_{-49}^{+31}$\footnoteref{Vigan}  & 12.4 & y\footnote{\label{Zuckerman}\cite{Zuckerman}}/- \\
\end{longtable}
\endgroup

\setcounter{footnote}{0}
\begin{landscape}
\begingroup
\begin{longtable}{lcc|cccccccccccccc}
\caption{Same as \Cref{tab_result_sco} for the $54$ stars in our  \sophie \  YNS RV survey for which RV could be computed  ($v\sin{i} <\SI{300}{\kilo\meter\per\second}$). 
}
\label{tab_result_s}\\
\\ \hline
\multicolumn{3}{c|}{Stellar characteristics} & \multicolumn{12}{c}{Survey results.}
 \\ \hline
Name & HIP & ST &  TBL & $N_{\rm m}$   & \multicolumn{3}{c}{RV}        & \multicolumn{3}{c}{BVS}      & RV-    & $<$FWHM$>$ & $<$\rhk$>$  & V   & T & B\\ 
 HD/BD/CD & &  & & & \multicolumn{3}{c}{\raisebox{.5\baselineskip}{$\overbrace{\hspace{2.7cm}}$}} & \multicolumn{3}{c}{\raisebox{.5\baselineskip}{$\overbrace{\hspace{2.7cm}}$}} & BVS & &($\sigma_{logR'_{\rm HK}}$)  & & &\\
&& &  &  & {\it A} &rms  & $<${\it U}$>$ & {\it A} & rms &$<${\it U}$>$ & corr.    &                 &  & & &  \\
&&&&    & \multicolumn{3}{c}{\raisebox{.5\baselineskip}{$\underbrace{\hspace{2.7cm}}$}} & \multicolumn{3}{c}{\raisebox{.5\baselineskip}{$\underbrace{\hspace{2.7cm}}$}} & &  & & &\\
&& &     (day)       && \multicolumn{3}{c}{($\si{\meter\per\second}$)}      & \multicolumn{3}{c}{($\si{\meter\per\second}$)}     &   &      ($\si{\kilo\meter\per\second}$)  &    & & &\\ \hline
\endfirsthead
\caption{continued.}\\
 \\ \hline
\multicolumn{3}{c|}{Stellar characteristics} & \multicolumn{12}{c}{Survey results.}
 \\ \hline
Name & HIP & ST &  TBL & $N_{\rm m}$   & \multicolumn{3}{c}{RV}        & \multicolumn{3}{c}{BVS}      & RV-    & $<$FWHM$>$ & $<$\rhk$>$ & V  & T & V \\ 
 HD/BD/CD & &  & & & \multicolumn{3}{c}{\raisebox{.5\baselineskip}{$\overbrace{\hspace{2.7cm}}$}} & \multicolumn{3}{c}{\raisebox{.5\baselineskip}{$\overbrace{\hspace{2.7cm}}$}} & BVS & &($\sigma_{logR'_{\rm HK}}$)  & & &\\
&& &  &  & {\it A} &rms  & $<${\it U}$>$ & {\it A} & rms &$<${\it U}$>$ & corr.    &                 &    & & &\\
&&&&    & \multicolumn{3}{c}{\raisebox{.5\baselineskip}{$\underbrace{\hspace{2.7cm}}$}} & \multicolumn{3}{c}{\raisebox{.5\baselineskip}{$\underbrace{\hspace{2.7cm}}$}} & & & & & \\
&& & (day)           && \multicolumn{3}{c}{($\si{\meter\per\second}$)}      & \multicolumn{3}{c}{($\si{\meter\per\second}$)}     &   &       ($\si{\kilo\meter\per\second}$) &  & & &\\ \hline
\endhead
\hline
\endfoot
HD377 & 682 & G2V & 747 & 17 & 652.3 & 139.1 & 6.3 & 302.6 & 15.0 & 73.6 & -0.84 & 22.8 & -4.39(0.04) & A & - &- \\ 
HD2454$^\dagger$ & 2235 & F6V & 805 & 29 & 545.6 & 154.9 & 4.2 & 105.4 & 11.0 & 28.4 & - & 14.6 & -4.89(0.08) & A & T &SB1 \\ 
HD7804 & 6061 & A3V & 740 & 92 & 8767.7 & 875.7 & 190.9 & 161539.7 & 447.1 & 15742.1 & - & 175.9 & -3.61(0.03) & P & - &- \\ 
HD13507 & 10321 & G0V & 886 & 24 & 100.0 & 24.5 & 2.2 & 42.7 & 6.5 & 10.8 & - & 9.9 & -4.45(0.04) & A & T &SB1 \\ 
HD13531 & 10339 & G0V & 882 & 33 & 103.1 & 28.6 & 2.7 & 52.9 & 7.7 & 11.6 & - & 11.9 & -4.44(0.04) & A & T &SB1 \\ 
HD17250 & 12925 & F8V & 5 & 5 & 1157.2 & 420.0 & 19.7 & 579.6 & 49.9 & 199.5 & - & 63.9 & -4.393(0.01) & - & T &SB1 \\ 
HD20630 & 15457 & G5V & 740 & 29 & 80.1 & 21.6 & 1.5 & 48.2 & 4.4 & 11.9 & -1.61 & 11.0 & -4.52(0.04) & A & - &- \\ 
HD25457 & 18859 & F5V & 498 & 10 & 121.0 & 39.7 & 6.3 & 158.5 & 14.8 & 54.8 & -0.59 & 27.1 & -4.48(0.02) & A & - &- \\
\end{longtable}
\endgroup
\end{landscape}

\clearpage
\subsection{\harps \ YNS sample}

\setcounter{footnote}{0}
\begingroup
\renewcommand\arraystretch{1.3}
\begin{longtable}{lcccccccc}
\caption{Same as \Cref{tab_carac_sco} for the $89$ stars in the initial sample of our \harps~YNS RV survey.
\label{tab_carac_h} }
 \\ \hline
Name & HIP & ST & \bv & Mass & Age  &\vsini &Rotation          &  IR/D\\ 
  HD/BD/CD & &&&(\Msun) & (\si{Myr}) &(\kms) & period ($\si{days}$)   & \\
\hline \hline
\endfirsthead
\caption{continued.}\\
 \\ \hline
Name & HIP & ST & \bv & Mass & Age  &\vsini &Rotation          &  IR/D\\ 
  HD/BD/CD & &&&(\Msun) & (\si{Myr})  &(\kms) & period ($\si{days}$)   & \\
 \\ \hline
  \hline
\endhead
\hline
\endfoot
HD105 & 490 &  G0V & 0.600 & 1.1\footnote{\label{Vigan_h}\cite{Vigan}} & $45_{-10}^{+5}$\footnoteref{Vigan_h}  & 10.4 & - & y\footnote{\label{Meyer}\cite{Meyer_105}}/y\footnote{\label{Donaldson}\cite{Donaldson}} \\ 
HD984 & 1134 &  F7V & 0.500 & 1.2\footnoteref{Vigan_h} & $42_{-7}^{+8}$\footnoteref{Vigan_h}  & 27.6 & - & n\footnote{\label{Carpenter_h}\cite{Carpenter}}/- \\ 
HD987 & 1113 &  G8V & 0.756 & 0.98\footnote{\label{Lagrange_h}\cite{Lagrange}} & $30_{-15}^{+15}$\footnote{\label{Weise}\cite{Weise}}  & 6.2 & $3.72\pm0.01$\footnote{\label{Messina}\cite{Messina}} & n\footnote{\label{Rebull_h}\cite{Rebull}}/- \\ 
HD1466 & 1481 &  F8V & 0.540 & 1.2\footnoteref{Vigan_h} & $45_{-10}^{+5}$\footnoteref{Vigan_h}  & 14.7 & - & y\footnote{\label{Mamajek_1466}\cite{Mamajek_1466}}/- \\ 
HD3221 & 2729 &  K4V & 1.226 & 0.9\footnoteref{Messina} & $45_{-10}^{+5}$\footnoteref{Vigan_h}  & 104.9 & $0.370\pm0.002$\footnoteref{Messina} & -/- \\ 
HD6569 & 5191 &  K1V & 0.830 & 0.8\footnoteref{Vigan_h} & $149_{-49}^{+31}$\footnoteref{Vigan_h}  & 4.6 & $7.13\pm0.05$\footnoteref{Messina} & -/- \\ 
HD7661 & 5938 &  K0V & 0.770 & 0.98\footnote{\label{Sylvano_mass_h} This work (\emph{cf.} \Cref{age_mass})} & $300_{-50}^{+50}$\footnoteref{Weise}  & 4.4 & 7.46\footnote{\label{Wright_prot}\cite{Wright_prot}} & n\footnote{\label{Lawler}\cite{Lawler}}/- \\ 
HD10008 & 7576 &  K0V & 0.803 & 0.8\footnoteref{Vigan_h} & $250_{-50}^{+50}$\footnoteref{Vigan_h}  & 3.5 & $7.15\pm0.10$\footnote{\label{Folsom}\cite{Folsom}} & y\footnote{\label{Plavchan_h}\cite{Plavchan}}/- \\
\end{longtable}
\endgroup

\setcounter{footnote}{0}
\begin{landscape}
\begingroup
\begin{longtable}{lcc|cccccccccccccc}
\caption{Same as \Cref{tab_result_sco} for the $89$ stars in the initial sample of our \harps~YNS RV survey.
\label{tab_result_h}}\\ 
\\ \hline
\multicolumn{3}{c|}{Stellar characteristics} & \multicolumn{12}{c}{Survey results.}
 \\ \hline
Name & HIP & ST &  TBL & $N_{\rm m}$   & \multicolumn{3}{c}{RV}        & \multicolumn{3}{c}{BVS}      & RV-    & $<$FWHM$>$ & $<$\rhk$>$  & V   & T & B\\ 
 HD/BD/CD & &  & & & \multicolumn{3}{c}{\raisebox{.5\baselineskip}{$\overbrace{\hspace{2.7cm}}$}} & \multicolumn{3}{c}{\raisebox{.5\baselineskip}{$\overbrace{\hspace{2.7cm}}$}} & BVS & & ($\sigma_{logR'_{\rm HK}}$) & & &\\
&& &  &  & {\it A} &rms  & $<${\it U}$>$ & {\it A} & rms &$<${\it U}$>$ & corr.    &                 &  & & &  \\
&&&&    & \multicolumn{3}{c}{\raisebox{.5\baselineskip}{$\underbrace{\hspace{2.7cm}}$}} & \multicolumn{3}{c}{\raisebox{.5\baselineskip}{$\underbrace{\hspace{2.7cm}}$}} & &  & & &\\
&& (day)&            && \multicolumn{3}{c}{($\si{\meter\per\second}$)}      & \multicolumn{3}{c}{($\si{\meter\per\second}$)}     &   &      ($\si{\kilo\meter\per\second}$) &    & & &\\ \hline
\endfirsthead
\caption{continued.}\\
 \\ \hline
\multicolumn{3}{c|}{Stellar characteristics} & \multicolumn{12}{c}{Survey results.}
 \\ \hline
Name & HIP & ST &  TBL & $N_{\rm m}$   & \multicolumn{3}{c}{RV}        & \multicolumn{3}{c}{BVS}      & RV-    & $<$FWHM$>$ & $<$\rhk$>$ & V  & T & V \\ 
 HD/BD/CD & &  & & & \multicolumn{3}{c}{\raisebox{.5\baselineskip}{$\overbrace{\hspace{2.7cm}}$}} & \multicolumn{3}{c}{\raisebox{.5\baselineskip}{$\overbrace{\hspace{2.7cm}}$}} & BVS & & ($\sigma_{logR'_{\rm HK}}$) & & &\\
&& &  &  & {\it A} &rms  & $<${\it U}$>$ & {\it A} & rms &$<${\it U}$>$ & corr.    &                 &    & & &\\
&&&&    & \multicolumn{3}{c}{\raisebox{.5\baselineskip}{$\underbrace{\hspace{2.7cm}}$}} & \multicolumn{3}{c}{\raisebox{.5\baselineskip}{$\underbrace{\hspace{2.7cm}}$}} & & & & & \\
&& (day)&            && \multicolumn{3}{c}{($\si{\meter\per\second}$)}      & \multicolumn{3}{c}{($\si{\meter\per\second}$)}     &   &      ($\si{\kilo\meter\per\second}$) &   & & &\\ \hline
\endhead
\hline
\endfoot
HD105 & 490 & G0V & 4606 & 36 & 236.8 & 61.1 & 4.2 & 310.7 & 9.8 & 72.9 & -0.67 & 22.6 & -4.32(0.02) & A & - &- \\ 
HD984 & 1134 & F7V & 867 & 21 & 301.9 & 84.6 & 16.4 & 571.8 & 38.2 & 137.4 & -0.46 & 59.7 & -4.37(0.02) & A & - &- \\ 
HD987 & 1113 & G8V & 2621 & 19 & 502.6 & 116.8 & 2.5 & 393.6 & 6.4 & 103.1 & -1.08 & 13.0 & -4.09(0.04) & A & - &- \\ 
HD1466 & 1481 & F8V & 4400 & 19 & 135.8 & 39.6 & 7.4 & 189.8 & 17.6 & 55.9 & -0.66 & 32.3 & -4.34(0.02) & A & - &- \\ 
HD3221 & 2729 & K4V & 4014 & 5 & 4793.4 & 1794.3 & 68.6 & 3115.1 & 172.5 & 1447.8 & -0.54 & 203.6 & -4.47(0.01) & - & - &- \\ 
HD6569 & 5191 & K1V & 8 & 4 & 24.0 & 11.4 & 1.5 & 17.8 & 3.8 & 7.5 & -1.49 & 9.6 & -4.229(0.003) & A & - &- \\ 
HD7661 & 5938 & K0V & 1525 & 29 & 96.0 & 28.1 & 1.2 & 50.5 & 3.3 & 12.5 & -1.42 & 9.2 & -4.33(0.05) & A & - &- \\ 
HD10008 & 7576 & K0V & 4021 & 17 & 20.5 & 6.2 & 1.0 & 17.5 & 2.7 & 5.2 & 0.44 & 7.5 & -4.36(0.03) & A & - &- \\
\end{longtable}
\endgroup
\end{landscape}

\clearpage

\twocolumn

\section{Age and mass estimations}

\label{age_mass}
First, the ages and masses of the targets of the \harps \ YNS, the \sophie \ YNS, and the \harps \ Sco-Cen surveys were taken from the literature when available.
We estimated the ages and mass of the remaining ones. Whenever  possible, we used the methods described in \cite{Desidera} and in \cite{Desidera_shine}, then in a homogeneous scale with the above works. Briefly, we considered membership to groups and associations (adopting the group ages from \cite{Bonavita}, also discussed in the upcoming paper), indirect indicators such as rotation, chromospheric and coronal activity, and $\SI{6708}{\angstrom}$  equivalent width, complemented by isochrone fitting. Preference was given to the moving-group criterion whenever possible (confirmed members). For field objects,  the weights assigned to the various methods depend on colors or spectral types and age range (e.g., saturation of chromospheric activity and coronal emission versus age below $100-\SI{150}{\mega\year}$; high sensitivity of lithium to age for K dwarfs younger than $300-\SI{500}{\mega\year}$, with limits only at older ages). Masses were derived using the PARAM interpolation code\footnote{\url{http://stev.oapd.inaf.it/cgi-bin/param_1.3} } \citep{Da_Silva} as in \cite{Desidera}.
For the remaining targets,  masses were estimated from an empirical $M_*=f(\bv)$ relation. The spectral types were first  estimated from  the \bv \ using the relation from \cite{Lang} (page 564),  and the masses were thereafter deduced from the spectral types by using the relation from \cite{Allen} (page 209).

\section{Star with companion data}

We present in this appendix the data for the stars with a companion  in the \harps \ Sco-Cen survey (\emph{cf.} \Cref{comp}): RV time series, BVS versus RV diagram, residuals of the companion fit, and bisectors or CCF.

\begin{figure*}[h!]
  \centering
\includegraphics[width=0.8\hsize]{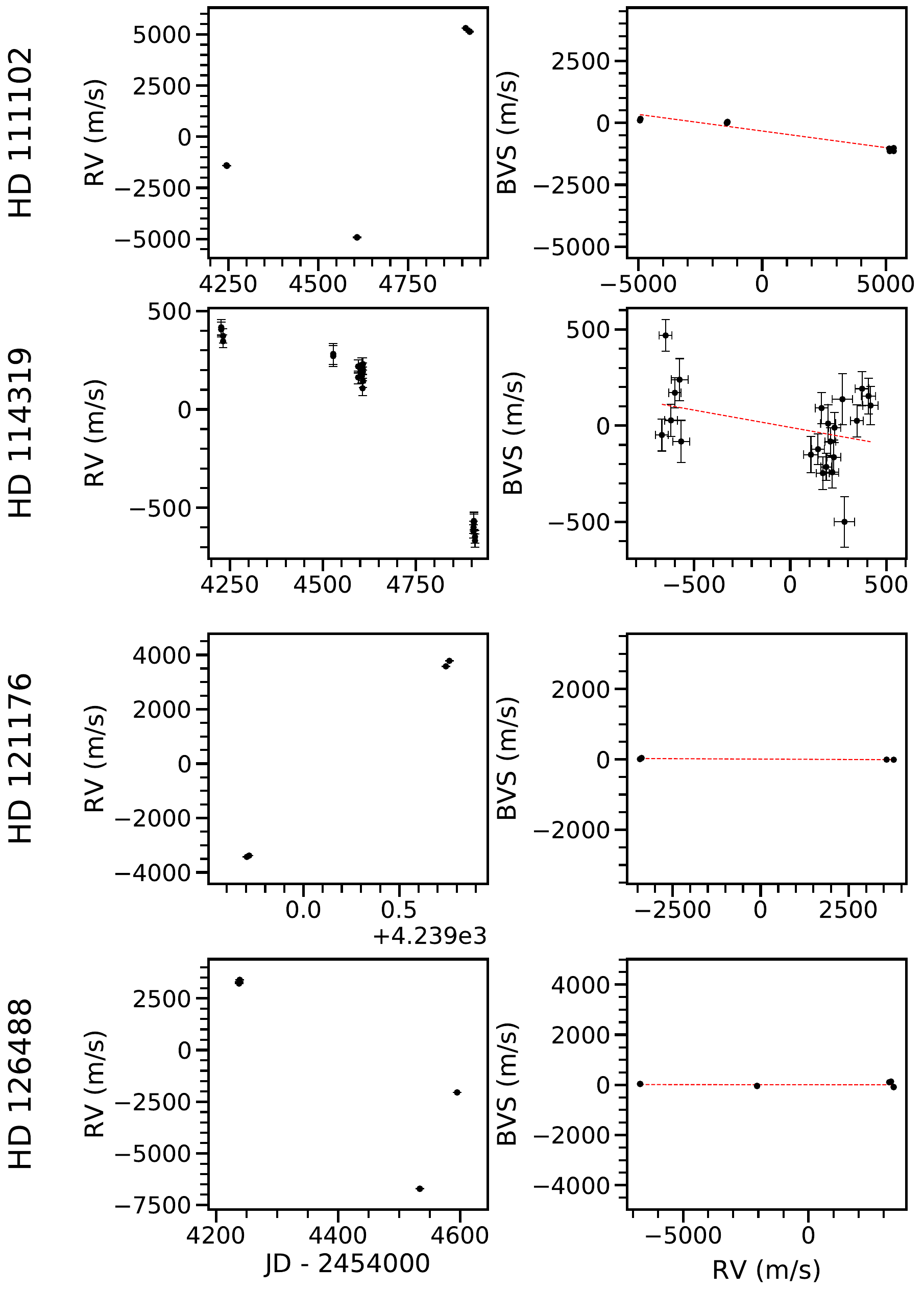}
\caption{Summary of binaries for which orbits could not be derived. \emph{First column}: RV time variations. \emph{Second column}: BVS vs. RV diagram (black) and its best linear model (red dashed line).} 
       \label{binary_noncarac}
\end{figure*}

\clearpage

\begin{figure*}[h]
\centering
\begin{subfigure}[t]{0.32\textwidth}
\centering
\includegraphics[width=1\hsize,valign=m]{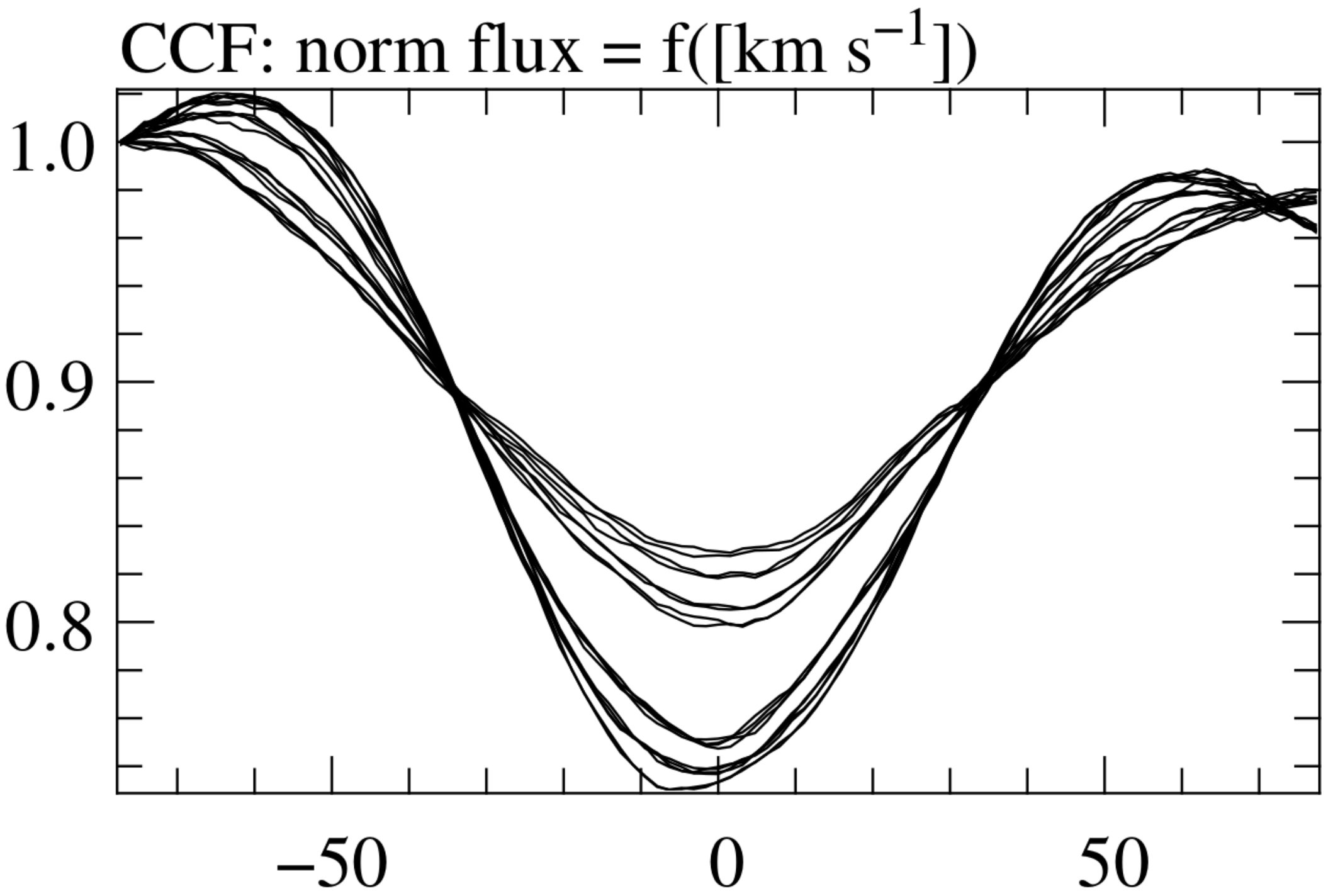}
\caption{\label{129590_ccf}}
\end{subfigure}
\begin{subfigure}[t]{0.32\textwidth}
\centering
\includegraphics[width=1\hsize,valign=m]{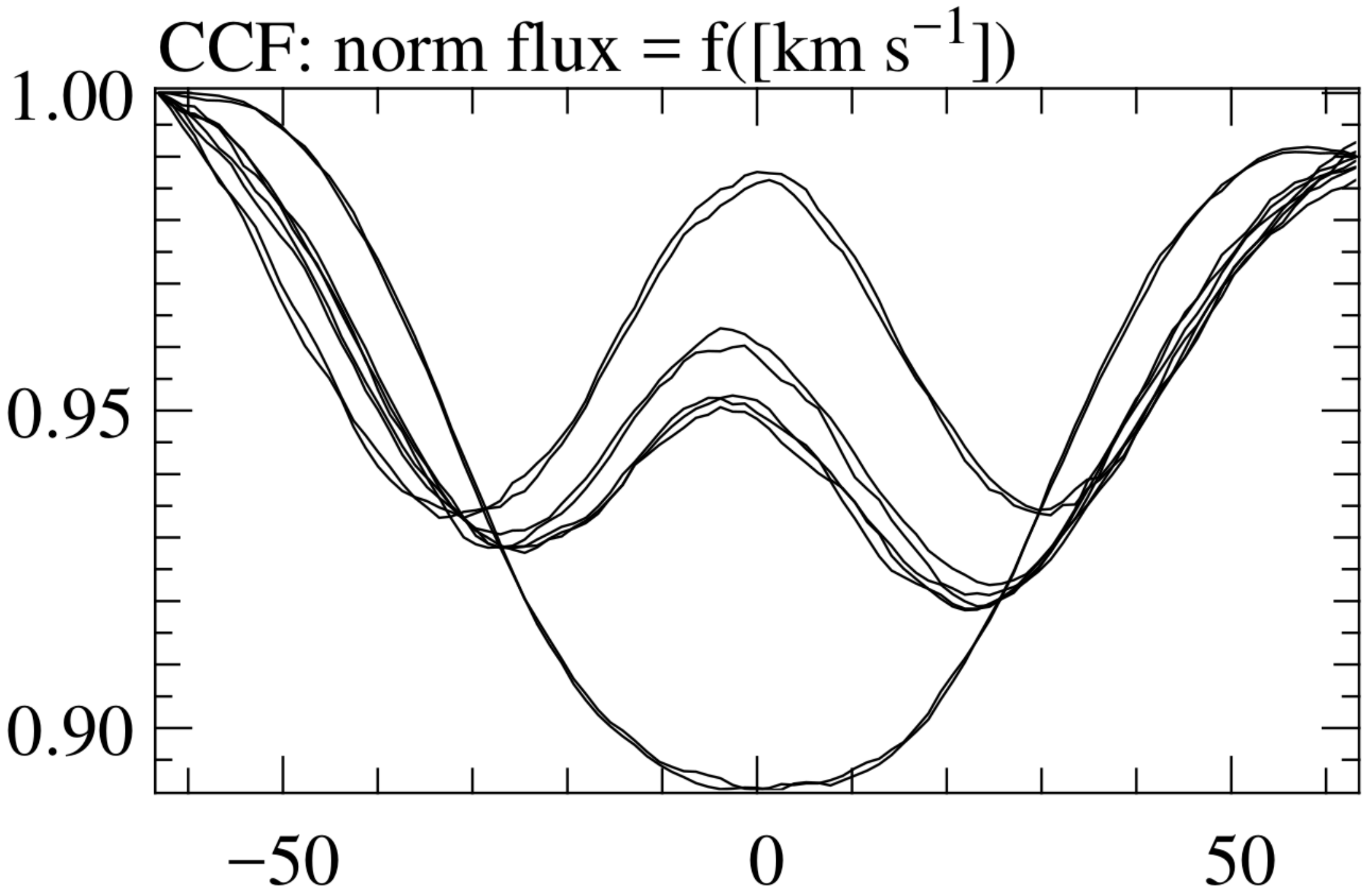}
\caption{\label{137057_ccf}}
\end{subfigure}
\begin{subfigure}[t]{0.32\textwidth}
\centering
\includegraphics[width=1\hsize,valign=m]{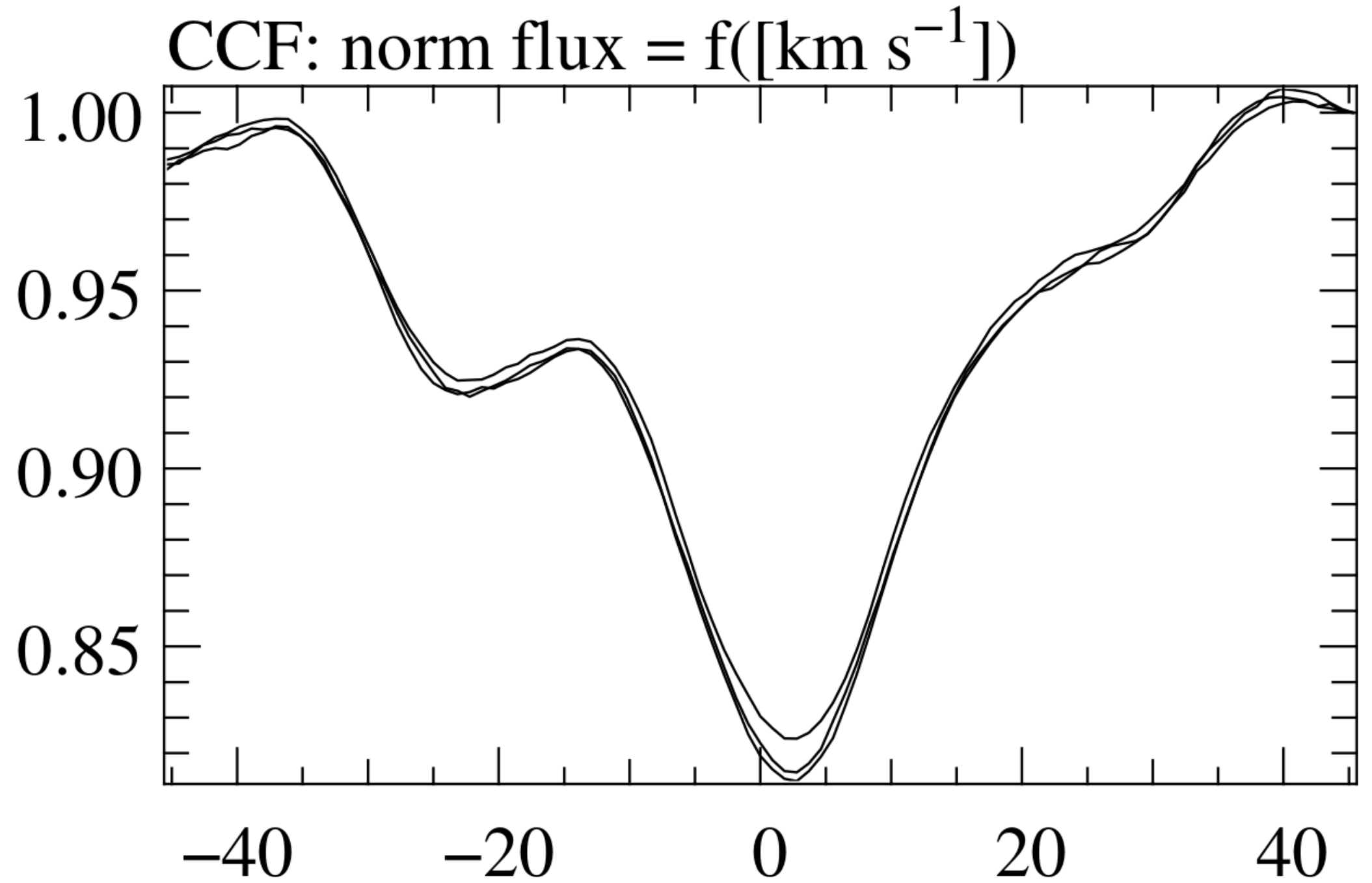}
\caption{\label{143811_ccf}}
\end{subfigure}

\caption{CCFs of HD  129590 (\subref{129590_ccf}), of HD 137057 (\subref{137057_ccf}), and of HD 143811 (\subref{143811_ccf}). For these three stars, each curve corresponds to a CCF derived from one of their spectra. \label{sco_ccf}}
\end{figure*}

\begin{figure*}[h]
\centering

\begin{subfigure}[t]{0.49\textwidth}
\centering
\includegraphics[width=0.8\hsize,valign=m]{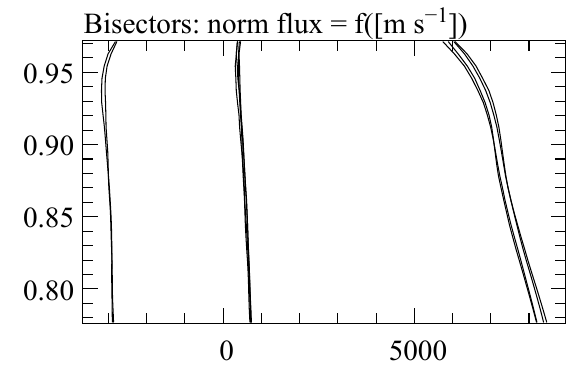}
\caption{\label{111102_bis}}
\end{subfigure}
\begin{subfigure}[t]{0.49\textwidth}
\centering
\includegraphics[width=0.8\hsize,valign=m]{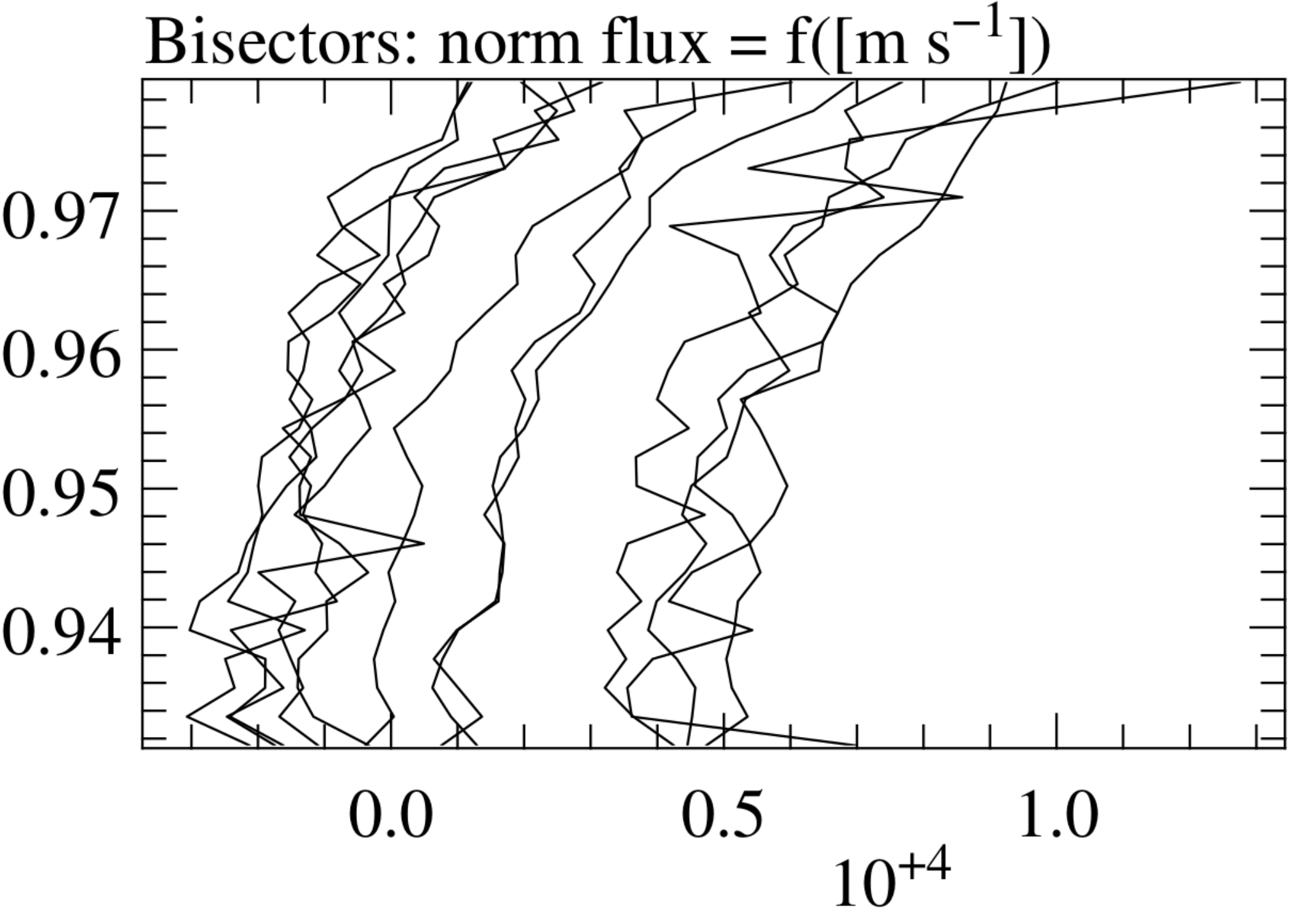}
\caption{\label{149790_bis}}
\end{subfigure}

\caption{Bisectors of HD 111102 (\subref{111102_bis}) and HD 149790 (\subref{149790_bis}).\label{sco_bis}}
\end{figure*}

\begin{figure*}[b]
  \centering
\includegraphics[width=1\hsize]{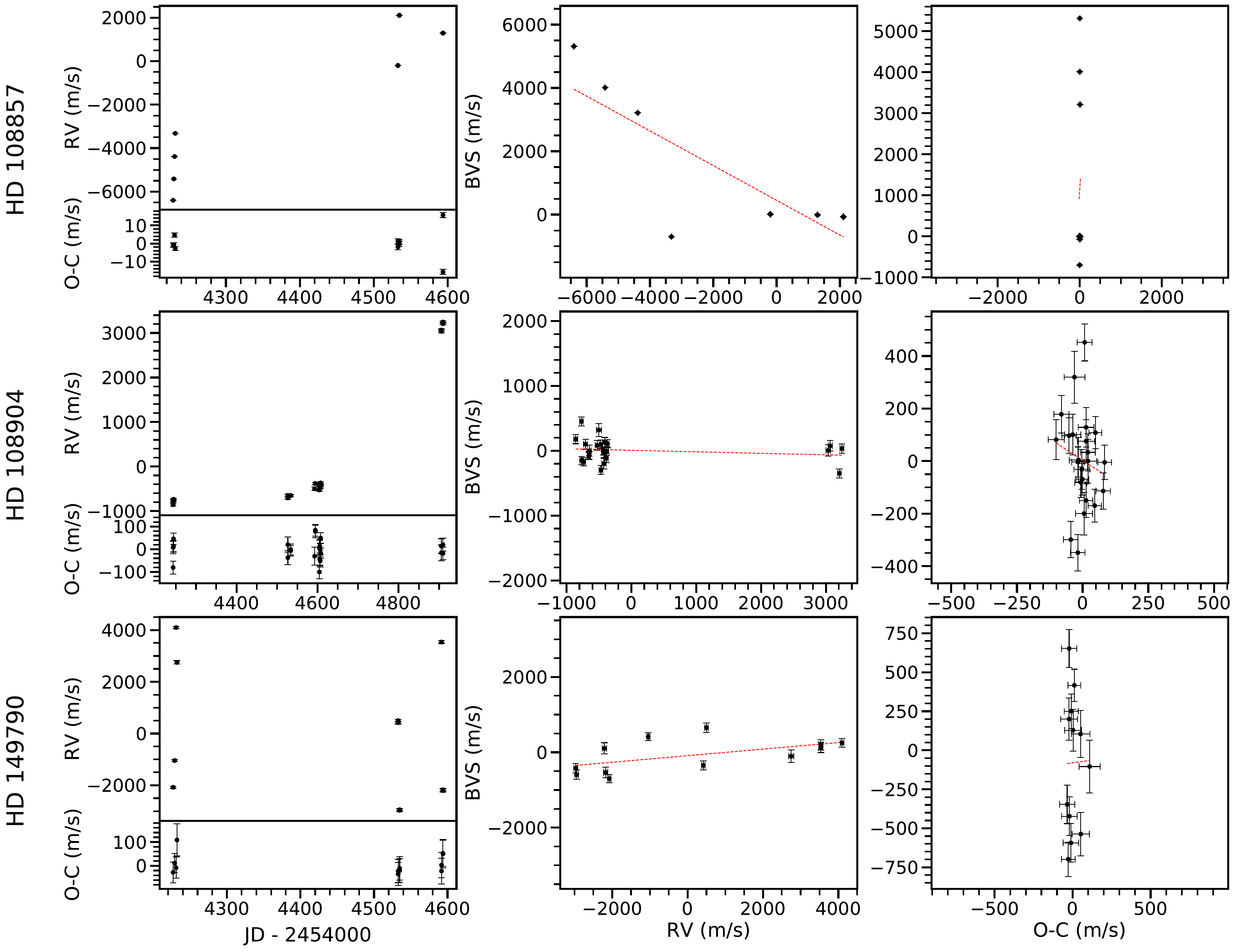}
\caption{RV of the characterized binaries. \emph{First column}: RV time variations  (\emph{top}) and  Keplerian fit residuals. The fits are presented in \Cref{108857} for HD 108857, in \Cref{108904} for HD 108904, and in \Cref{149790} for HD 149790. \emph{Second column}: BVS vs. RV diagram (black) and its best linear model (dashed red line).  \emph{Third column}: BVS vs. RV residuals diagram and its best linear model (dashed red line).} 
       \label{binary}
\end{figure*}

\begin{figure*}[t]
  \centering
\includegraphics[width=1\hsize]{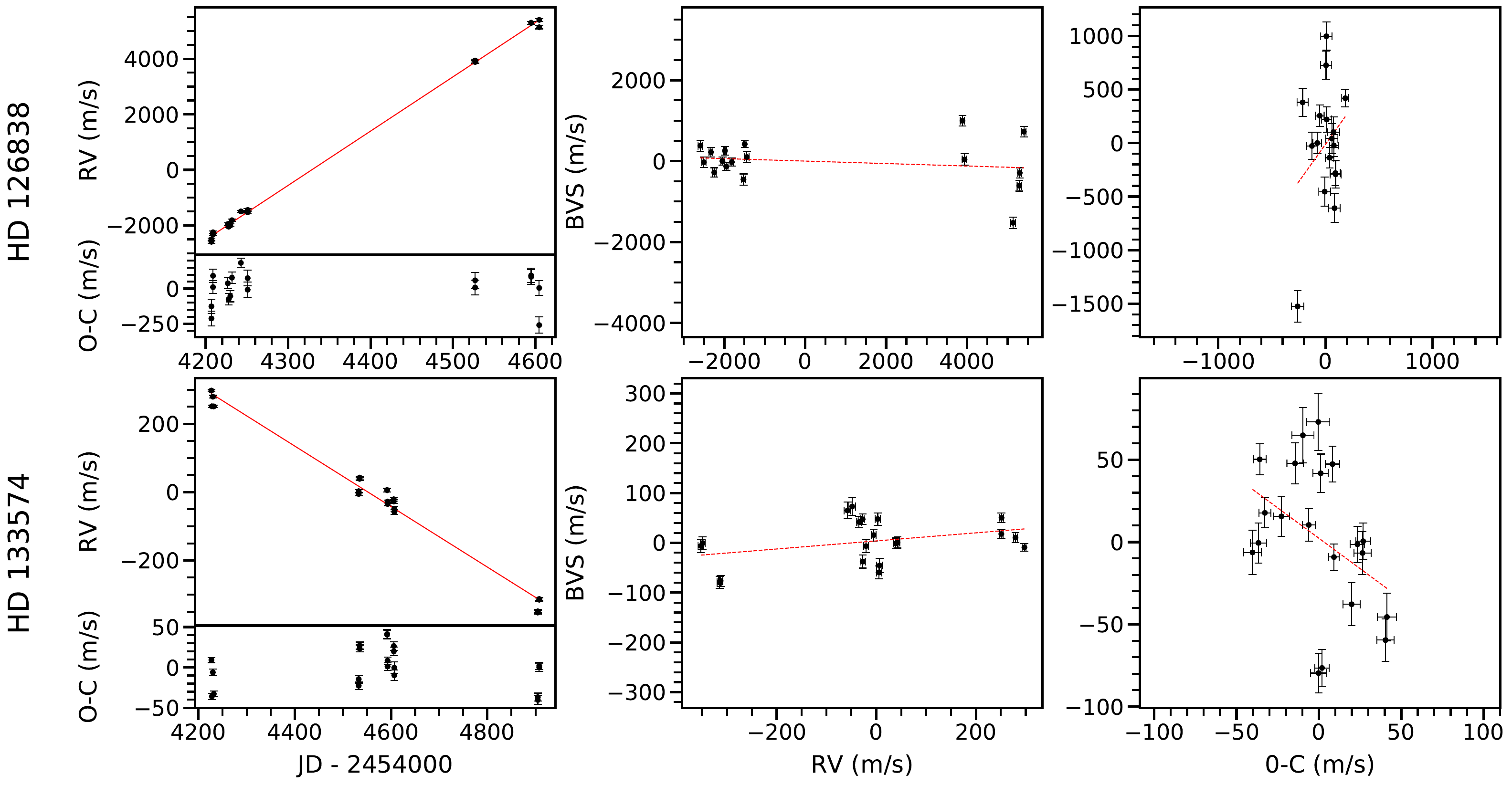}
\caption{Stars with RV trend  for which a linear regression was performed. \emph{First column}: \emph{top}: RV time variations (black) with the model of its linear regression (red line).  \emph{Bottom}: Residuals of the linear regression.  \emph{Second column}: BVS vs. RV diagram (black) and its best linear model (dashed red line).  \emph{Third column}: BVS vs. RV residuals diagram and its best linear model (dashed red line). \label{trend}} 
\end{figure*}

\begin{figure*}[t]
  \centering
\includegraphics[width=1\hsize]{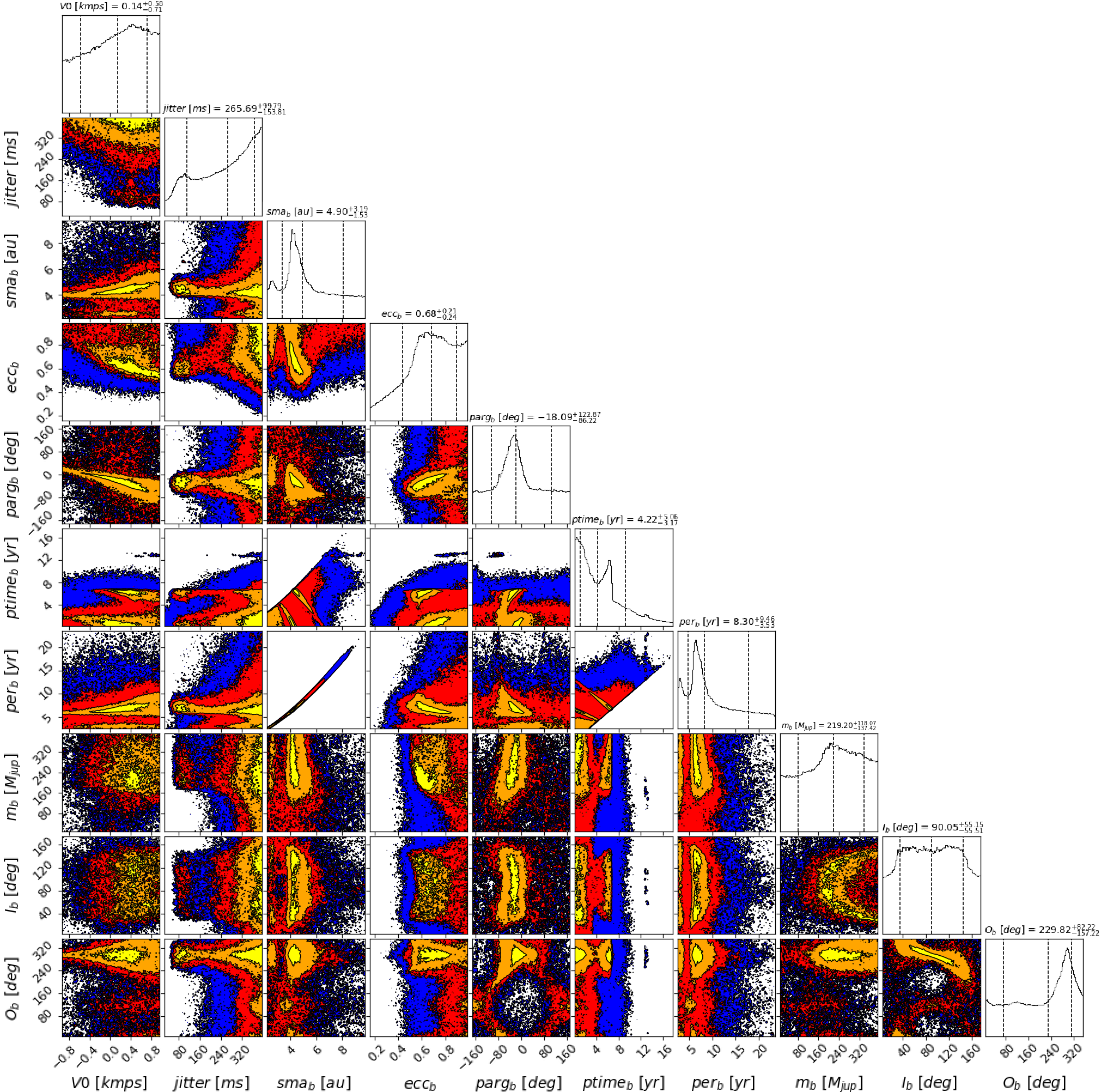}
\caption{Posterior distributions of the MCMC fit performed on the HD 108904 RV and HD 108904 B relative astrometry simultaneously. \emph{From top to bottom:} RV offset to take into account that the RVs are relative to a reference ($V_0$), RV jitter amplitude ($jitter$), semimajor axis ($sma_b$), eccentricity ($ecc_b$), argument of periastron ($parg_b$), period ($per_b$), mass ($m_b$), inclination ($I_b$), and longitude of the ascending node ($O_b$) of HD 108904 B. \label{corner_108904}} 
\end{figure*}

\clearpage

\section{Statistical tests}
\label{test_stat}

We present the two statistical tests  we used to determine whether the populations of companions studied by different surveys are statistically different (and therefore whether their occurrence rates are statistically different): the p-value test, and the pooled version of the two-proportion Z-test. 

These tests are evaluated under a null hypothesis; if a test is validated, then the null hypothesis can be rejected. For these tests, we used the following null hypothesis: the occurrence rate of the two compared companion populations in the studied ($P$, $M_c\sin{i}$) box are identical. Thus, if a test is successful and therefore the null hypothesis is rejected, we can deduce that it is unlikely that the occurrence rates of the  two compared populations are identical.

For these tests, we assimilated the surveys performed to constrain the companion occurrence rates to Bernoulli tests  $\mathcal{B}(n,p)$, with  $n$ the number of  targets for a given survey and  $p$ the actual companion occurrence rate.

\subsection{p-value test}

The $p_{value}$ test is a statistical test that evaluates the probability of obtaining  the number of observed successes by chance for a given null hypothesis. If this probability is below a validation criterion $\alpha$, the test is validated.

 In our case, we compared two Bernoulli tests ($\mathcal{B}_1(n_1,p_1)$ and $\mathcal{B}_2(n_2,p_2)$) under the following null hypothesis: $p_1 = p_2$. To perform this test, we computed the probability of obtaining the results of the first test ($k_1$ success among $n_1$ observations), knowing the probability $p_1 = p_2$:

\begin{displaymath}
p_{value} = \dbinom{n}{k}p_2^{k_1}(1-p_2)^{n_1-k_1}
\end{displaymath}
with  $\dbinom{n}{k}$ the  binomial coefficient of $k$ among $n$.

If $p_{value} <\alpha$, the probability of observing the results of the first Bernouilli test under the null hypothesis by chance is below $\alpha,$ and the latter can therefore be rejected with a confidence level of $1-\alpha$.

\subsection{Pooled version of the two-proportion Z-test}

The pooled version of the two-proportion Z-test is a parametric test that allows testing a difference in proportion of successes between two Bernoulli process ($\mathcal{B}_1(n_1,p_1)$ and $\mathcal{B}_2(n_2,p_2)$, respectively). 

This test can be used only if the two compared samples are large enough for their probability of success to follow a normal law (as the binomial distribution tends toward a normal distribution when $n$ tends toward infinity).
Many criteria are used to ensure it.  The most widely used is $n_{1,2}>30$, but stricter criteria are also used, for example, $n_{1,2}p_{1,2}>5; n_{1,2}(1-p_{1,2})>5$, or $n_{1,2}p_{1,2}(1-p_{1,2})>9$. In our case, only the more relaxed criterion, $n_{1,2}>30$, is validated. 

Then, if these criteria are validated,  the difference in the proportion of successes between two Bernoulli process also follows a normal distribution ($\mathcal{N}(\mu,\sigma^2)$).
Moreover, in the case of our null hypothesis $p_1=p_2$, the mean of the normal distribution equals $0$.
Thus, to test this null hypothesis, we computed the actual difference of the proportion of success between the two compared Bernoulli processes, and we tested whether it actually followed a centered normal distribution ($\mathcal{N}(0,\sigma^2)$). This is the case when

\begin{equation}
\frac{|p_1-p_2|}{\sqrt{p(1-p)(\frac{1}{n_1}+\frac{1}{n_2}})} > \mathcal{U}_{1 - \frac{\alpha}{2}}
\label{ztest}
,\end{equation}

with $p = \frac{n_1 f_1 + n_2 f_2}{n_1+n_2}$ and with $\mathcal{U}_{1 - \frac{\alpha}{2}}$ the quantile of order $1 - \frac{\alpha}{2}$ of the reduced centered normal distribution.

If \Cref{ztest} is verified, the test is validated and the null hypothesis ($p_1=p_2$) can be rejected with a confidence level of $1-\alpha$.

\end{appendix}

\end{document}